\documentclass[aps,prl,twocolumn,amsmath,amssymb,floatfix,superscriptaddress]{revtex4-1}
\usepackage{graphicx}
\usepackage{bm}
\usepackage[hidelinks]{hyperref}

\usepackage{braket,bbold,color,enumerate,multirow}

\newcommand{\bs}[1]{\boldsymbol{#1}}
\begin{document}

\title{Non-Compact Atomic Insulators}

\author{Frank Schindler}
\affiliation{Princeton Center for Theoretical Science, Princeton University, Princeton, NJ 08544, USA}

\author{B. Andrei Bernevig}
\affiliation{Department of Physics, Princeton University, Princeton, NJ 08544, USA}

\begin{abstract}
We study the conditions for Bloch bands to be spanned by symmetric and strictly compact Wannier states that have zero overlap with all lattice sites beyond a certain range. Similar to the characterization of topological insulators in terms of an algebraic (rather than exponential) localization of Wannier states, we find that there may be impediments to the compact localization even of topologically ``trivial" obstructed atomic insulators. These insulators admit exponentially-localized Wannier states centered at unoccupied orbitals of the crystalline lattice. First, we establish a sufficient condition for an insulator to have a compact representative. Second, for $\mathcal{C}_2$ rotational symmetry, we prove that the complement of fragile topological bands cannot be compact, even if it is an atomic insulator. Third, for $\mathcal{C}_4$ symmetry, our findings imply that there exist fragile bands with zero correlation length. Fourth, for a $\mathcal{C}_3$-symmetric atomic insulator, we explicitly derive that there are no compact Wannier states overlapping with less than $18$ lattice sites. We conjecture that this obstruction generalizes to all finite Wannier sizes. Our results can be regarded as the stepping stone to a generalized theory of Wannier states beyond dipole or quadrupole polarization.
\end{abstract}

\maketitle

In band theory, Wannier states are the Fourier transforms of Bloch states. They have multifold applications ranging from chemical bonding to ab-initio calculations. The gauge freedom in defining Bloch states can be exploited to construct maximally localized Wannier states~\cite{Wannier37,Kohn59,Cloizeaux63,VanderbiltReview}. Recently, the notion of topological insulators was reformulated in terms of an obstruction to exponentially-localized Wannier states, which allowed for a systematic classification of topological band structures in all symmetry classes~\cite{Po_2017,Bradlyn17,Watanabeeaat8685,elcoro2020magnetic}. In this Letter, we study Wannier states satisfying an even more stringent localization requirement. These are \emph{compact Wannier states} that are strictly local and have zero overlap with all lattice sites outside of a finite domain. They are symmetric when they share the symmetries of the lattice restricted to the site-symmetry group that leaves their Wannier center invariant~\cite{ZakEBR1,ZakEBR2,Slager17,Po_2017,Bradlyn17,Song20,Alexandradinata20}. In the following, we assume that all Wannier states are symmetric. Moreover, we require that compact states originating from different unit cells are orthogonal~\cite{Sathe_2021}. This criterion was not enforced in previous works, which instead studied compact localized states (compact Wannier-type states)~\cite{Aoki96,Leon08,DubailRead15,Read17,Flach17,Maimaiti17,Rontgen18,FlachReview,Lazarides19,Rhim19,Zhang20} that need not be orthogonal.

A Bloch band induced from a delta-function Wannier state at any atomic site of the unit cell -- resulting in a trivial or \emph{unobstructed} atomic insulator~\cite{Po_2017,Bradlyn17} -- can always be adiabatically transformed to have compact Wannier states. Conversely, a topological band cannot -- by definition -- be written in terms of exponentially-localized Wannier states, much less compact ones~\cite{Brouder2007Exponential,Soluyanov2011Wannier,Taherinejad14,Budich14,Bradlyn17,Schindler20}. The same holds for fragile topological bands that can be trivialized upon mixing with non-topological bands~\cite{AshvinFragile,AdrianFragile,BarryFragile,ZhidaFragileTwist2,Else19,Ahn19,Hwang19,ZhidaFragileAffine}.

There is so far one known category of insulators allowing for exponentially-localized Wannier states which are necessarily not delta-function-like: delicate topological insulators~\cite{nelson2020multicellularity}, which are characterized by Hopf invariants and returning Thouless pumps. Here, we explore a second category of non-delta-function insulators that are \emph{obstructed} atomic insulators (OAIs)~\cite{Alexandradinata14,Benalcazar16,Benalcazar17,Song17,Po_2017,Bradlyn17,benalcazar2018quantization}, whose Wannier states may only be exponentially-localized away from the atomic orbitals. Surprisingly, we find that not all OAIs have a compact representation: there are topological obstructions to compactness. We call the resulting phases non-compact atomic insulators. The condition of non-compactness is stronger than the ``multicellularity" of delicate topological insulators, meaning that the Wannier states cannot be completely localized in a primitive unit cell: non-compact Wannier states cannot be completely localized in any, potentially non-primitive, unit cell. (Presently, it is not known if delicate topological insulators ultimately satisfy the stronger condition.) While the general theory of non-compact atomic insulators is still outstanding, our paper proves their existence.

\begin{figure}[t]
\centering
\includegraphics[width=0.48\textwidth]{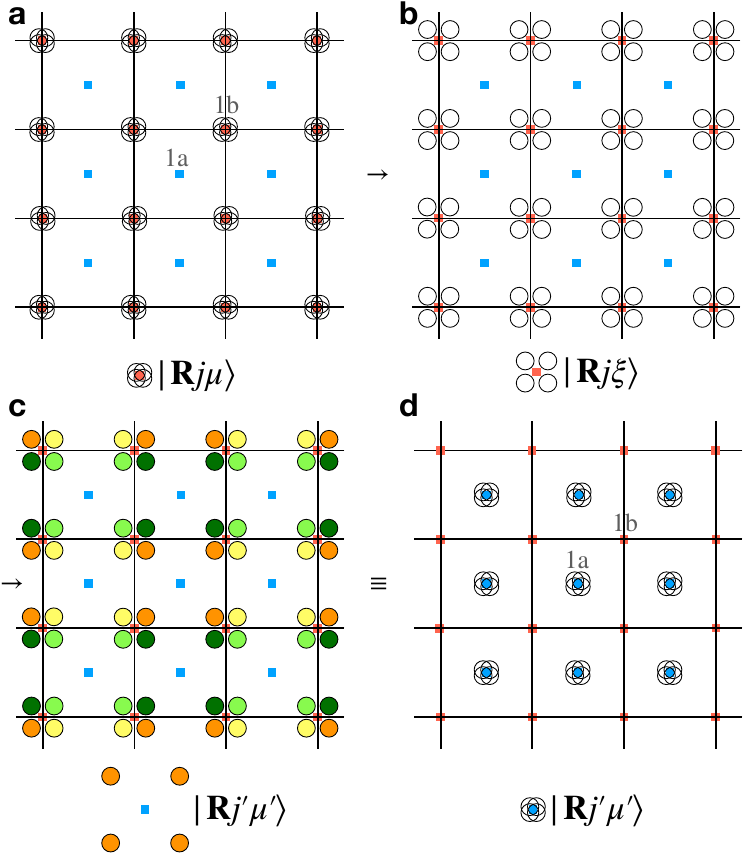}
\caption{Mobile clusters with $\mathcal{C}_4$ symmetry. A mobile cluster is a minimal set of physical orbitals $\ket{\bs{R} j \mu}$ whose transformation behavior under the crystalline symmetry is compatible with being located at any (possibly non-maximal) Wyckoff position. For spinless $\mathcal{C}_4$ symmetry, the mobile clusters on maximal Wyckoff positions 1a and 1b contain four orbitals with $\mathcal{C}_4$ eigenvalues $\{1,-1,\mathrm{i},-\mathrm{i}\}$. (a)~We begin with mobile clusters centered at the $1b$ Wyckoff position. (b)~Next, we locally change bases to obtain the states $\ket{\bs{R} j \xi}$ that do not have well-defined $\mathcal{C}_4$ eigenvalues, but instead are cyclically permuted by the action of $\mathcal{C}_4$. (c,~d)~These states can be used to construct new mobile clusters $\ket{\bs{R} j' \mu'}$ that are centered around Wyckoff position $1a$.}
\label{fig: mobile_cluster_transform}
\end{figure}

\begin{figure}[t]
\centering
\includegraphics[width=0.48\textwidth]{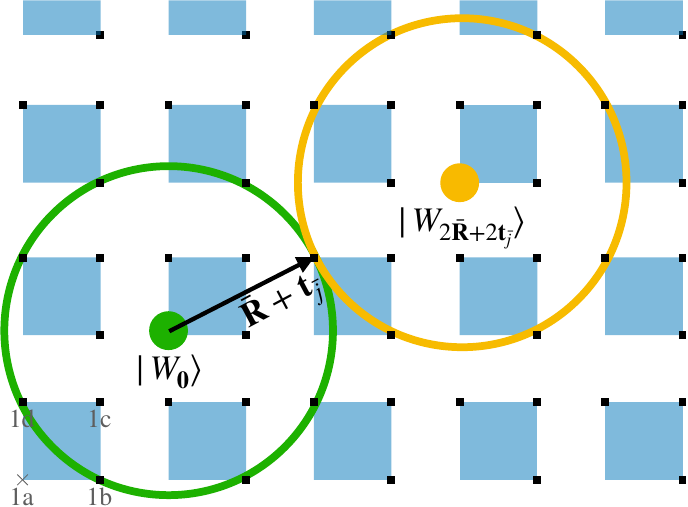}
\caption{Overlap of compact trial Wannier states with $\mathcal{C}_2$ rotational symmetry. For any compact state $\ket{W_1} = \ket{W_{\bs{0}}}$ of an OAI, there is another translated Wannier state $\ket{W_2} = \ket{W_{2\bar{\bs{R}}+2\bs{t}_{\bar{j}}}}$ that shares a single lattice site of non-zero overlap. Orthogonality $\braket{W_1 | W_2} = 0$ is impossible when this site carries a single orbital. (Here, the OAI is centered at Wyckoff position 1a, while the atomic orbitals locate at 1b, 1c, 1d.)}
\label{fig: compactInvObstruction}
\end{figure}

\emph{Compact Wannier states---} We denote the atomic orbitals on a lattice with space group $G$ by $\ket{\bs{R} j \mu}$. Here, $\bs{R}$ indicates the unit cell coordinate, while $j$ labels the atomic site $\bs{t}_j \in A$ within the unit cell. We assume all atomic sites contained in $A$ to be maximal Wyckoff positions~\cite{Bradlyn17,Cano17-2}. The index $\mu$ labels the orbitals at a given site, which respect the site-symmetry group. To form Wannier states for an OAI, we construct obstructed orbitals at the positions $\bs{t}_\alpha \in B$:
\begin{equation} \label{eq: firstAnsatzWannier}
\ket{W_{\bs{R}, \alpha}} = \sum_{\bs{R}' j \mu} S_{j \mu, \alpha} (\bs{R}-\bs{R}') \ket{\bs{R}' j \mu}.
\end{equation} 
If $B \cap A =\emptyset$, the OAI has a \emph{spatial} obstruction, in that its Wannier states are centered at empty positions of the crystalline lattice. If $B \cap A \neq \emptyset$, the OAI has a \emph{representation} obstruction, in that the transformation behavior of its Wannier states under the crystalline symmetry differs from that of all atomic orbitals present at the same site. While spatial and representation obstructions were treated on equal footing in previous works~\cite{Po_2017,Bradlyn17,AshvinFragile,ZhidaFragileTwist2,ZhidaFragileAffine}, we must distinguish between them when studying the real-space structure of Wannier states. The functions $S_{j \mu, \alpha}(\bs{R}-\bs{R}') \in \mathbb{C}$ must respect the space group symmetry. Furthermore, the states $\ket{W_{\bs{R}, \alpha}}$ have compact support when $S_{j \mu, \alpha}(\bs{R}-\bs{R}')$ is strictly zero for all $|\bs{R}+\bs{t}_\alpha-\bs{R}'-\bs{t}_j|$ greater than a certain distance. For the obstructed orbitals $\ket{W_{\bs{R}, \alpha}}$ to form a Wannier basis, they must also be orthonormal:
\begin{equation} \label{eq: fullcompactnessconstraints}
\braket{W_{\bs{R},\alpha} | W_{\bs{R'},\beta}} = \delta_{\bs{R}\bs{R'}} \delta_{\alpha \beta}.
\end{equation}
The interplay between orthogonality, symmetry, and compact support is already nontrivial in two-dimensional systems where $G$ contains a single $\mathcal{C}_n$ rotation (and translations), which we focus on in the following.
\begin{figure*}[t]
\centering
\includegraphics[width=0.99\textwidth]{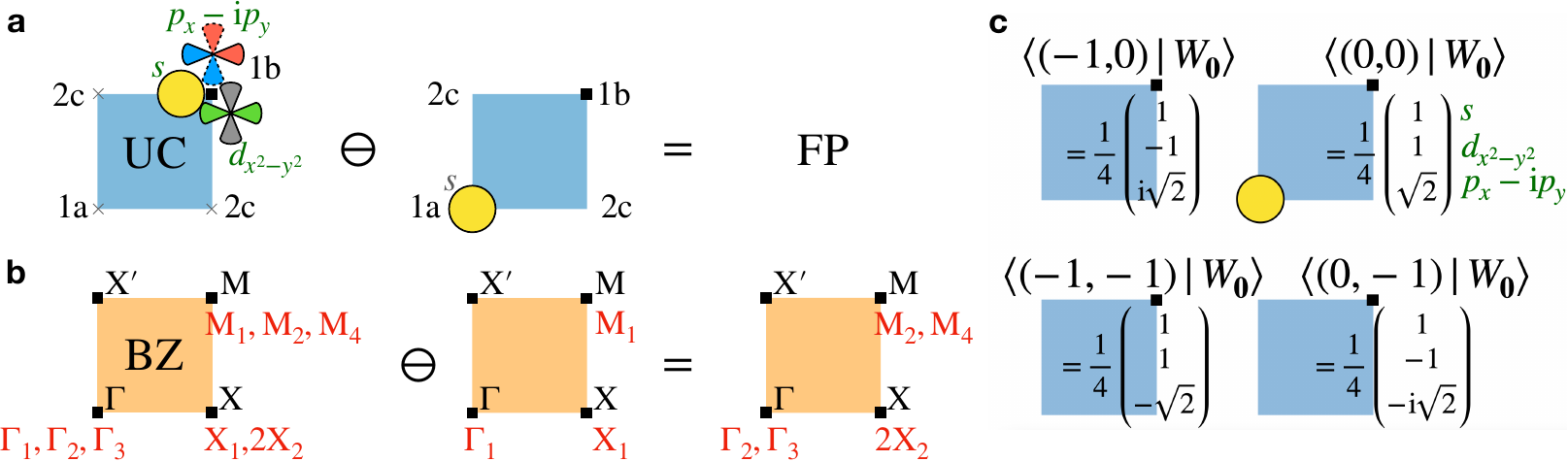}
\caption{Compact OAI with $\mathcal{C}_4$ rotational symmetry and fragile complement. (a)~Real-space illustration of the fragile state. The unit cell (UC) contains three orbitals at Wyckoff position 1b ($s$, $d_{x^2-y^2}$, and $p_x-\mathrm{i}p_y$ orbitals carry $\mathcal{C}_4$ eigenvalues $1$, $-1$, and $\mathrm{i}$, respectively). (b)~Brillouin zone (BZ) decomposition of the corresponding Bloch bands into irreducible representations. (Representation labels follow the Bilbao Crystallographic Server~\cite{Aroyo2011183}.) (c)~Compact Wannier state for the OAI, supported on four unit cells (blue) surrounding the 1a position (yellow) of the unit cell at $\bs{R} = \bs{0}$. The vectors $\braket{\bs{R}|W_{\bs{0}}}$ at each site contain the overlaps $\left(\braket{\bs{R}|W_{\bs{0}}}\right)_\mu \equiv \braket{\bs{R},\mathrm{1b},\mu|W_{\bs{0}}}$ of the Wannier state $\ket{W_{\bs{0}}}$ with the on-site orbitals at Wyckoff position 1b, which are indexed by $\mu$ (orbital labels are shown in green).
}
\label{fig: compactC4counterexample}
\end{figure*}
Assuming spinless rotational symmetries, so that $(\mathcal{C}_n)^n = 1$, the rotation eigenvalues $\gamma_\mu$ take values
\begin{equation} \label{eq: mc_orb_eigvals}
\gamma_\mu = e^{\mathrm{i}\frac{2\pi}{n}l}, \quad l = 0 \dots n-1.
\end{equation}
We call a \emph{mobile cluster} a configuration of orbitals whose $\mathcal{C}_n$ eigenvalues exhaust all $l = 0 \dots n-1$, with each $l$ appearing exactly once. These configurations are special in that they can be used to construct compact basis states at any Wyckoff position, not just at the atomic positions hosting the mobile cluster orbitals~\cite{Bradlyn17, Cano17-2}. For instance, given that $\ket{\bs{R} j \mu}$, $\mu = 1 \dots n$ is a mobile cluster, there exists a strictly local unitary effecting 
\begin{equation} \label{eq: mobileclustermagic}
\ket{\bs{R} j \mu} \quad \rightarrow \quad \ket{\bs{R} j' \mu'},
\end{equation}
where $j$, $j'$ label two Wyckoff positions with $\mathcal{C}_n$ symmetry, and $\mu'$ labels a new set of orbitals that also forms a mobile cluster~\footnote{See Supplemental Material at [URL will be inserted by publisher] for supporting derivations.}. (See Fig.~\ref{fig: mobile_cluster_transform}.)

Fragile phases are the band complements of other fragile phases or OAIs~\cite{AshvinFragile,AdrianFragile,BarryFragile,ZhidaFragileTwist2,Else19,Ahn19,Hwang19,ZhidaFragileAffine}. In the latter case, they are a difference of atomic insulators:
\begin{equation} \label{eq: first-model-subtraction}
\mathrm{FP} = \mathrm{AI} \ominus \mathrm{OAI},
\end{equation}
where $\mathrm{FP}$ denotes the fragile phase, and $\mathrm{AI}$ is the (unobstructed) atomic insulator induced from the lattice.
Now, let $N(\mathrm{AI})$ count the number of mobile clusters in the unit cell. That is, for every group of $n$ orbitals containing all eigenvalues in Eq.~\eqref{eq: mc_orb_eigvals} present in the unit cell of $\mathrm{AI}$, we increase $N(\mathrm{AI})$ by one, starting from zero. If only a part of the orbitals required for a mobile cluster is present (in addition to the orbitals already counted), $N(\mathrm{AI})$ is unaffected and remains integer-valued. For example, for the unit cell in Fig.~\ref{fig: mobile_cluster_transform}a, we have $N(\mathrm{AI})=1$. Furthermore, let $\bar{N}(\mathrm{OAI})$ count the minimal number of mobile clusters containing all orbitals of the OAI. That is, we envision an atomic limit $\tilde{\mathrm{AI}}$ whose unit cell contains all orbitals of $\mathrm{OAI}$ just once \emph{and} whose full set of orbitals can be grouped into mobile clusters without any missing or remaining orbitals: then, $\bar{N}(\mathrm{OAI}) = N(\tilde{\mathrm{AI}})$. Now, for $\mathrm{FP}$ to be fragile, we need $N(\mathrm{AI}) < \bar{N}(\mathrm{OAI})$: otherwise, the OAI could be built from a subset of the mobile cluster orbitals, potentially using Eq.~\eqref{eq: mobileclustermagic}, while the remaining orbitals form a compact and symmetric Wannier basis for $\mathrm{FP}$. Conversely, we see that $N(\mathrm{AI}) \geq \bar{N}(\mathrm{OAI})$ is a sufficient condition for the OAI to be compact \emph{and} to have an OAI (not fragile) complement. For instance, in wallpaper group $p2$, we have $N[((A)_\mathrm{1a} \oplus (B)_\mathrm{1a})\uparrow G] = \bar{N}[(A)_\mathrm{1b} \uparrow G] = 1$~\footnote{Here, we use the labelling conventions of the Bilbao Crystallographic Server for wallpaper group $p2$~\cite{Aroyo2011183}, so that $(A/B)_\mathrm{1x}$ is an $s/p$ orbital at Wyckoff position $\mathrm{1x}$.}, implying that the OAI $(A)_\mathrm{1b} \uparrow G$ has a compact representative, and so does its complement $(B)_\mathrm{1b} \uparrow G$.

\emph{$\mathcal{C}_2$ symmetry---} OAIs with $\mathcal{C}_2$ symmetry and a fragile complement are non-compact. Consider the OAI induced from an $s$ orbital on Wyckoff position 1a of wallpaper group $p2$ ($\bs{t}_{\mathrm{1a}} = \bs{0}$), where the lattice hosts $s$ orbitals located at Wyckoff positions 1b, 1c and 1d [so that $\bs{t}_j \in \{(1/2,0), (1/2,1/2), (0,1/2)\}$ is the position of the $j$th atomic $s$ orbital]. The complement of the OAI is~\cite{Note2}
\begin{equation} \label{eq: c2archetypicalSubtraction}
\left[(A)_\mathrm{1b} \oplus (A)_\mathrm{1c} \oplus (A)_\mathrm{1d} \right] \uparrow G \ominus (A)_\mathrm{1a} \uparrow G = \mathrm{FP}.
\end{equation}
We first note that $N[((A)_\mathrm{1b} \oplus (A)_\mathrm{1c} \oplus (A)_\mathrm{1d})\uparrow G] = 0$ (the unit cell does not contain a full mobile cluster) and $\bar{N}[(A)_\mathrm{1a} \uparrow G] = 1$ (we need at least one mobile cluster to reproduce the OAI). Therefore, the necessary condition $N(\mathrm{AI}) < \bar{N}(\mathrm{OAI})$ for a non-compact OAI with fragile complement is satisfied.
And indeed, $\mathrm{FP}$ in Eq.~\eqref{eq: c2archetypicalSubtraction} is the simplest possible fragile state, requiring the smallest crystalline symmetry (wallpaper group $p2$), and the smallest number of bands (two occupied and one empty band)~\cite{ZhidaFragileTwist2}. Eq.~\eqref{eq: c2archetypicalSubtraction} does not involve complex representations and is therefore compatible with (spinless) time-reversal symmetry (TRS). We will next show that the OAI $(A)_\mathrm{1a} \uparrow G$  is non-compact.
Let us assume that $\ket{W_{\bs{R}}}$ are compact Wannier states: then, the overlap $\braket{\bs{R}' j | W_{\bs{R}}}$ is nonzero only for a finite number of separations $|\bs{R}-\bs{R}'-\bs{t}_j|$. Moreover, $\mathcal{C}_2$ symmetry implies $C_2 \ket{W_{\bs{0}}} = \ket{W_{\bs{0}}}$, where $C_2$ represents a $\mathcal{C}_2$ rotation about Wyckoff position 1a of the unit cell at $\bs{R} = \bs{0}$. Now, consider an orbital $\ket{\bar{\bs{R}} \bar{j}}$ at maximal distance $|\bar{\bs{R}} + \bs{t}_{\bar{j}}|$ from the origin which still has a nonzero overlap $\braket{\bar{\bs{R}} \bar{j} | W_{\bs{0}}} \neq 0$ with $\ket{W_{\bs{0}}}$. Then, by $\mathcal{C}_2$ symmetry, it follows that
\begin{equation}
0 \neq \braket{\bar{\bs{R}} \bar{j} | C_2^\dagger C_2 | W_{\bs{0}}} = \braket{(-\bar{\bs{R}}-2\bs{t}_{\bar{j}}) \bar{j} | W_{\bs{0}}}
\end{equation}
is also nonzero. But this implies that $\braket{W_{2\bar{\bs{R}}+2\bs{t}_{\bar{j}}} | W_{\bs{0}}} \neq 0$, because these two Wannier functions have finite overlap on exactly one $s$ orbital, located at $\bar{\bs{R}} + \bs{t}_{\bar{j}}$. (See Fig.~\ref{fig: compactInvObstruction}.) We conclude that a compact set of Wannier states $\ket{W_{\bs{R}}}$ satisfying Eq.~\eqref{eq: fullcompactnessconstraints} cannot exist. This argument does not make any assumptions on the size of the Wannier states, as long as it is finite. Therefore, the OAI $(A)_\mathrm{1a} \uparrow G$ in Eq.~\eqref{eq: c2archetypicalSubtraction} is non-compact. In the Supplemental Material (SM)~\cite{Note1}, we show that in fact \emph{all} $\mathcal{C}_2$-symmetric OAIs with fragile complement are non-compact.

\emph{$\mathcal{C}_4$ symmetry---} Any OAI that is non-compact with $\mathcal{C}_2$ symmetry remains non-compact when the symmetry group is enlarged to contain $\mathcal{C}_4$ rotations: because $(\mathcal{C}_4)^2 = \mathcal{C}_2$, $\mathcal{C}_4$-symmetric compact Wannier states inherit the constraints imposed by $\mathcal{C}_2$, and additionally need to form a representation under $\mathcal{C}_4$. In the SM~\cite{Note1}, we moreover explicitly construct $\mathcal{C}_4$-symmetric compact Wannier states for all spatially-obstructed OAIs that have a compact representation when $\mathcal{C}_4$ symmetry is relaxed to $\mathcal{C}_2$ symmetry. As a consequence, there exist $\mathcal{C}_4$-protected fragile phases -- these are necessarily trivial with respect to $\mathcal{C}_2$ symmetry -- that have a compact OAI complement. Consider the fragile state~\footnote{Here, we use the labelling conventions of the Bilbao Crystallographic Server for wallpaper group $p4$~\cite{Aroyo2011183}.}
\begin{equation} \label{eq: c4_1band_subtraction}
\left[(A)_\mathrm{1b} \oplus (B)_\mathrm{1b} \oplus ({}^{2}E)_\mathrm{1b}\right] \uparrow G \ominus (A)_\mathrm{1a} \uparrow G = \mathrm{FP},\end{equation}
which is illustrated in Fig.~\ref{fig: compactC4counterexample}a,b. The $1$-band OAI $(A)_\mathrm{1a} \uparrow G$ admits a compact Wannier basis $\ket{W_{\bs{R}}}$, with $\ket{W_{\bs{0}}}$ shown in Fig.~\ref{fig: compactC4counterexample}c. These compact states can be used to build a strictly local Hamiltonian whose ground state is $\mathrm{FP}$, $H = \sum_{\bs{R}} \ket{W_{\bs{R}}}$ $\bra{W_{\bs{R}}}$. $H$ has zero correlation length and a flat band spectrum~\cite{DasSarma12,Parameswaran13,Bergholtz13,DubailRead15,Thomale16,Read17,Maimaiti17,Rontgen18,Rhim19,FlachReview,Hatsugai20,Chiu20,Ma20,Peri21,skurativska2021flat}. We note that Eq.~\eqref{eq: c4_1band_subtraction} involves the unpaired complex representation ${}^{2}E$ and hence assumes broken TRS, implying that its realization requires magnetism.

In contrast, some \emph{representation-obstructed} (not spatially-obstructed) OAIs are non-compact only due to constraints imposed by $\mathcal{C}_4$ symmetry. Consider
\begin{equation} \label{eq: c4simpleREPobstruction}
\left[(A)_\mathrm{1b} \oplus (B)_\mathrm{1b} \oplus (B)_\mathrm{1a}\right] \uparrow G \ominus (A)_\mathrm{1a} \uparrow G= \mathrm{FP},
\end{equation}
which unlike Eq.~\eqref{eq: c4_1band_subtraction} is compatible with TRS and therefore non-magnetic.
In the SM~\cite{Note1}, we prove that the representation-obstructed OAI $(A)_\mathrm{1a} \uparrow G$ is non-compact. Nevertheless, it becomes unobstructed (and thereby compact) when $\mathcal{C}_4$ symmetry is relaxed to $\mathcal{C}_2$ symmetry: the $\mathcal{C}_4$ representations $(A)$ and $(B)$ both map into the same $\mathcal{C}_2$ representation $(A)$.

\emph{$\mathcal{C}_3$ symmetry---}All spatially obstructed $1$- and $2$-band OAIs with $\mathcal{C}_3$ symmetry are compact, irrespective of whether their band complement is another OAI or a fragile state. For lattices where $N(\mathrm{AI}) \geq \bar{N}(\mathrm{OAI})$, compactness of the OAI (and its complement) follows directly from the reasoning below Eq.~\eqref{eq: first-model-subtraction}. More nontrivially, consider the following TRS-broken fragile states in wallpaper group $p3$: 
\begin{equation}
\left[(\gamma)_\mathrm{1b} \oplus (e^{\mathrm{i}\frac{2\pi}{3}} \gamma)_\mathrm{1b} \oplus (e^{\mathrm{i}\frac{2\pi}{3}} \gamma)_\mathrm{1c}\right] \uparrow G \ominus (\gamma)_\mathrm{1a} \uparrow G= \mathrm{FP},
\end{equation}
where $(\mu)_{\mathrm{1x}}$ is an orbital with $\mathcal{C}_3$ eigenvalue $\mu$ at Wyckoff position 1x, and $\gamma \in \{1,e^{\mathrm{i}\frac{2\pi}{3}},e^{-\mathrm{i}\frac{2\pi}{3}}\}$ is a free parameter.
The compact Wannier state for the OAI at $\bs{R} = \bs{0}$ is
\begin{equation} \label{eq: compactC3state1} \begin{aligned}
\ket{W_{\bs{0} \gamma}} &= \frac{1}{3} \left[\ket{w_{\bs{0} \gamma}} + \gamma^* C_3 \ket{w_{\bs{0} \gamma}} + (\gamma^* C_3)^2 \ket{w_{\bs{0} \gamma}} \right], \\
\ket{w_{\bs{0} \gamma}} &= \ket{\bs{0}, \mathrm{1b}, \gamma} + \ket{\bs{0}, \mathrm{1b}, e^{\mathrm{i}\frac{2\pi}{3}} \gamma} + \ket{\bs{0}, \mathrm{1c}, e^{\mathrm{i}\frac{2\pi}{3}} \gamma},
\end{aligned} \end{equation}
where $C_3$ rotates about Wyckoff position 1a of the unit cell at $\bs{R} = \bs{0}$. Similarly, we construct the compact states of all further $\mathcal{C}_3$-symmetric $1$- and $2$-band OAIs with spatial obstruction in the SM~\cite{Note1} (there, we also discuss $\mathcal{C}_3$-symmetric OAIs with a representation-obstruction).

\begin{figure}[t]
\centering
\includegraphics[width=0.48\textwidth]{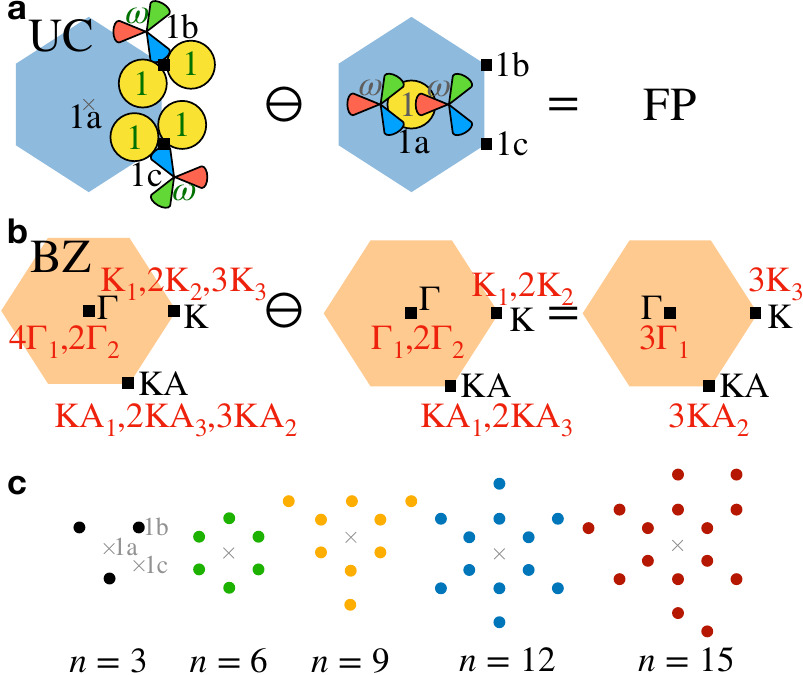}
\caption{Non-compact OAI with $\mathcal{C}_3$ rotational symmetry and fragile complement. (a)~Real-space illustration of the fragile state. The unit cell (UC) contains three orbitals each at Wyckoff positions 1b and 1c (orbitals are labelled by their $\mathcal{C}_3$ eigenvalue, $\omega=e^{\mathrm{i}\frac{2\pi}{3}}$). (b)~Brillouin zone (BZ) decomposition of the corresponding Bloch bands into irreducible representations. (Representation labels follow the Bilbao Crystallographic Server~\cite{Aroyo2011183}.) (c)~Trial state support for the OAI, labelled by size. Each colored atomic site indicates that the trial states for all three bands may have non-zero overlap with orbitals on that site. For all support sizes shown, there are no trial states that satisfy the requirements for an orthonormal Wannier basis [Eq.~\eqref{eq: fullcompactnessconstraints}].}
\label{fig: noncompactC3}
\end{figure}

In contrast, ascertaining the compactness properties of $3$-band OAIs with $\mathcal{C}_3$ symmetry is a challenging yet unsolved problem. Consider the TRS-broken fragile state~\footnote{Here, we use the labelling conventions of the Bilbao Crystallographic Server for wallpaper group $p3$~\cite{Aroyo2011183}.}
\begin{equation}
\begin{aligned}
\left[2(A_1)_\mathrm{1b} \oplus ({}^{2}E)_\mathrm{1b} \oplus 2(A_1)_\mathrm{1c} \oplus ({}^{2}E)_\mathrm{1c}\right] \uparrow &G \\ \ominus 
\left[(A_1)_\mathrm{1a} \oplus 2({}^{2}E)_\mathrm{1a} \right] \uparrow &G = \mathrm{FP}.
\end{aligned}
\end{equation}
Here, $\mathrm{FP}$ is obtained as the complement of a $3$-band OAI built from Wannier states at Wyckoff position 1a that have $\mathcal{C}_3$ eigenvalues $\lambda_1 = 1$, $\lambda_2 = \lambda_3 = e^{\mathrm{i}\frac{2\pi}{3}}$. (See Fig.~\ref{fig: noncompactC3}a,b for an illustration.) To obtain a compact basis, we must impose the constraints in Eq.~\eqref{eq: fullcompactnessconstraints}, where $\alpha = 1,2,3$ belongs to the obstructed orbital with $\mathcal{C}_3$ eigenvalue $\lambda_\alpha$.
For $\mathcal{C}_3$-symmetric trial Wannier states that have overlap with $n$ lattice sites (located at the 1b and 1c Wyckoff positions and carrying three orbitals each), Eq.~\eqref{eq: fullcompactnessconstraints} is a system of coupled quadratic equations in $N = 6n$ complex variables. The problem of determining whether solutions to general systems of quadratic equations exist is NP-complete~\cite{BSS89}, and the runtime of all (currently known) algorithms scales exponentially in $N$. For OAIs with $\mathcal{C}_2$ and $\mathcal{C}_4$ symmetry, we were able to circumvent this difficulty: for $\mathcal{C}_2$-symmetric OAIs with fragile complement, a solution to Eq.~\eqref{eq: fullcompactnessconstraints} can be ruled out by a single translation (Fig.~\ref{fig: compactInvObstruction}), proving non-compactness. For all spatially-obstructed $\mathcal{C}_4$-symmetric OAIs, and likewise all spatially-obstructed $\mathcal{C}_3$-symmetric OAIs with $1$ and $2$ bands, we found explicit solutions to Eq.~\eqref{eq: fullcompactnessconstraints}, proving compactness. In the present case, however, both strategies fail~\footnote{As shown in the SM~\cite{Note1}, this difficulty also arises for a representation-obstructed $2$-band OAI with $\mathcal{C}_3$ symmetry.}. Nevertheless, we prove in the SM~\cite{Note1} that Eq.~\eqref{eq: fullcompactnessconstraints} cannot be solved by states overlapping with $n < 18$ lattice sites (see Fig.~\ref{fig: noncompactC3}c). We conjecture that there is also no solution for $n \geq 18$.

\emph{Discussion---} 
The existence of non-compact atomic insulators suggests to explore non-compactness as a new ordering principle for gapped phases. Promising directions of future study are the generalization of our analysis to arbitrary finite Wannier state sizes, larger symmetry groups, and higher dimensions. Moreover, it is fruitful to investigate the observable consequences of non-compactness. In particular, both the superfluid weight~\cite{Peotta15,JonahUpcoming} and the conductivity in presence of disorder~\cite{FrankUpcoming} of a set of bands directly depends on Wannier spread. Hence, we expect that both are enhanced in the non-compact case.

\begin{acknowledgments}
We thank Nicolas Regnault, Luis Elcoro, and Zhida Song for helpful discussions. FS was supported by a fellowship at the Princeton Center for Theoretical Science. BAB was supported by the DOE Grant No. DE-SC0016239, the Schmidt Fund for Innovative Research, Simons Investigator Grant No. 404513, the Packard Foundation, the Gordon and Betty Moore Foundation through Grant No. GBMF8685 towards the Princeton theory program, and a Guggenheim Fellowship from the John Simon Guggenheim Memorial Foundation. Further support was provided by the NSF-EAGER Grant No. DMR 1643312, NSF-MRSEC Grant No. DMR-1420541 and DMR-2011750, ONR Grant No. N00014-20-1-2303, BSF Israel US foundation Grant No. 2018226, and the Princeton Global Network Funds.
\end{acknowledgments}

\bibliography{references}

\end{document}


\title{Supplemental Material for ``Non-Compact Atomic Insulators"}

\author{Frank Schindler}
\affiliation{Princeton Center for Theoretical Science, Princeton University, Princeton, NJ 08544, USA}

\author{B. Andrei Bernevig}
\affiliation{Department of Physics, Princeton University, Princeton, NJ 08544, USA}

\maketitle

\tableofcontents

\newpage
\section{Basic notions} \label{sec: introduction}
This section pedagogically introduces the concept of compact Wannier states, and serves as a basis for all further derivations.

\subsection{Real space construction}
We begin by constructing compact trial states for an obstructed atomic insulator (OAI) in real space.
Throughout this section, we denote the physical atomic orbitals on a lattice of sites $\bs{R}$ with space group $G$ by $\ket{\bs{R} j}$, $j \in A$, where $A$ is the set of orbitals in the unit cell. These orbitals respect the symmetries of the little group of their respective Wyckoff position.
 
Out of the atomic orbitals, we construct obstructed orbitals at the \emph{unoccupied} orbitals $\alpha \in B$, $B \cap A =\emptyset$ of the lattice:
\begin{equation} \label{eq: firstAnsatzWannier}
\ket{W_{\bs{R}, \alpha}} = \sum_{\bs{R}', j\in A} \ket{\bs{R}' j} S_{j \alpha} (\bs{R}-\bs{R}').
\end{equation} 
Translational invariance guarantees that $S$ depends only on $\bs{R}-\bs{R}'$. The matrix $S(\bs{R}-\bs{R}')$, whose elements are complex numbers, also must respect the space group symmetry. Note that $S(\bs{R}-\bs{R}')$ is generically \emph{not} a unitary transformation, as the cardinality (number of orbitals) of $B$ is usually smaller than that of $A$. Also, generally, $S(\bs{R}-\bs{R}')$ is chosen by requiring compact support, so that it is strictly zero for any $|\bs{R}-\bs{R}'|$ greater than a certain distance.
 
In general, the obstructed orbitals $\ket{W_{\bs{R}, \alpha}}$ may not be orthonormal and hence do not, in general, form a Wannier basis. An orthonormalization procedure renders the orbitals without compact support. We will hence see that there is a tradeoff between orthogonalizing the obstructed orbitals and keeping $S(\bs{R}-\bs{R}')$ finite-range. A systematic method to obtain an orthonormal and translationally invariant Wannier basis (which may or may not be compact) is to exploit the orthogonality of momentum eigenstates in Fourier space, which we turn to next.

\subsection{Momentum space orthogonalization}
We perform a Fourier transform to obtain the momentum-space representation
\begin{equation} \label{eq: SkDef}
S_{j \alpha}(\bs{k}) =\sum_{\bs{R}} S_{j \alpha} (\bs{R}) e^{\mathrm{i} \bs{k}\cdot(\bs{R}+ \bs{t}_\alpha - \bs{t}_j)},
\end{equation} 
where $\bs{t}_j$, $\bs{t}_{\alpha}$ are the positions of the physical ($\in A$) and obstructed ($\in B$) orbitals in the unit cell, respectively. Introducing the momentum basis states
\begin{equation} \label{eq: fourrier}
\ket{\bs{k}j} = \frac{1}{\sqrt{V}} \sum_{\bs{R}} e^{\mathrm{i} \bs{k} \cdot (\bs{R}+\bs{t}_j)} \ket{\bs{R} j}, \quad \ket{\bs{R} j} = \frac{1}{\sqrt{V}} \sum_{\bs{k}} e^{-\mathrm{i} \bs{k} \cdot (\bs{R}+\bs{t}_j)} \ket{\bs{k}j},
\end{equation} 
where $V$ is the (dimensionless) volume, the projected Bloch state reads
\begin{equation} \label{eq: gammaprojDEF}
\ket{u_{\bs{k}, \alpha}}= e^{\mathrm{i} \bs{k} \cdot \bs{t}_\alpha} \sum_{j} \ket{\bs{k}j}\braket{\bs{k}j | W_{\bs{0}, \alpha}} =  \sum_{j} \ket{\bs{k}j} S_{j \alpha}(\bs{k}).
\end{equation}
The projected Bloch states are orthogonal at different momenta by default, but not yet fully orthornormal. We can compute their equal-momentum overlap
\begin{equation}
\braket{u_{\bs{k}, \alpha} | u_{\bs{k}, \beta}}= M_{\alpha\beta}(\bs{k}) = [S^\dagger (\bs{k}) S  (\bs{k})]_{\alpha \beta}.
\end{equation}
The trial states $\ket{W_{\bs{R}, \alpha}}$ have to be chosen such that $M(\bs{k})$ is invertible at every $\bs{k}$ (does not have zero eigenvalues at all $\bs{k}$). This is always possible if the state we want to construct is non-topological, and if the number of unoccupied positions $\alpha$ is smaller than the number of occupied positions $j$, as is the case for an OAI. Note that $M(\bs{k})$ is also positive semidefinite by construction, and hence, since it does not have zero eigenvalues, it is positive definite. Then, computing its square-root proceeds straightforwardly by diagonalizing $M(\bs{k})$:
\begin{equation}
M(\bs{k}) = U(\bs{k}) D(\bs{k}) U^\dagger(\bs{k}), \quad D(\bs{k})>0,
 \end{equation} 
so that the orthonormalized eigenstates of the OAI are
\begin{equation} \label{eq: orthonormalBloch}
\ket{\tilde{u}_{\bs{k}, \alpha}} = \sum_{\beta} \ket{u_{\bs{k}, \beta}} [U(\bs{k}) D^{-\frac{1}{2}} (\bs{k})]_{\beta \alpha} \equiv \sum_j \ket{\bs{k}j} \tilde{S}_{j \alpha}(\bs{k}),
\end{equation}
where $\tilde{S}(\bs{k}) = S(\bs{k}) U(\bs{k}) D^{-\frac{1}{2}} (\bs{k})$.

\subsection{Compact support} \label{sec: constraint_on_blochnorm}
The Wannier functions $\ket{\tilde{W}_{\bs{R}, \alpha}}$ of a generic OAI, obtained as the Fourier transforms of Eq.~\eqref{eq: orthonormalBloch}, although exponentially localized in real space, do \emph{not} in general have compact support. We now track down the cause of non-compact support. The matrix $S(\bs{R})$ has compact support by construction. Hence, the entries of $S(\bs{k})$ are finite-degree Laurent polynomials in $e^{\mathrm{i} \bs{k}}$. However, generically $D^{-1/2} (\bs{k})$ is \emph{not} such a polynomial. For example, for the case of one obstructed band $|B|=1$, we have that 
\begin{equation}
D^{-\frac{1}{2}} (\bs{k}) = \frac{1}{\sqrt{S^\dagger (\bs{k}) S(\bs{k})}}.
\end{equation}
We see that normalization is the source of non-compactness of $\ket{\tilde{W}_{\bs{R}, \alpha}}$: The inverse square root of a Laurent matrix polynomial is in general not any Laurent matrix polynomial of finite degree. In the Fourier transform back to real space, this infinite sum of harmonics will give rise to infinite range contributions, and hence does not yield compact-support Wannier states. 

However, this immediately suggests a strategy for constructing compact states: $S (\bs{R})$ needs to be chosen such that $S^\dagger (\bs{k}) S(\bs{k}) = \mathbb{1}$. This choice, if possible, guarantees that $\tilde{S} (\bs{R}) = S(\bs{R})$ has compact support, and so the states $\ket{\tilde{W}_{\bs{R}, \alpha}}$ span an obstructed atomic \emph{limit}.

We show in the following that whether such a construction is possible for a given OAI depends on both the OAI (on the set $B$), and on the physical lattice hosting it (on the set $A$). This is because $A$ determines the Hilbert space in which we must find local, symmetric, and orthogonal Wannier states for all elements of $B$. If $A$ is much larger than $B$, such compact Wannier states can usually be constructed in a straightforward manner, because the available Hilbert space is large enough to accommodate for all constraints imposed by compactness. However, on lattices $A$ with few atomic orbitals per unit cell, the problem may be overconstrained. In this case, any Wannier states for $B$ are necessarily non-compact. These notions will be made precise in the following. 

We note that there is so far one known category of insulators allowing for exponentially-localized Wannier states which are necessarily not delta-function-like: delicate topological insulators~\cite{nelson2020multicellularity}, which are characterized by Hopf invariants and returning Thouless pumps. In our work, we explore OAIs as a second category of non-delta-function insulators. Surprisingly, we find that not all OAIs have a compact representation: there are topological obstructions to compactness. We call the resulting phases non-compact atomic insulators. The condition of non-compactness is stronger than the ``multicellularity" of delicate topological insulators, meaning that the Wannier states cannot be completely localized in a primitive unit cell: non-compact Wannier states cannot be completely localized in any, potentially non-primitive, unit cell. (Presently, it is not known if delicate topological insulators ultimately satisfy the stronger condition.) Non-compact atomic insulators and delicate topological insulators satisfy qualitatively similar stability conditions: their non-localizability can be nullified by the addition of particular trivial bands to the conduction subspace. In fact, this has to be true according a theorem that all band representations can be represented by delta-function Wannier states if the conduction subspace can be arbitrarily enlarged~\cite{Alexandradinata20}. One distinction is that  the conduction subspace of non-compact atomic insulators is necessarily fragile topological, but this is not true of delicate topological insulators.

\section{Obstructed atomic insulators and their complements} \label{sec: OAI_mc_generalities}
We first derive some general results for OAIs with rotational symmetry. Unless otherwise mentioned, we assume the Wannier states of all OAIs to be centered at empty Wyckoff positions of the lattice, resulting in a \emph{spatial} obstruction. The case of a \emph{representation} obstruction, where the OAI Wannier centers coincide with the positions of physical orbitals, while the OAI symmetry representation does not, is discussed separately in Sec.~\ref{sec: repob}.

\subsection{Mobile clusters of physical orbitals} \label{subsec: mobile_clusters}
We begin by introducing the notion of mobile clusters, which are sets of orbitals that characterize the underlying physical lattice of a band structure in presence of crystalline symmetry. This notion will provide us with a real-space condition for fragile topology, in addition to allowing for general statements about the compactness of atomic bands.

\begin{figure}[t]
\centering
\includegraphics[width=1\textwidth,page=8]{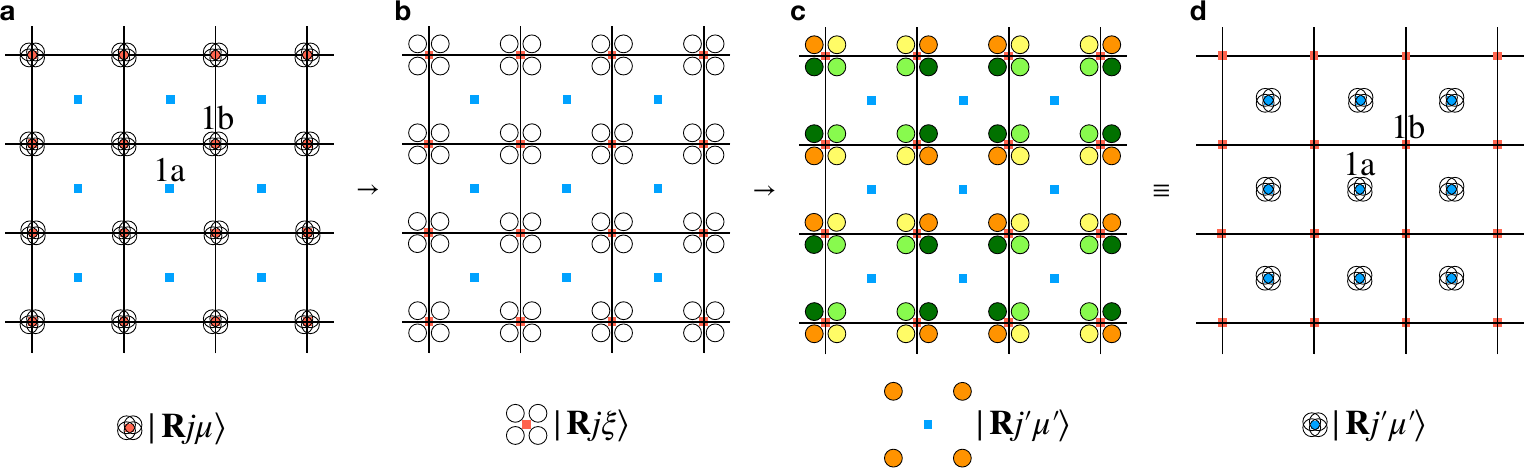}
\caption{Mobile clusters with $\mathcal{C}_4$ rotational symmetry. A mobile cluster is defined as a minimal set of physical orbitals $\ket{\bs{R} j \mu}$ whose transformation behavior under the crystalline symmetry group is compatible with being located at any (possibly non-maximal) Wyckoff position. For spinless $\mathcal{C}_4$ symmetry, the mobile clusters on maximal Wyckoff positions 1a and 1b contain $4$ orbitals with eigenvalues $\gamma_{1 \dots 4} = (1,-1,\mathrm{i},-\mathrm{i})$. (a)~We begin with mobile clusters centered at the $1b$ Wyckoff position. (b)~Next, we locally change bases to obtain the states $\ket{\bs{R} j \xi}$ [Eq.~\eqref{eq: local_rot_into_featureless}] that do not have well-defined $\mathcal{C}_4$ eigenvalues, but instead are cyclically permuted by the action of $\mathcal{C}_4$. (c,~d)~These states can be used to construct new mobile clusters $\ket{\bs{R} j' \mu'}$ that are centered around Wyckoff position $1a$. Note that the transformation fully preserves translational symmetry.}
\label{fig: mobile_cluster_transform}
\end{figure}

In the following, we denote the orbitals of the unit cell at position $\bs{R}$ by $\ket{\bs{R} j \mu}$, $j \in A$. In comparison to the states $\ket{\bs{R} j}$ appearing in Sec.~\ref{sec: introduction}, we have introduced the refined orbital label $\mu$, so that $j$ now only distinguishes between inequivalent Wyckoff positions in the unit cell, while $\mu$ labels the orbitals at a given position. Without loss of generality, we assume that all orbitals are located at maximal Wyckoff positions. Indeed, if the states $\ket{\bs{R} \tilde{j} \tilde{\mu}}$ are located at non-maximal positions, they must form an orbit under the space group symmetry. It is then always possible to find a basis transformation $O_{j \mu, \tilde{j} \tilde{\mu}}$ so that the states 
\begin{equation}
\ket{\bs{R} j \mu} = \sum_{\tilde{j} \tilde{\mu}} O_{j \mu, \tilde{j} \tilde{\mu}} \ket{\bs{R} \tilde{j} \tilde{\mu}}
\end{equation}
are centered at maximal Wyckoff positions~\cite{Cano17-2}. Because $O_{j \mu, \tilde{j} \tilde{\mu}}$ is a strictly local unitary transformation (it does not induce mixing between different unit cells) that preserves the crystalline symmetry, any such basis change does not affect our conclusions on compact support: applying $O_{j \mu, \tilde{j} \tilde{\mu}}$ does not change the range of compact Wannier states (when defined with respect to full unit cells), and preserves their symmetry and orthonormality. The orbitals $\ket{\bs{R} j \mu}$ can then be chosen to form a representation of the site-symmetry group. In presence of $\mathcal{C}_n$ rotational symmetry, they are eigenstates:
\begin{equation}
C_n|_{\bs{R},j} \ket{\bs{R} j \mu} = \gamma_\mu \ket{\bs{R} j \mu},
\end{equation}
where $C_n|_{\bs{R},j}$ rotates about the $\mathcal{C}_n$-symmetric Wyckoff position $j$ of the unit cell at $\bs{R}$. Assuming spinless rotational symmetries, so that $(C_n|_{\bs{R},j})^n = 1$, the $\mathcal{C}_n$ eigenvalues $\gamma_\mu$ take values in the $n$-th roots of unity:
\begin{equation}
\gamma_\mu = e^{\mathrm{i}\frac{2\pi}{n}l}, \quad l = 0 \dots n-1.
\end{equation}
We call a \emph{mobile cluster} a configuration of orbitals whose $\mathcal{C}_n$ eigenvalues exhaust all $l = 0 \dots n-1$, with each $l$ appearing exactly once. These configurations are special in that they can be used to construct local basis states at any particular maximal Wyckoff position, not just at the atomic positions hosting the mobile cluster orbitals~\cite{Cano17-2}. That is, given that $\ket{\bs{R} j \mu}$, $\mu = 1 \dots n$ is a mobile cluster, there exist strictly local unitary transformations that effect a transformation
\begin{equation} \label{eq: mobileclustermagic}
\ket{\bs{R} j \mu} \quad \rightarrow \quad \ket{\bs{R} j' \mu'},
\end{equation}
where $j$, $j'$ are two (sets of) Wyckoff positions with isomorphic site-symmetry groups, and $\mu'$ labels a new set of orbitals that forms a mobile cluster as well. We pictorially illustrate the unitary transformation between Wyckoff positions for the case of $\mathcal{C}_4$ rotational symmetry in Fig.~\ref{fig: mobile_cluster_transform}. To prove Eq.~\eqref{eq: mobileclustermagic} in general, we first note that mobile clusters have the defining property that they can be transformed into a set of featureless orbitals (without local symmetry)
\begin{equation} \label{eq: local_rot_into_featureless}
\ket{\bs{R} j \xi} = \sum_{\mu} U_{\xi \mu} \ket{\bs{R} j \mu}, \quad C_n|_{\bs{R},j} \ket{\bs{R} j \xi} = \ket{\bs{R} j (\xi+1 \mod n)},
\end{equation}
by an on-site unitary
\begin{equation}
U_{\xi \mu} = (\gamma_\mu)^{\xi}, \quad \xi = 0 \dots n-1.
\end{equation}
Instead of being eigenstates of the site-symmetry operation $C_n|_{\bs{R},j}$, the orbitals $\ket{\bs{R} j \xi}$ are cyclically permuted by its action. We can always also arrange them so that 
\begin{equation}
C_n|_{\bs{R},j'} \ket{\bs{R} j \xi} = \ket{\bs{R}' j (\xi+1 \mod n)}
\end{equation}
holds for a suitable choice of $\bs{R}' \neq \bs{R}$ (see Fig.~\ref{fig: mobile_cluster_transform}b,c). But this implies that we may now form orbitals $\mu'$ centered around $j'$ as follows:
\begin{equation} \label{eq: shiftedMCorbitals}
\begin{aligned}
&\ket{\bs{R} j' \mu'} = \sum_{m = 0}^{n-1} \left(\mu'^* C_n|_{\bs{R},j'}\right)^m \ket{\bs{R} j (\xi=0)} \equiv \sum_{\bs{R}'\xi} V_{\bs{R} j'\mu',\bs{R}'j\xi} \ket{\bs{R}' j \xi}. 
\end{aligned}
\end{equation}
(Note that $j$ is not summed over, and we have used that all $\mathcal{C}_n$ eigenvalues lie on the unit circle so that $|\mu'|^2 = 1$.) Now, the unitary transformations $V_{\bs{R} j'\mu',\bs{R}'j\xi}$ at different unit cells $\bs{R}$ act on disjoint sets of orbitals: for example, in Fig.~\ref{fig: mobile_cluster_transform}c, $V_{\bs{R} j'\mu',\bs{R}'j\xi}$ only acts on the orbitals drawn within the unit cell at $\bs{R}$ (with different unit cells demarcated by black lines). Therefore, the unitary transformations $V_{\bs{R} j'\mu',\bs{R}'j\xi}$ at different unit cells $\bs{R}$ all commute with each other and give rise to a strictly local unitary acting on the entire lattice. In conclusion, we have shown that given a mobile cluster $\ket{\bs{R} j \mu}$, the basis change
\begin{equation} \label{eq: final_mc_transform}
\forall \bs{R}:\quad \ket{\bs{R} j \mu} \quad \rightarrow \quad \ket{\bs{R} j' \mu'} = \sum_{\bs{R}' \mu} \left(\sum_\xi V_{\bs{R} j'\mu',\bs{R}'j\xi} U_{\xi \mu}\right) \ket{\bs{R}' j \mu}
\end{equation}
can be effected by a strictly local unitary transformation $V_{\bs{R} j'\mu',\bs{R}'j\xi} U_{\xi \mu}$. This means that from a mobile cluster, we can construct strictly local and symmetric basis states at any Wyckoff position of the unit cell; moreover, their union will also form a mobile cluster. 

\subsection{Necessary condition for fragile bands} \label{subsec: realspace_fragile}
Fragile topological phases are obtained as the band complements of other fragile phases or OAIs~\cite{AshvinFragile,Cano17-2}. In the former case, both sets of bands cannot be expressed in terms of exponentially localized symmetric Wannier functions, so that compact Wannier states are impossible. In the latter case, the fragile set of bands can be formally expressed as a difference of atomic insulators~\cite{ZhidaFragileTwist2}:
\begin{equation} \label{eq: first-model-subtraction}
\mathrm{FP} = \mathrm{AI} \ominus \mathrm{OAI},
\end{equation}
where $\mathrm{FP}$ denotes the fragile phase, and $\mathrm{AI}$ is the (unobstructed) atomic limit corresponding to the physical lattice. We will often refer to a relationship of this form as an OAI \emph{subtraction}.
Now, let $N(\mathrm{AI})$ count the number of mobile clusters in the unit cell (in the set $A$ introduced in Sec.~\ref{sec: introduction}). For instance, the inversion-symmetric SSH model in 1D, when defined on two sites that lie at general Wyckoff positions of the unit cell, or on a lattice hosting an $s$ and a $p$ orbital at either of the two maximal Wyckoff positions, has $N(\mathrm{AI}) = 1$. Conversely, let $\bar{N}(\mathrm{OAI})$ count the minimal number of mobile clusters that contains all orbitals of the OAI [potentially making use of the transformation in Eq.~\eqref{eq: mobileclustermagic}]. For instance, in wallpaper group $p2$, we have $\bar{N}[(A)_\mathrm{1a}\uparrow G] = 1$, because we need one mobile cluster $[(A)_\mathrm{1a} \oplus (B)_\mathrm{1a}]\uparrow G$ to supply $(A)_\mathrm{1a}\uparrow G$. Next, $\bar{N}[(A)_\mathrm{1a}\uparrow G \oplus (A)_\mathrm{1b}\uparrow G] = 2$, because we need two mobile clusters, $[(A)_\mathrm{1a} \oplus (B)_\mathrm{1a}]\uparrow G$ and $[(A)_\mathrm{1b} \oplus (B)_\mathrm{1b}]\uparrow G$, to support both orbitals. Moreover, $\bar{N}[(A)_\mathrm{1a}\uparrow G \oplus (A)_\mathrm{1a}\uparrow G] = 2$, because we again need the two mobile clusters, $2[(A)_\mathrm{1a} \oplus (B)_\mathrm{1a}]\uparrow G$, this time at the same Wyckoff position. Finally, $\bar{N}[(A)_\mathrm{1a}\uparrow G \oplus (B)_\mathrm{1a}\uparrow G] = 1$, because this configuration already corresponds to a single mobile cluster (this situation is equivalent to the occupied and unoccupied band of the SSH model discussed above). Here, we have labelled the site-symmetry representations of the OAIs following the conventions of the Bilbao crystallographic server~\cite{Aroyo2011183}.

Then, for $\mathrm{FP}$ to be a fragile phase, that is, to be inexpressible in terms of localized Wannier functions, we clearly need 
\begin{equation}
N(\mathrm{AI}) < \bar{N}(\mathrm{OAI}).
\end{equation}
Otherwise, the OAI can be built from Wannier states that are equal to a subset of the mobile cluster orbitals [Eq.~\eqref{eq: shiftedMCorbitals}] of each unit cell. Moreover, the remaining orbitals -- those that are not used up to build the OAI -- would form a compact and symmetric basis for the bands of $\mathrm{FP}$ (since the transformation in Eq.~\eqref{eq: final_mc_transform} is unitary, all orbitals of a mobile cluster are mutually orthogonal even when they are not located at the original atomic Wyckoff position). 

\subsection{Sufficient condition for compact bands} \label{sec: oai_oai_compactness}
We have seen that $N(\mathrm{AI}) \geq \bar{N}(\mathrm{OAI})$ leads to
\begin{equation} \label{eq: second-model-subtraction}
\mathrm{AI} \ominus \mathrm{OAI} = \mathrm{OAI'},
\end{equation}
where $\mathrm{OAI'}$ is another obstructed atomic insulator. Moreover, by the basis transformation in Eq.~\eqref{eq: final_mc_transform}, we can always find a compact set of Wannier states for the $\mathrm{OAI}$ set of bands. Conversely, by rewriting Eq.~\eqref{eq: second-model-subtraction} as 
\begin{equation}
\mathrm{AI} \ominus \mathrm{OAI'} = \mathrm{OAI},
\end{equation}
we see that $\mathrm{OAI'}$ also admits a compact Wannier representation. In conclusion, when $N(\mathrm{AI}) \geq \bar{N}(\mathrm{OAI})$ holds, the OAI is the band complement of another OAI and is guaranteed to have a compact representative. On the other hand, we will see in the following that OAIs that are the band complements of fragile phases may or may not admit a compact representation.

\section{Compactness constraints}
\begin{figure}[t]
\centering
\includegraphics[width=0.4\textwidth,page=12]{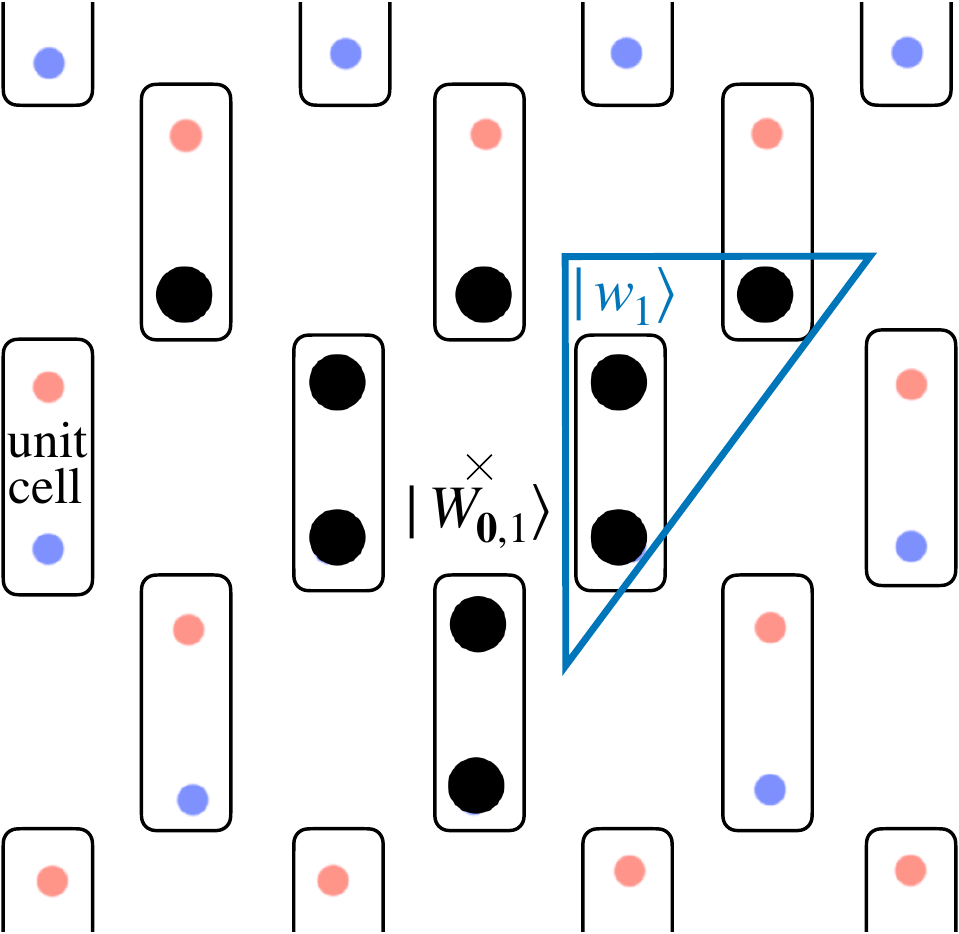}
\caption{Support of the asymmetric part $\ket{w_{1}}$ [defined in Eq.~\eqref{eq: symmWannierAnsatz}] of a $\mathcal{C}_3$-symmetric trial Wannier state $\ket{W_{\bs{0}, 1}}$, centered at Wyckoff position $\mathrm{1a}$ of the unit cell at $\bs{R} = \bs{0}$. The full trial state support (indicated in black) is obtained by successively applying $\mathcal{C}_3$ rotations on $\ket{w_{\mathrm{1a}}}$. Importantly, the support of neither $\ket{W_{\bs{0}, 1}}$ nor $\ket{w_{\mathrm{1a}}}$ needs to be confined to a unit cell of the lattice, or coincide with an integer multiple of unit cells (in the present example, the primitive unit cell hosts two lattice sites).}
\label{fig: AU_def}
\end{figure}

We next discuss the necessary and sufficient conditions for compactness, that is, for atomic bands to be spanned by strictly local, symmetric, and orthonormal Wannier states. For this, we start with a set of parametrized trial Wannier states $\ket{W_{\bs{R}, \alpha}}$ [Eq.~\eqref{eq: firstAnsatzWannier}], and then derive the quadratic constraint equations for compactness. We formulate these equations in both real and momentum space -- the momentum space constraints are ultimately equivalent to the real space constraints, but give rise to an insightful geometric interpretation.

Although the task of establishing whether solutions to systems of quadratic equations of the form discussed here exist is in general exponentially difficult in the total number of variables, we systematically solve the compactness constraints (in the case of spatial obstructions) for all OAIs protected by $\mathcal{C}_2$ and $\mathcal{C}_4$ symmetry in Secs.~\ref{sec: nogo_p2symmetry} and~\ref{sec: overarching_p4symmetry}, respectively, as well as for $1$- and $2$-band OAIs with $\mathcal{C}_3$ symmetry in Sec.~\ref{sec: overarching_c3}. We conjecture that the symmetry of the point groups makes the problem solvable for all space groups, but have not yet been able to find the general formalism.

\subsection{Real space} \label{sec: real_space_constraints_general}
We derive the real-space compactness constraints for systems with $\mathcal{C}_n$ rotational symmetry. The orthonormality conditions on a set of trial Wannier states are given by 
\begin{equation} \label{eq: CompactConstraintWannierREAL}
\braket{W_{\bs{0}, \alpha} | W_{\bs{R}, \beta}} = \braket{W_{\bs{0}, \alpha} | T_{\bs{R}} | W_{\bs{0}, \beta}} = \delta_{\alpha \beta} \delta_{\bs{R},\bs{0}},
\end{equation} 
where $T_{\bs{R}}$ translates all sites of the unit cell by the lattice vector $\bs{R}$.

In presence of $\mathcal{C}_n$ symmetry, the Wannier states at $\mathcal{C}_n$-symmetric maximal Wyckoff positions can be chosen as eigenstates of the rotation
\begin{equation}
C_n \ket{W_{\bs{0}, \alpha}} = \lambda_\alpha \ket{W_{\bs{0}, \alpha}},
\end{equation}
where $C_n$ is the rotation operator around the obstructed site. They can then be expressed in a manifestly symmetric fashion:
\begin{equation} \label{eq: symmWannierAnsatz}
\ket{W_{\bs{0}, \alpha}} = \sum_{m = 0}^{n-1} \left(\lambda^*_\alpha C_n \right)^{m} \ket{w_\alpha}.
\end{equation}
Here, $\ket{w_\alpha}$ denotes the asymmetric part of the Wannier state that in $\mathcal{C}_n$-symmetric groups has overlap with $(1/n)$-th of the sites of $\ket{W_{\bs{0}, \alpha}}$, an example is shown in Fig.~\ref{fig: AU_def} (we assume $\ket{W_{\bs{0}, \alpha}}$ to form an OAI centered on an empty Wyckoff position of the lattice). We have also used that $\mathcal{C}_n$ eigenvalues lie on the unit circle, so that $|\lambda_\alpha|^2 = 1$.

Inserting the symmetric decomposition of Eq.~\eqref{eq: symmWannierAnsatz} into Eq.~\eqref{eq: CompactConstraintWannierREAL} yields
\begin{equation}
\bra{w_\alpha} \tilde{T}_{\bs{R},\alpha \beta} \ket{w_\beta} = \delta_{\alpha\beta} \delta_{\bs{R},\bs{0}}, \quad \tilde{T}_{\bs{R},\alpha \beta} = \sum_{l,m = 0}^{n-1} \left(\lambda^*_\alpha C_n \right)^{\dagger l} T_{\bs{R}} \left(\lambda^*_\beta C_n \right)^{m}.
\end{equation}
Now, due to the normalization 
\begin{equation} \label{eq: trial_asymmetric_unit_normalization}
\braket{W_{\bs{0}, \alpha} |W_{\bs{0}, \beta}} = \delta_{\alpha \beta} \quad \rightarrow \quad \braket{w_{\alpha} |w_{\alpha}} = \frac{1}{n},
\end{equation}
we obtain
\begin{equation} \label{eq: realspaceconstraints_general}
\bra{w_\alpha}\left(\tilde{T}_{\bs{R},\alpha \beta} - n \delta_{\alpha\beta} \delta_{\bs{R},\bs{0}} \mathbb{1} \right) \ket{w_\beta} = 0.
\end{equation}
To simplify the algebra, it is instructive to to view Eq.~\eqref{eq: realspaceconstraints_general} as a linear constraint
\begin{equation}
\sum_{ij} N_{\bs{R},ij}^{\alpha \beta} \left(w^*_{\alpha i} w_{\beta j} \right) = 0, \quad N_{\bs{R},ij}^{\alpha \beta} = \bra{i} \left(\tilde{T}_{\bs{R},\alpha \beta} - n \delta_{\alpha\beta} \delta_{\bs{R},\bs{0}} \mathbb{1} \right) \ket{j},
\end{equation}
on the space of element-wise products $\left(w^*_{\alpha i} w_{\beta j} \right)$, where we use the components $w_{\alpha i} = \braket{i |w_\alpha}$. Here, $\ket{i}$ are the orbitals of the asymmetric part of the trial state, i.e., the on-site orbitals used to create the Wannier state. Now, numerically, the minimal set of constraints can be generated by applying row reduction (Gaussian elimination) to the set of matrices $(N^{\alpha \beta})_{\bs{R},ij} = N_{\bs{R},ij}^{\alpha \beta}$ [where rows are labelled by $\bs{R}$ and columns by the composite index $(ij)$], after which we arrive at equations of the form
\begin{equation} \label{eq: real_space_constraints_reduced}
\sum_{ij} N_{\lambda,ij}^{\alpha \beta} \left(w^*_{\alpha i} w_{\beta j} \right) = 0, \quad \lambda = 1 \dots r_{\alpha \beta},
\end{equation}
where the number of irreducible equations $r_{\alpha \beta} \in \mathbb{N}$ depends on the problem in question.

\subsection{Momentum space}
We next re-derive the compactness constraints in momentum-space.
It was shown in Sec.~\ref{sec: constraint_on_blochnorm} that for the trial states $\ket{W_{\bs{R}, \alpha}}$ to be compact Wannier states, their momentum space projections must satisfy
\begin{equation} \label{eq: WannierCompactMomentumConstraint}
V \braket{u_{\bs{k}, \alpha} | u_{\bs{k}, \beta}} e^{\mathrm{i} \bs{k} \cdot (\bs{t}_\alpha-\bs{t}_\beta)} = V \bra{W_{\bs{0}, \alpha}}P_{\bs{k}} \ket{W_{\bs{0}, \beta}} = \delta_{\alpha\beta}, \quad P_{\bs{k}} = \sum_{j} \ket{\bs{k}j} \bra{\bs{k}j},
\end{equation}
for all values of $\bs{k}$ in the Brillouin zone, where $V$ is the volume appearing in the Fourier transform in Eq.~\eqref{eq: fourrier}.
Inserting the symmetric decomposition of Eq.~\eqref{eq: symmWannierAnsatz} into Eq.~\eqref{eq: WannierCompactMomentumConstraint} yields
\begin{equation} \label{eq: ptildedef_kspace}
\bra{w_\alpha} \tilde{P}_{\bs{k},\alpha \beta} \ket{w_\beta} = \delta_{\alpha\beta}, \quad \tilde{P}_{\bs{k},\alpha \beta} = V \sum_{l,m = 0}^{n-1} \left(\lambda^*_\alpha C_n \right)^{\dagger l} P_{\bs{k}} \left(\lambda^*_\beta C_n \right)^{m}.
\end{equation}
Now, due to the normalization choice of Eq.~\eqref{eq: trial_asymmetric_unit_normalization}, we obtain
\begin{equation}
\bra{w_\alpha}\left(\tilde{P}_{\bs{k},\alpha \beta} - n \delta_{\alpha\beta} \mathbb{1} \right) \ket{w_\beta} = 0.
\end{equation}
When this set of equations is enforced at all momenta $\bs{k}$, the $\ket{W_{\bs{R}, \alpha}}$ that result as translates of $\ket{W_{\bs{0}, \alpha}}$ in Eq.~\eqref{eq: symmWannierAnsatz} form a compact Wannier basis.

It is again fruitful to view the constraints as linear equations 
\begin{equation} \label{eq: constraintmat_def}
\sum_{ij} M_{\bs{k},ij}^{\alpha \beta} \left(w^*_{\alpha i} w_{\beta j} \right) = 0, \quad M_{\bs{k},ij}^{\alpha \beta} = \bra{i} \left(\tilde{P}_{\bs{k},\alpha \beta} - n \delta_{\alpha\beta} \mathbb{1} \right) \ket{j}
\end{equation}
on the space of element-wise products $\left(w^*_{\alpha i} w_{\beta j} \right)$. Then, a necessary criterion for compactness is that the vectors $(\vec{M}_{\bs{k}}^{\alpha \beta})_{ij} = M_{\bs{k},ij}^{\alpha \beta}$, with entries labelled by the composite index $(ij)$, must share at least one common normal vector [so that they lie in a (hyper-)plane]: Since $\left(w^*_{\alpha i} w_{\beta j} \right)$ does not depend on $\bs{k}$, it must be such a normal vector in order to satisfy the compactness constraints. When this assumption holds, the constraints reduce to equations of the form 
\begin{equation} \label{eq: final_mom_reduced_constraints}
\sum_{ij} M_{\lambda,ij}^{\alpha \beta} \left(w^*_{\alpha i} w_{\beta j} \right) = 0, \quad \lambda = 1 \dots s_{\alpha \beta},
\end{equation}
where the vectors $\vec{M}_{\lambda}^{\alpha \beta}$ span the (hyper-)plane containing all $\vec{M}_{\bs{k}}^{\alpha \beta}$, and $s_{\alpha \beta} \in \mathbb{N}$ is its dimension. These constraints are equivalent to the real-space constraints, Eq.~\eqref{eq: real_space_constraints_reduced}.

\begin{figure}[t]
\centering
\includegraphics[width=1\textwidth,page=7]{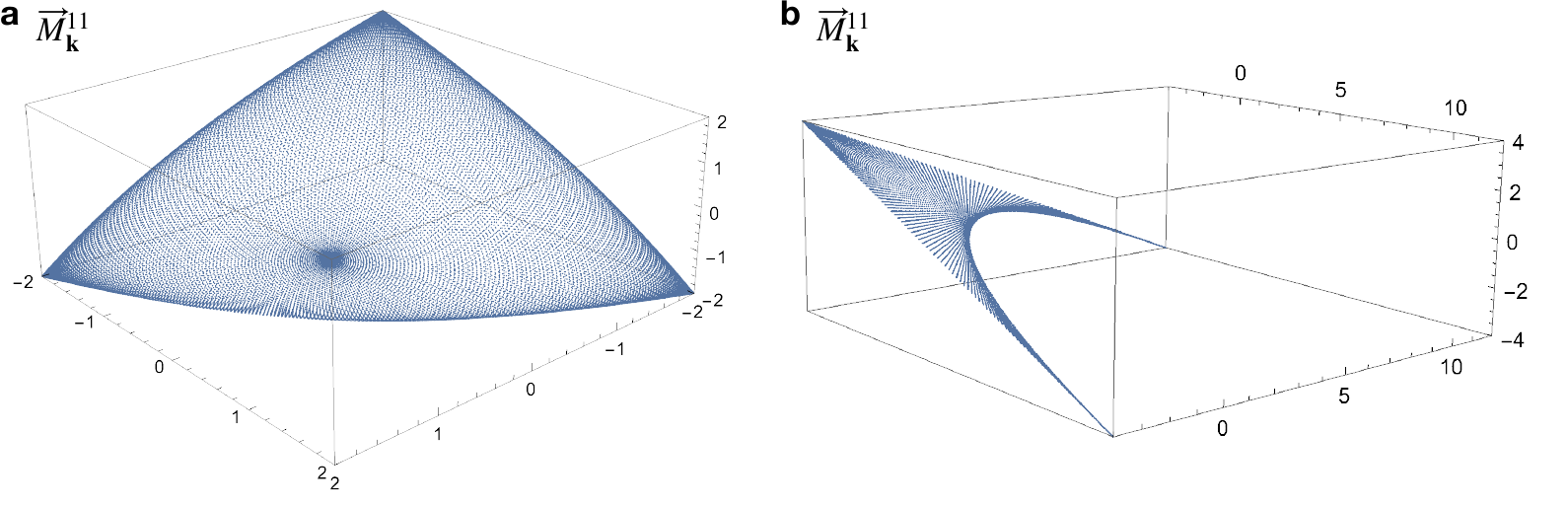}
\caption{Compactness manifolds for OAIs with $\mathcal{C}_2$ and $\mathcal{C}_4$ symmetry. (a)~The set of all vectors $\vec{M}_{\bs{k}}^{1 1}$ for the constraint matrix in Eq.~\eqref{eq: c2_constraintmatrix}. Since this set does not lie in any 2D plane, there is no single vector that is orthogonal to all of the $\vec{M}_{\bs{k}}^{1 1}$, implying that the OAI $(A)_{\mathrm{1a}}\uparrow G$ in Eq.~\eqref{eq: c2simplesubtraction} can never be made compact. (b)~The set of all vectors $\vec{M}_{\bs{k}}^{1 1}$ for the constraint matrix in Eq.~\eqref{eq: c4_constraintmatrix}. The vectors $\vec{M}_{\bs{k}}^{1 1}$ form a 2D plane in 3D space, implying that there is a compact representative of the OAI $(A)_{\mathrm{a}}$ in Eq.~\eqref{eq: c4simplesubtraction}: according to Eq.~\eqref{eq: final_mom_reduced_constraints}, the squared overlap of $\ket{w_{1}}$ [Eq.~\eqref{eq: symmWannierAnsatz}] with the three lattice orbitals is given by the plane normal vector $\vec{n} = (1, 1, 2)^\mathrm{T}$ (see also Fig.~\ref{fig: compactC4counterexample}b).}
\label{fig: c2_manifold}
\end{figure}

\subsubsection{Example: $\mathcal{C}_2$ symmetry} \label{subsec: momconstraints_c2example}
Consider a lattice in wallpaper group $p2$ that has Wyckoff positions 1b, 1c, 1d occupied with a single $s$ orbital (our conventions for wallpaper group $p2$ are explicitly stated at the beginning of Sec.~\ref{sec: nogo_p2symmetry}). We now form the fragile root~\cite{ZhidaFragileTwist2} state
\begin{equation} \label{eq: c2simplesubtraction}
\mathrm{FP} = \left[(A)_{\mathrm{1b}} \oplus (A)_{\mathrm{1c}} \oplus (A)_{\mathrm{1d}}\right]\uparrow G \ominus (A)_{\mathrm{1a}}\uparrow G,
\end{equation}
where the $\mathcal{C}_2$ site-symmetry representation labels follow the Bilbao crystallographic server~\cite{Aroyo2011183}. The fragile phase $\mathrm{FP}$ is the band complement of an OAI at Wyckoff position 1a with $\mathcal{C}_2$ eigenvalue $\lambda_1 = 1$ (we label the Wannier state by $\alpha = 1$). 
Let us assume that the trial Wannier states of this OAI are supported on $\mathcal{C}_2$-related pairs of $\mathrm{1b}$, $\mathrm{1c}$, and $\mathrm{1d}$ positions surrounding $\mathrm{1a}$ (so that $\ket{w_1}$, defined in Eq.~\eqref{eq: symmWannierAnsatz}, overlaps with $3$ lattice sites). On this support, the momentum basis states $\ket{\bs{k}j}$, where $j=1,2,3$ denote the $s$-orbital at 1b, 1c, and 1d, respectively, are given by
\begin{equation}
\begin{aligned}
\ket{\bs{k},1} &= \frac{1}{\sqrt{2}} \left(e^{\mathrm{i} k_x/2} \ket{(0,0),1} + e^{-\mathrm{i} k_x/2} \ket{(-1,0),1} \right), \\
\ket{\bs{k},2} &= \frac{1}{\sqrt{2}} \left(e^{\mathrm{i} (k_x+k_y)/2} \ket{(0,0),2} + e^{-\mathrm{i} (k_x+k_y)/2} \ket{(-1,-1),2} \right), \\
\ket{\bs{k},3} &= \frac{1}{\sqrt{2}} \left(e^{\mathrm{i} k_y/2} \ket{(0,0),3} + e^{-\mathrm{i} k_y/2} \ket{(0,-1),3} \right).
\end{aligned}
\end{equation}
Correspondingly, in the basis $\ket{\bs{R} j} \in \{\ket{(0,0),1},\ket{(0,0),2},\ket{(0,0),3},\ket{(-1,0),1},\ket{(-1,-1),2},\ket{(0,-1),3}\}$, we have
\begin{equation}
P_{\bs{k}} = \frac{1}{2} \begin{pmatrix} 
1 & 0 & 0 & e^{\mathrm{i}k_x} & 0 & 0 \\ 
0 & 1 & 0 & 0 & e^{\mathrm{i}(k_x+k_y)} & 0 \\
0 & 0 & 1 & 0 & 0 & e^{\mathrm{i}k_y} \\
e^{-\mathrm{i}k_x} & 0 & 0 & 1 & 0 & 0 \\
0 & e^{-\mathrm{i}(k_x+k_y)} & 0 & 0 & 1 & 0 \\
0 & 0 & e^{-\mathrm{i}k_y} & 0 & 0 & 1
\end{pmatrix}.
\end{equation}
$\mathcal{C}_2$ symmetry is represented by
\begin{equation}
C_2 = \begin{pmatrix}
0 & 0 & 0 & 1 & 0 & 0 \\
0 & 0 & 0 & 0 & 1 & 0 \\
0 & 0 & 0 & 0 & 0 & 1 \\
1 & 0 & 0 & 0 & 0 & 0 \\
0 & 1 & 0 & 0 & 0 & 0 \\
0 & 0 & 1 & 0 & 0 & 0
\end{pmatrix}.
\end{equation}
Then, for the OAI $(A)_{\mathrm{1a}}\uparrow G$ in Eq.~\eqref{eq: c2simplesubtraction}, we find from Eq.~\eqref{eq: ptildedef_kspace} that ($V = 2$)
\begin{equation}
\tilde{P}_{\bs{k},11} = 2 \begin{pmatrix} 
1 + \cos k_x & 0 & 0 & 1 + \cos k_x & 0 & 0 \\ 
0 & 1 + \cos (k_x + k_y) & 0 & 0 & 1 + \cos (k_x + k_y) & 0 \\
0 & 0 & 1 + \cos k_y & 0 & 0 & 1 + \cos k_y \\
1 + \cos k_x & 0 & 0 & 1 + \cos k_x & 0 & 0 \\ 
0 & 1 + \cos (k_x + k_y) & 0 & 0 & 1 + \cos (k_x + k_y) & 0 \\
0 & 0 & 1 + \cos k_y & 0 & 0 & 1 + \cos k_y \\
\end{pmatrix}.
\end{equation}
Consulting Eq.~\eqref{eq: constraintmat_def}, the constraint matrix in the basis $\ket{i} \in \{\ket{(0,0),1},\ket{(0,0),2},\ket{(0,0),3}\}$ is given by 
\begin{equation} \label{eq: c2_constraintmatrix}
M_{\bs{k},ij}^{1 1} = \bra{i} \left(\tilde{P}_{\bs{k},1 1} - 2 \mathbb{1} \right) \ket{j} = 2\begin{pmatrix} \cos k_x & 0 & 0 \\ 0 & \cos{(k_x + k_y)} & 0 \\ 0 & 0 & \cos k_y \end{pmatrix}_{ij}.
\end{equation}
Because only the diagonal entries are nonzero, the manifold of vectors $(\vec{M}_{\bs{k}}^{1 1})_{i} = M_{\bs{k},ii}^{1 1}$ can be visualized in 3D space and is shown in Fig.~\ref{fig: c2_manifold}a. The resulting manifold is not a (hyper-)plane, implying that there is no choice of $\ket{w_{1}}$ in Eq.~\eqref{eq: symmWannierAnsatz} that has $w^*_{1 i} w_{1 i} \equiv (\vec{w}_{1 1})_i$ orthogonal to all $\vec{M}^{1 1}_{\bs{k}}$. We conclude that there is no compact Wannier basis with support on $6$ lattice sites. In fact, as we will see in Sec.~\ref{sec: nogo_p2symmetry}, the absence of a compact representation is a general property of $\mathcal{C}_2$-protected fragile band complements.

\subsubsection{Example: $\mathcal{C}_4$ symmetry}

\begin{figure}[t]
\centering
\includegraphics[width=1\textwidth,page=2]{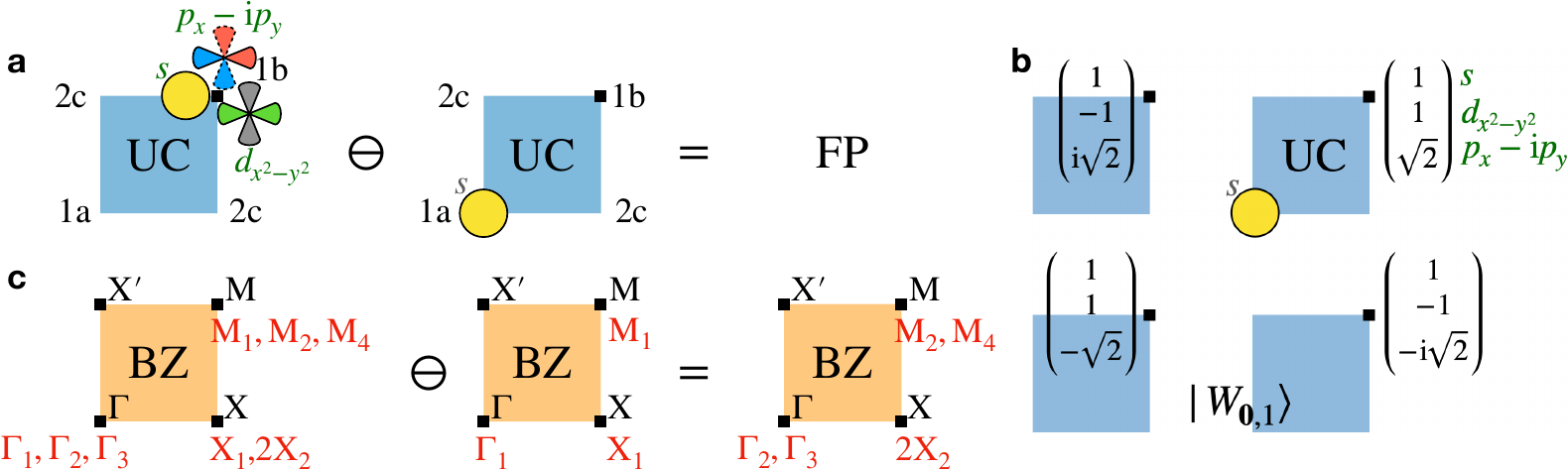}
\caption{(a)~Subtracting an OAI to obtain a fragile root with $\mathcal{C}_4$ rotational symmetry. (b)~Compact Wannier state for the OAI [also obtained from Eq.~\eqref{eq: compactC4state1} for $\lambda=1$], having support on four unit cells (indicated in blue) surrounding the 1a position (indicated in yellow) of the home unit cell at $\bs{R} = \bs{0}$. The vectors shown at each site contain the (non-normalized up to a factor of $4$) overlaps of the Wannier state $\ket{W_{\bs{0},1}}$ with the on-site orbitals (the corresponding orbital labels are indicated for the home unit cell). (c)~Brillouin zone decomposition of the relevant bands into irreducible representations. (Representation labels follow the Bilbao Crystallographic Server~\cite{Aroyo2011183}.)}
\label{fig: compactC4counterexample}
\end{figure}

Consider a lattice in wallpaper group $p4$ that has Wyckoff position 1b occupied with three physical orbitals: $s$, $d_{x^2-y^2}$, and $p_x -\mathrm{i} p_y$ (our conventions for wallpaper group $p4$ are explicitly stated at the beginning of Sec.~\ref{sec: overarching_p4symmetry}). We can form a fragile root state via the subtraction
\begin{equation} \label{eq: c4simplesubtraction}
\mathrm{FP} = \left[(A)_{\mathrm{1b}} \oplus (B)_{\mathrm{1b}} \oplus ({}^2E)_{\mathrm{1b}}\right]\uparrow G \ominus (A)_{\mathrm{1a}}\uparrow G,
\end{equation}
where we again used the Bilbao crystallographic server~\cite{Aroyo2011183} conventions. The same subtraction is depicted in Fig.~\ref{fig: compactC4counterexample}a and~c. The fragile phase $\mathrm{FP}$ is the band complement of an OAI at Wyckoff position 1a with $\mathcal{C}_4$ eigenvalue $\lambda = 1$. Let us assume that the trial Wannier states of this OAI are supported on four lattice sites (which are the four $\mathcal{C}_4$-related $\mathrm{1b}$ positions surrounding $\mathrm{1a}$, so that $\ket{w_{1}}$ has support on $1$ lattice site). On this support, the momentum basis states $\ket{\bs{k}j}$, where $j=1,2,3$ denote the $s$, $d_{x^2-y^2}$, and $p_x -\mathrm{i} p_y$ orbital at 1b, respectively, are given by
\begin{equation}
\begin{aligned}
\ket{\bs{k},1} &= \frac{1}{\sqrt{2}} \left(e^{\mathrm{i} (k_x+k_y)/2} \ket{(0,0),1} + e^{\mathrm{i} (-k_x+k_y)/2} \ket{(-1,0),1} + e^{-\mathrm{i} (k_x+k_y)/2} \ket{(-1,-1),1} + e^{\mathrm{i} (k_x-k_y)/2} \ket{(0,-1),1} \right), \\
\ket{\bs{k},2} &= \frac{1}{\sqrt{2}} \left(e^{\mathrm{i} (k_x+k_y)/2} \ket{(0,0),2} + e^{\mathrm{i} (-k_x+k_y)/2} \ket{(-1,0),2} + e^{-\mathrm{i} (k_x+k_y)/2} \ket{(-1,-1),2} + e^{\mathrm{i} (k_x-k_y)/2} \ket{(0,-1),2} \right), \\
\ket{\bs{k},3} &= \frac{1}{\sqrt{2}} \left(e^{\mathrm{i} (k_x+k_y)/2} \ket{(0,0),3} + e^{\mathrm{i} (-k_x+k_y)/2} \ket{(-1,0),3} + e^{-\mathrm{i} (k_x+k_y)/2} \ket{(-1,-1),3} + e^{\mathrm{i} (k_x-k_y)/2} \ket{(0,-1),3} \right). \\
\end{aligned}
\end{equation}
Correspondingly, in the basis 
\begin{equation} \begin{aligned}
\ket{\bs{R} j} \in \{&\ket{(0,0),1},\ket{(0,0),2},\ket{(0,0),3},\ket{(-1,0),1},\ket{(-1,0),2},\ket{(-1,0),3}, \\ &\ket{(-1,-1),1},\ket{(-1,-1),2},\ket{(-1,-1),3},\ket{(0,-1),1},\ket{(0,-1),2},\ket{(0,-1),3}\},
\end{aligned} \end{equation}
we can find $P_{\bs{k}}$, $C_4$, and $\tilde{P}_{\bs{k},11}$ exactly as was done in the $\mathcal{C}_2$ example of Sec.~\ref{subsec: momconstraints_c2example} (using that the orbitals labelled by $j=1,2,3$ have $\mathcal{C}_4$ eigenvalues $1$, $-1$, and $\mathrm{i}$, respectively). The resulting constraint matrix in the unit cell basis $\ket{i} \in \{\ket{(0,0),1},\ket{(0,0),2},\ket{(0,0),3}\}$ is then given by 
\begin{equation} \label{eq: c4_constraintmatrix}
M_{\bs{k},ij}^{1 1} = \bra{i} \left(\tilde{P}_{\bs{k},1 1} - 4 \mathbb{1} \right) \ket{j} = 4\begin{pmatrix} \cos k_x + \cos k_y + \cos k_x \cos k_y & 0 & 0 \\ 0 & -\cos k_x - \cos k_y + \cos k_x \cos k_y & 0 \\ 0 & 0 & - \cos k_x \cos k_y \end{pmatrix}_{ij}.
\end{equation}
The manifold of vectors $(\vec{M}_{\bs{k}}^{\alpha \beta})_{i} = M_{\bs{k},ii}^{\alpha \beta}$ can be visualized in 3D space and is shown in Fig.~\ref{fig: c2_manifold}b. In contrast to the example of Sec.~\ref{subsec: momconstraints_c2example}, it lies in a 2D plane, implying that a compact representative of the OAI $(A)_{\mathrm{1a}}\uparrow G$ corresponding to the plane normal direction $\vec{n} = (1, 1, 2)^\mathrm{T}$ exists. The corresponding compact Wannier state is obtained from the plane normal and is depicted in Fig.~\ref{fig: compactC4counterexample}b. In fact, as we will show in Sec.~\ref{sec: overarching_p4symmetry}, all (spatially-obstructed) $\mathcal{C}_4$-protected fragile complements have a compact representative as long as they are not already non-compact due to $\mathcal{C}_2$ symmetry.

\section{$\mathcal{C}_2$ symmetry} \label{sec: nogo_p2symmetry}
A fragile state cannot be written as any atomic insulator. It is the band complement either of another fragile state, or of an OAI. In this section, we prove that OAIs in wallpaper group $p2$ which have a fragile complement cannot be expressed in terms of compactly supported Wannier functions. Our proof can be straightforwardly generalized to $d$ spatial dimensions with inversion symmetry.

We use conventions where the lattice vectors of the $\mathcal{C}_2$-symmetric wallpaper group $p2$ are given by $\bs{a}_1 = \hat{x}$ ($\hat{x}$ is the unit vector in $x$-direction) and $\bs{a}_2 = \hat{y}$. Correspondingly, the reciprocal lattice vectors are given by $\bs{b}_1 = 2\pi \hat{x}$, $\bs{b}_2 = 2\pi \hat{y}$. The maximal Wyckoff positions of the unit cell have coordinates $\bs{t}_\mathrm{1a} = \bs{0}$, $\bs{t}_\mathrm{1b} = \hat{x}/2$, $\bs{t}_\mathrm{1c} = (\hat{x}+\hat{y})/2$, and $\bs{t}_\mathrm{1d} = \hat{y}/2$. The high-symmetry momenta of the Brillouin zone are defined as $\bs{\Gamma} = \bs{0}$, $\bs{X} = \bs{b}_1/2$, $\bs{M} = \bs{b}_1/2 + \bs{b}_2/2$, and $\bs{Y} = \bs{b}_2/2$. For future reference, we list the elementary band representations (EBRs) of wallpaper group $p2$ in Tab.~\ref{tab: c2_ebrs}.
\begin{table}[h]
\begin{tabular}{|c|c|c|c|c|c|}
\hline
$\lambda^{(2)}$              & WP   & $\mathbf{\Gamma}$ & $\mathbf{X}$ & $\mathbf{M}$ & $\mathbf{Y}$ \\ \hline
$1$        & 1a & $1$                  & $1$             & $1$       & $1$                    \\
              & 1b & $1$                  & $-1$             & $-1$  & $1$                         \\ 
              & 1c & $1$                  & $-1$             & $1$  & $-1$                         \\ 
              & 1d & $1$                  & $1$             & $-1$  & $-1$                         \\  \hline
$-1$        & 1a & $-1$                  & $-1$             & $-1$       & $-1$                    \\
              & 1b & $-1$                  & $1$             & $1$  & $-1$                         \\ 
              & 1c & $-1$                  & $1$             & $-1$  & $1$                         \\ 
              & 1d & $-1$                  & $-1$             & $1$  & $1$                         \\  \hline
\end{tabular}
\caption{\label{tab: c2_ebrs}Wyckoff position-resolved EBRs of wallpaper group $p2$. The site-symmetry representations are labelled by their (spinless) $\mathcal{C}_2$ eigenvalue $\lambda^{(2)}$. Depending on which Wyckoff position (WP) they are placed at, they give rise to different Bloch band $\mathcal{C}_2$ eigenvalues at the high-symmetry momenta $\mathbf{\Gamma},\mathbf{X},\mathbf{M}$, and $\mathbf{Y}$, respectively.}
\end{table}

\subsection{Overview}
In order to prove that fragile complements in wallpaper group $p2$ are non-compact, we first show in real space that the maximal number of compact Wannier states of the same $\mathcal{C}_2$ eigenvalue ($\pm 1$) is equal to the number of $s \oplus p$ pairs of physical orbitals in the unit cell. 
Then, in momentum space, we go on to prove that the maximal number of obstructed bands with the same $\mathcal{C}_2$ eigenvalue at $\bs{\Gamma}$ that has an atomic (in contrast to a fragile) complement is also equal to the number of $s \oplus p$ pairs in the unit cell.
By applying the necessariy condition $N(\mathrm{AI}) < \bar{N}(\mathrm{OAI})$ for fragile bands that was derived in Sec.~\ref{subsec: realspace_fragile}, we therefore conclude that fragile bands cannot be the complements of compact bands, or else they would be atomic.

\begin{figure}[t]
\centering
\includegraphics[width=0.7\textwidth,page=1]{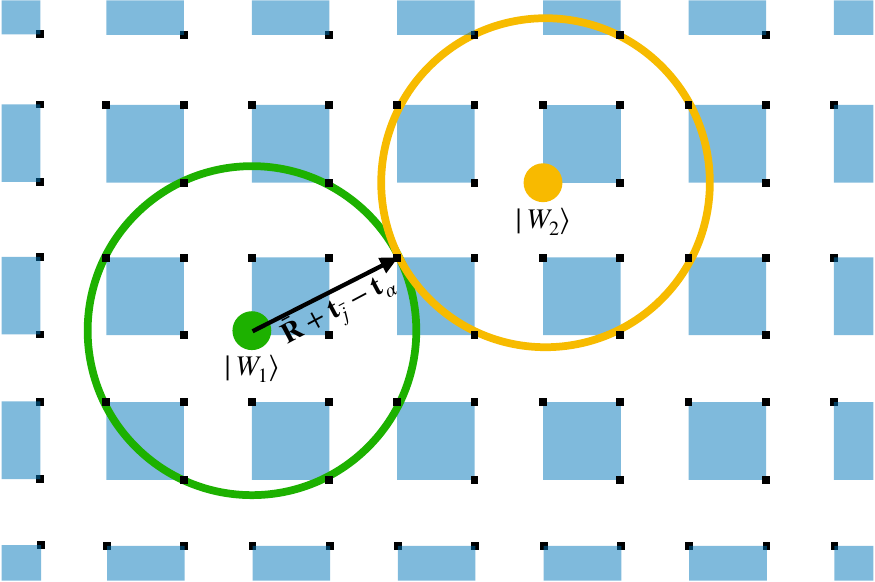}
\caption{$\mathcal{C}_2$ rotational symmetry ensures that for a compact Wannier state $\ket{W_1}$, there is another translated Wannier state $\ket{W_2}$ that shares a single lattice site of nonzero overlap. Orthogonality $\braket{W_1 | W_2} = 0$ then implies that this site needs to host at least one $s \oplus p$ pair of orbitals. Note that here we have chosen the lattice to contain orbitals at the $\mathrm{1b}$, $\mathrm{1c}$, and $\mathrm{1d}$ Wyckoff positions, so that the Wannier states at $\mathrm{1a}$ belong to an OAI.}
\label{fig: compactInvObstruction}
\end{figure}

\subsection{Compactly localizing Wannier states} \label{subsec: c2_sp_realspace_crit}
We begin by analyzing the conditions for compact and obstructed Wannier states in real space. 
We again denote the basis states by $\ket{\bs{R} i \mu}$, where the index $\mu = 1 \dots D$ identifies a particular orbital at site $i = 1 \dots N$ in the unit cell at $\bs{R}$. Let $\bs{t}_i \in \{(0,0), (1/2,0), (1/2,1/2), (0,1/2)\}$ denote the position of the $i$th orbital as measured from the center of the unit cell. Our restriction to maximal Wyckoff positions (explained at the beginning of Sec.~\ref{subsec: mobile_clusters}) means that each basis state is mapped to a translated copy of itself when acted upon by $\mathcal{C}_2$ symmetry:
\begin{equation}
C_2|_{\bs{r}} \ket{\bs{R}' i \mu} = \gamma_\mu \ket{\bs{R} i \mu} \text{ whenever } \bs{R}' + \bs{t}_i = \bs{r} - (\bs{R} + \bs{t}_i - \bs{r}).
\end{equation}
Here, $\bs{r}$ is any $\mathcal{C}_2$ center of the infinite lattice (we can choose it to be $\bs{r} = \bs{0}$), and $\gamma_\mu = \pm 1$ is the $\mathcal{C}_2$ eigenvalue of the orbital labelled by $\mu$.

We now consider an arbitrary set of OAI bands on this lattice. We denote their Wannier states by $\ket{W_{\bs{R}\alpha}}$, $\alpha = 1 \dots M$, where $M \leq D N$ is the number of bands that make up the OAI. It is always possible to choose these Wannier states in an $\mathcal{C}_2$-symmetric fashion, so that 
\begin{equation}
C_2|_{\bs{t}_\alpha} \ket{W_{\bs{0}\alpha}} = \lambda_\alpha \ket{W_{\bs{0}\alpha}}
\end{equation}
holds for the $\mathcal{C}_2$ center at $\bs{r} = \bs{t}_\alpha$, where $\lambda_\alpha = \pm 1$ is the $\mathcal{C}_2$ eigenvalue of the respective Wannier function, and $\bs{t}_\alpha$ is the Wannier center. By translational symmetry, all further Wannier states $\ket{W_{\bs{R}\alpha}}$ are $\mathcal{C}_2$-symmetric about their centers $\bs{t}_{\bs{R}\alpha} \equiv \bs{R} + \bs{t}_\alpha$.

For compactness to be a nontrivial property, it is crucial that the Wannier states belong to an OAI and not to a trivial (unobstructed) atomic insulator: otherwise we could just choose a subset of the basis states $\ket{\bs{R} i \mu}$ as orthonormal and compact Wannier states. For a spatially-obstructed OAI, the Wannier centers
\begin{equation}
\bs{t}_{\bs{R}\alpha} = \braket{W_{\bs{R}\alpha}| \hat{\bs{x}} |W_{\bs{R}\alpha}} = \sum_{\bs{R}i\mu} (\bs{R}+\bs{t}_i) \left|\braket{\bs{R}i\mu|W_{\bs{R}\alpha}}\right|^2,
\end{equation}
where $\hat{\bs{x}}$ is the position operator, take values that are not lattice sites: $\bs{t}_{\bs{R}\alpha} \neq (\bs{R}'+ \bs{t}_i)$ for any choice of $\bs{R}'$,$i$. Conversely, if the OAI is representation-obstructed, lattice orbitals with $\mathcal{C}_2$ eigenvalue $\gamma_\mu = - \lambda_\alpha$ might be present at $\bs{t}_{\bs{R}\alpha}$, however, these cannot lie in the support of $\ket{W_{\bs{R}\alpha}}$ as they have a different $\mathcal{C}_2$ eigenvalue. Since the Wannier states are $\mathcal{C}_2$-symmetric, $\bs{t}_{\bs{0}\alpha} \equiv \bs{t}_{\alpha}$ must necessarily be a maximal Wyckoff position of multiplicity $1$, and thereby a $\mathcal{C}_2$ center of the infinite system.

Let us now assume that $\ket{W_{\bs{0}\alpha}}$ is compact and has overlap at most with orbitals at $\bar{\bs{R}} + \bs{t}_{\bar{j}}$, that is, $|\bar{\bs{R}} + \bs{t}_{\bar{j}} - \bs{t}_{\alpha}|$ is the maximal radius of each $\ket{W_{\bs{R}\alpha}}$ (here, $\bar{\bs{R}}$ is a given lattice vector, and $\bs{t}_{\bar{j}}$ is a given site in the unit cell). This radius is nonzero because we assume the Wannier states to be obstructed, so even for $\bar{\bs{R}} = \bs{0}$ we have $\bs{t}_{\bar{j}} \neq \bs{t}_{\alpha}$. Then, $\bar{\bs{R}} + \bs{t}_{\bar{j}}$ is the only site of overlap of $\ket{W_{\bs{0}\alpha}}$ with $\ket{W_{\bs{R}'\alpha}}$, where 
\begin{equation}
\bs{R}' = 2\bar{\bs{R}}+2\bs{t}_{\bar{j}}-2\bs{t}_{\alpha}.
\end{equation}
This is because translational symmetry implies that
\begin{equation}
\bs{t}_{\bs{R}'\alpha} - (\bar{\bs{R}}+\bs{t}_{\bar{j}}) = (\bar{\bs{R}}+\bs{t}_{\bar{j}}) - \bs{t}_{\alpha},
\end{equation}
while $\mathcal{C}_2$ symmetry implies the shared overlap.
If the two Wannier states had more than one site in common, our assumption that $\ket{W_{\bs{R}\alpha}}$ is supported only up to a distance $|\bar{\bs{R}} + \bs{t}_{\bar{j}} - \bs{t}_{\alpha}|$ would be violated. When the choice of $\bar{\bs{R}} + \bs{t}_{\bar{j}}$ is not unique (even after excluding the $\mathcal{C}_2$-related partner), we may choose any maximal $\bar{\bs{R}} + \bs{t}_{\bar{j}}$: since all equivalent choices lie on a circle of radius $|\bar{\bs{R}} + \bs{t}_{\bar{j}} - \bs{t}_{\alpha}|$ around $\bs{t}_{\alpha}$ (shown in green in Fig.~\ref{fig: compactInvObstruction}), the state $\ket{W_{\bs{R}'\alpha}}$ will only overlap with $\ket{W_{\bs{0}\alpha}}$ on a single site. See Fig.~\ref{fig: compactInvObstruction} for a visual representation in the case where $|\bar{\bs{R}} + \bs{t}_{\bar{j}} - \bs{t}_{\alpha}|$ extends to next-next-next-nearest neighbor sites.

We then evaluate the overlap
\begin{equation} \label{eq: c2_sp_requirement}
\begin{aligned}
\braket{W_{\bs{R}'\alpha}|W_{\bs{0}\alpha}} &= \sum_\mu \braket{W_{\bs{R}'\alpha} |\bar{\bs{R}}\bar{j}\mu} \braket{\bar{\bs{R}}\bar{j}\mu | W_{\bs{0}\alpha}} = \sum_\mu \braket{W_{\bs{0}\alpha} |(\bar{\bs{R}}-\bs{R}')\bar{j}\mu} \braket{\bar{\bs{R}}\bar{j}\mu | W_{\bs{0}\alpha}} \quad (\text{by translational symmetry}) \\
&= \sum_\mu \gamma_\mu \braket{W_{\bs{0}\alpha} | C_2|_{\bs{t}_\alpha} | \bar{\bs{R}}\bar{j}\mu} \braket{\bar{\bs{R}}\bar{j}\mu | W_{\bs{0}\alpha}} = \lambda_\alpha \sum_\mu \gamma_\mu \left|\braket{\bar{\bs{R}}\bar{j}\mu | W_{\bs{0}\alpha}}\right|^2.
\end{aligned}
\end{equation}
This is a sum of strictly positive numbers $\left|\braket{\bar{\bs{R}}\bar{j}\mu | W_{\bs{0}\alpha}}\right|^2$ that are weighted by the $\mathcal{C}_2$ eigenvalues $\gamma_\mu = \pm 1$ ($s$ or $p$) of physical orbitals. For $\ket{W_{\bs{R}\alpha}}$ to be compact and orthonormal $\braket{W_{\bs{R}'\alpha}|W_{\bs{0}\alpha}} = 0$, we then need at least one $s \oplus p$ pair of physical orbitals, so that $\braket{W_{\bs{R}'\alpha}|W_{\bs{0}\alpha}}$ can be made to vanish (a Wannier basis is required to be orthonormal). 

Moreover, when the unit cell indeed hosts an $s \oplus p$ pair at $\bs{t}_{\bar{j}}$, we can immediately find a compact representation for both $\ket{W_{\bs{R}\alpha}}$ and another set of Wannier states $\ket{W_{\bs{R}\tilde{\alpha}}}$, also localized at Wyckoff position $\bs{t}_\alpha$ but with $\mathcal{C}_2$ eigenvalue $\lambda_{\tilde{\alpha}} = -\lambda_{\alpha}$: the $s \oplus p$ pair forms a mobile cluster (see Sec.~\ref{subsec: mobile_clusters}) and can be used to construct compact states of $s \oplus p$ character at $\bs{t}_\alpha$ (or, in fact, any maximal Wyckoff position). More concretely, we form the covalent bond states
\begin{equation} \label{eq: c2compact_covalents}
\begin{aligned}
\ket{W_{\bs{R}\alpha}} &= \frac{1}{2} \left[\left(\ket{\bs{R},\bar{j},1} + \ket{\bs{R},\bar{j},2}\right) + \lambda_\alpha \left(\ket{\bs{R}+2\bs{t}_\alpha,\bar{j},1} - \ket{\bs{R}+2\bs{t}_\alpha,\bar{j},2}\right) \right], \\
\ket{W_{\bs{R}\tilde{\alpha}}} &= \frac{1}{2} \left[\left(\ket{\bs{R},\bar{j},1} + \ket{\bs{R},\bar{j},2}\right) - \lambda_\alpha \left(\ket{\bs{R}+2\bs{t}_\alpha,\bar{j},1} - \ket{\bs{R}+2\bs{t}_\alpha,\bar{j},2}\right) \right],
\end{aligned}
\end{equation}
where $\mu = 1,2$ labels the $s$ and the $p$ orbital at site $\bs{t}_{\bar{j}}$, respectively. These states satisfy
\begin{equation}
\braket{W_{\bs{R}\alpha} |W_{\bs{R}'\alpha}} = \braket{W_{\bs{R}\tilde{\alpha}} |W_{\bs{R}'\tilde{\alpha}}} = \delta_{\bs{R},\bs{R}'}, \quad \braket{W_{\bs{R}\alpha} |W_{\bs{R}'\tilde{\alpha}}} = 0,
\end{equation}
as required for a compact Wannier basis.
Any further compact Wannier states cannot have support on the same $s \oplus p$ pair: since for each unit cell, we have constructed two compact Wannier states ($\ket{W_{\bs{R}\alpha}}$ and $\ket{W_{\bs{R}\tilde{\alpha}}}$) out of two physical orbitals, the Hilbert space provided by the $s \oplus p$ pair is fully exhausted. Hence, in order to find additional compact Wannier states that are orthogonal to all $\ket{W_{\bs{R}\alpha}}$ and $\ket{W_{\bs{R}\tilde{\alpha}}}$, more $s \oplus p$ pairs of orbitals are required. We conclude that, in order to construct $M$ compact obstructed Wannier states of the same $\mathcal{C}_2$ eigenvalue $\lambda$, we require a unit cell hosting (at least) $M$ $s \oplus p$ pairs of physical orbitals. Given $M$ $s \oplus p$ pairs, we may then additionally construct $M$ compact and obstructed Wannier states of $\mathcal{C}_2$ eigenvalue $-\lambda$.

\subsection{Trivializing obstructed band complements}
We now examine the momentum-space conditions for a given set of obstructed bands to have a fragile complement. The same conditions can also be obtained from arguments based on real-space indicators (RSIs)~\cite{ZhidaFragileTwist2} or Wilson loops~\cite{Alexandradinata14}. We begin by recalling that an atomic insulator can be decomposed into EBRs that are induced from atomic orbitals at maximal Wyckoff positions in the unit cell~\cite{Bradlyn17}. The relationship between unit cell (UC) orbitals and Brillouin zone (BZ) symmetry eigenvalues is encoded in the EBR matrix~\cite{Song20,ZhidaFragileTwist2}, whose columns contain the BZ eigenvalue content of a given atomic orbital ($s$ or $p$) at a given Wyckoff position (1a, 1b, 1c, or 1d). In particular, let $n^\pm_{\bs{k}}$ denote the multiplicity of $\pm 1$ $\mathcal{C}_2$ eigenvalues at the $\mathcal{C}_2$-symmetric momentum $\bs{k}$ in the Brillouin zone, and let $n^+_{\mathrm{1x}}$ ($n^-_{\mathrm{1x}}$) count the $s$ ($p$) orbitals at maximal Wyckoff position $1x$ in the unit cell. Consulting Tab.~\ref{tab: c2_ebrs}, we have
\begin{equation} \label{eq: UCtoBZ_mapping}
\begin{aligned}
\begin{pmatrix}n^+_{\bs{\Gamma}}\\n^+_{\bs{X}}\\n^+_{\bs{Y}}\\n^+_{\bs{M}}\\n^-_{\bs{\Gamma}}\\n^-_{\bs{X}}\\n^-_{\bs{Y}}\\n^-_{\bs{M}}\end{pmatrix} = 
\begin{pmatrix} 
1 & 1 & 1 & 1 & 0 & 0 & 0 & 0 \\ 
1 & 0 & 0 & 1 & 0 & 1 & 1 & 0 \\
1 & 1 & 0 & 0 & 0 & 0 & 1 & 1 \\
1 & 0 & 1 & 0 & 0 & 1 & 0 & 1 \\
0 & 0 & 0 & 0 & 1 & 1 & 1 & 1 \\
0 & 1 & 1 & 0 & 1 & 0 & 0 & 1 \\
0 & 0 & 1 & 1 & 1 & 1 & 0 & 0 \\
0 & 1 & 0 & 1 & 1 & 0 & 1 & 0
\end{pmatrix}
\begin{pmatrix}n^+_{1a}\\n^+_{1b}\\n^+_{1c}\\n^+_{1d}\\n^-_{1a}\\n^-_{1b}\\n^-_{1c}\\n^-_{1d}\end{pmatrix},
\end{aligned}
\end{equation}
which we abbreviate by 
\begin{equation} \label{eq: realtomomentumspace_invariants}
\bs{n}_{\mathrm{BZ}} = \Xi \bs{n}_{\mathrm{UC}}.
\end{equation}
Here, $\Xi$ is the EBR matrix of wallpaper group $p2$~\cite{Bradlyn17}.
All insulators with equal symmetry indicators $\bs{n}_{\mathrm{BZ}}$ can be adiabatically transformed into one another. Insulators exhibiting symmetry indicators that are not of the form of Eq.~\eqref{eq: realtomomentumspace_invariants} are topological, in that they cannot be deformed to any atomic limit.

When the crystal is built from a certain set of orbitals at maximal Wyckoff positions of the lattice, these together give rise to a specific multiplicity vector $\bs{n}_{\mathrm{UC}}$ formed of non-negative integers. Importantly $\det(\Xi) = 0$, and so $\Xi$ is in general not invertible: there are multiple real-space configurations $\bs{n}_{\mathrm{UC}}$ that correspond to the same momentum-space eigenvalues $\bs{n}_{\mathrm{BZ}}$. The null space of $\Xi$ is spanned by the three vectors
\begin{equation}
\bs{n}^{(1)}_{\mathrm{ker} \Xi} = \begin{pmatrix} -1 & 0 & 0 & 1 & -1 & 0 & 0 & 1 \end{pmatrix}^\mathrm{T}, \quad \bs{n}^{(2)}_{\mathrm{ker} \Xi} = \begin{pmatrix} -1 & 0 & 1 & 0 & -1 & 0 & 1 & 0 \end{pmatrix}^\mathrm{T}, \quad \bs{n}^{(3)}_{\mathrm{ker} \Xi} = \begin{pmatrix} -1 & 1 & 0 & 0 & -1 & 1 & 0 & 0 \end{pmatrix}^\mathrm{T}.
\end{equation}
Their physical interpretation is that any $s \oplus p$ pair of orbitals (a mobile cluster) can be moved around the unit cell adiabatically, and so can equally well (with respect to the resulting $\bs{n}_{\mathrm{BZ}}$) locate at any maximal Wyckoff position. This is just the statement that an $s \oplus p$ pair forms a mobile cluster for $\mathcal{C}_2$ symmetry as discussed in Sec.~\ref{subsec: mobile_clusters}. Correspondingly, subtracting an $s$ (a $p$) orbital from any site is equivalent to adding a $p$ (an $s$) orbital to the same site and subtracting an $s \oplus p$ pair from any other site. 

The complement of an OAI band is fragile topological when its $\bs{n}_{\mathrm{BZ}}$ cannot be expressed in terms of a $\bs{n}_{\mathrm{UC}}$ corresponding to a positive sum of EBRs~\cite{AshvinFragile}. All $\bs{n}_{\mathrm{UC}}$ vectors belonging to physically realizable atomic insulators are comprised of non-negative integers, while a fragile $\bs{n}_{\mathrm{UC}}$ necessarily involves negative entries. We can exclude the case of strong topology, where the entries of $\bs{n}_{\mathrm{UC}}$ cannot be chosen integer-valued at all, because the complement of any strong topological set of bands is itself strong topological~\cite{AshvinFragile}. We now formally write
\begin{equation}
\mathrm{C} = \mathrm{AI} \ominus \mathrm{OAI},
\end{equation}
where $\mathrm{AI}$ denotes the atomic insulator formed of the unobstructed lattice bands, $\mathrm{OAI}$ is the obstructed atomic insulator, and $\mathrm{C}$ its band complement.
Taking the band complement corresponds to a subtraction on both sides of Eq.~\eqref{eq: realtomomentumspace_invariants}, so that we have
\begin{equation}
\bs{n}_{\mathrm{BZ}}(\mathrm{C}) = \bs{n}_{\mathrm{BZ}}(\mathrm{AI}) - \bs{n}_{\mathrm{BZ}}(\mathrm{OAI}) = \Xi [\bs{n}_{\mathrm{UC}}(\mathrm{AI}) - \bs{n}_{\mathrm{UC}}(\mathrm{OAI})]
\end{equation}
Note that $\bs{n}_{\mathrm{UC}}(\mathrm{AI}) - \bs{n}_{\mathrm{UC}}(\mathrm{OAI})$ necessarily has negative elements by virtue of the OAI being obstructed. The complement is trivial (equivalent to an atomic insulator) if and only if all elements of this vector can be made non-negative under the equivalence relation of adding linear superpositions of $\bs{n}^{(1),(2),(3)}_{\mathrm{ker} \Xi}$ -- otherwise it is fragile. For wallpaper group $p2$, this condition is equivalent to the fragile criteria developed in Ref.~\onlinecite{ZhidaFragileTwist2}.

Consider now the complement of $M$ obstructed bands with equal $\mathcal{C}_2$ eigenvalue at $\bs{\Gamma}$, these have Wannier states of equal $\mathcal{C}_2$ eigenvalue for rotations about their respective Wannier center. (We choose $\bs{\Gamma}$ for convenience, the same analysis can be carried out with respect to any other of the $\mathcal{C}_2$-symmetric momenta in the Brillouin zone.) In order for $\bs{n}_{\mathrm{BZ}}(\mathrm{C})$ to be equivalent to an OAI, $\bs{n}_{\mathrm{UC}}(\mathrm{AI}) - \bs{n}_{\mathrm{UC}}(\mathrm{OAI})$ should be transformable into a vector with non-negative entries under the equivalence relation of adding linear superpositions of $\bs{n}^{(1),(2),(3)}_{\mathrm{ker} \Xi}$. Since each addition of $\bs{n}^{(1),(2),(3)}_{\mathrm{ker} \Xi}$ removes an $s \oplus p$ pair at one Wyckoff position and adds it to another, this is only possible if $\bs{n}_{\mathrm{UC}}$ has at least $M$ $s \oplus p$ pairs available. We conclude that the maximal number of obstructed bands (with equal $\mathcal{C}_2$ eigenvalue at $\bs{\Gamma}$) that has an atomic (in contrast to a fragile) complement is equal to the number of $s \oplus p$ pairs in the unit cell. Moreover, in presence of $M$ $s \oplus p$ pairs, the complement of $M$ obstructed bands with equal $\mathcal{C}_2$ eigenvalue at $\bs{\Gamma}$ will contain $M$ obstructed bands exhibiting the opposite $\mathcal{C}_2$ eigenvalue at $\bs{\Gamma}$.

\subsection{No-go theorem}
We have seen that the unit cell of a crystal in wallpaper group $p2$ needs to host at least $M$ $s \oplus p$ pairs of physical orbitals in order to support $M$ compact and obstructed Wannier states of the same $\mathcal{C}_2$ eigenvalue. Moreover, in Eq.~\eqref{eq: c2compact_covalents}, we provided an explicit construction of compact Wannier states in the case where a sufficient number of $s \oplus p$ pairs of orbitals is available.

At the same time, we have shown that at least $M$ $s \oplus p$ pairs of physical orbitals per unit cell are required in order for the complement of $M$ obstructed bands of the same $\mathcal{C}_2$ eigenvalue at $\bs{\Gamma}$, or in fact at any other high-symmetry point of the Brillouin zone, to realize an atomic insulator.

A fragile set of bands is by definition not trivializable (not expressible as an atomic insulator) without enlarging the occupied subspace~\cite{AshvinFragile}. We therefore conclude that with $\mathcal{C}_2$ symmetry, all OAIs having a fragile band complement cannot be expressed in terms of compactly localized Wannier states (although their Wannier functions can be made to decay exponentially, because they are topologically trivial~\cite{Brouder2007Exponential,Soluyanov2011Wannier,Bradlyn17}). Otherwise, the lattice would host enough $s \oplus p$ pairs to also trivialize the fragile bands.

Since fragile bands with $\mathcal{C}_2$ symmetry either have OAIs or fragile bands as their complement~\cite{AshvinFragile} (the latter of which do not even support exponentially localized Wannier functions), we deduce that the complement of \emph{any} fragile set of bands in wallpaper group $p2$ necessarily has non-compact Wannier functions.

\section{$\mathcal{C}_4$ symmetry} \label{sec: overarching_p4symmetry}
In this section, we prove that $\mathcal{C}_4$-protected OAIs with spatial obstruction are compact if and only if they are compact when $\mathcal{C}_4$ symmetry is relaxed to $\mathcal{C}_2$ symmetry. (We have already fully characterized the sets of compact and non-compact $\mathcal{C}_2$-protected OAIs in Sec.~\ref{sec: nogo_p2symmetry}.) Clearly, an OAI that is non-compact with $\mathcal{C}_2$ symmetry must remain non-compact with $\mathcal{C}_4$ symmetry: because $(\mathcal{C}_4)^2 = \mathcal{C}_2$ holds, this symmetry only poses additional constraints on any compact basis candidates (these now also have to respect $\mathcal{C}_4$ symmetry in addition to $\mathcal{C}_2$ symmetry and orthonormality). To prove our claim, we must therefore only show that all $\mathcal{C}_4$-symmetric, $\mathcal{C}_2$-compact OAIs -- those built from lattices containing a sufficient number of $s \oplus p$ pairs of physical orbitals, as explained in Sec.~\ref{sec: nogo_p2symmetry} -- are also guaranteed to have a $\mathcal{C}_4$-symmetric compact Wannier basis.

We use conventions where the lattice vectors of the $\mathcal{C}_4$-symmetric wallpaper group $p4$ are given by $\bs{a}_1 = \hat{x}$, $\bs{a}_2 = \hat{y}$. Correspondingly, the reciprocal lattice vectors are given by $\bs{b}_1 = 2\pi \hat{x}$, $\bs{b}_2 = 2\pi \hat{y}$. The maximal Wyckoff positions of the unit cell are located at $\bs{t}_\mathrm{1a} = \bs{0}$, $\bs{t}_\mathrm{1b} = (\hat{x}+\hat{y})/2$, and $\bs{t}_\mathrm{2c} = \{\hat{x}/2,\hat{y}/2\}$. The high-symmetry momenta of the Brillouin zone are defined as $\bs{\Gamma} = \bs{0}$, $\bs{X} = \bs{b}_1/2$, $\bs{X}' = \bs{b}_2/2$, and $\bs{M} = \bs{b}_1/2 + \bs{b}_2/2$. Since spinless $\mathcal{C}_4$ symmetry satisfies $(\mathcal{C}_4)^4 = \mathbb{1}$, its eigenvalues are taken from the set $\{1, -1, \mathrm{i}, -\mathrm{i}\}$. For future reference, we list the EBRs of wallpaper group $p4$ in Tabs.~\ref{tab: c4_ebrs},~\ref{tab: c4_ebrs2}~\cite{Aroyo2011183}.

\begin{table}[h]
\begin{tabular}{|c|c|c|c|c|}
\hline
$\lambda^{(4)}$              & WP   & $\mathbf{\Gamma}$ & $\mathbf{M}$ & $\mathbf{X}$,$\mathbf{X}'$ \\ \hline
$1$           & 1a & $1$                  & $1$             & $1$                           \\
              & 1b & $1$                  & $-1$             & $-1$                           \\ \hline
$-1$          & 1a & $-1$                  & $-1$             & $1$                           \\
              & 1b & $-1$                  & $1$             & $-1$                           \\ \hline
$\mathrm{i}$  & 1a & $\mathrm{i}$                  & $\mathrm{i}$             & $-1$                           \\
              & 1b & $\mathrm{i}$                  & $-\mathrm{i}$             & $+1$                           \\ \hline
$-\mathrm{i}$ & 1a & $-\mathrm{i}$                  & $-\mathrm{i}$             & $-1$                           \\
              & 1b & $-\mathrm{i}$                  & $\mathrm{i}$             & $+1$                          \\ \hline
\end{tabular}
\caption{\label{tab: c4_ebrs}Multiplicity-$1$ Wyckoff position-resolved EBRs of wallpaper group $p4$. The site-symmetry representations are labelled by their $\mathcal{C}_4$ eigenvalue $\lambda^{(4)}$. Depending on which Wyckoff position (WP) they are placed at, they give rise to different Bloch band $\mathcal{C}_4$ and $\mathcal{C}_2$ eigenvalues at the high-symmetry momenta $\mathbf{\Gamma},\mathbf{M}$ and $\mathbf{X}, \mathbf{X}'$, respectively.}
\end{table}

\begin{table}[h]
\begin{tabular}{|c|c|c|c|c|}
\hline
$\lambda^{(2)}$              & WP   & $\mathbf{\Gamma}$ & $\mathbf{M}$ & $\mathbf{X}$,$\mathbf{X}'$ \\ \hline
$1$           & 2c & $1, -1$                  & $\mathrm{i}, -\mathrm{i}$             & $1, -1$                           \\
$-1$          & 2c & $\mathrm{i}, -\mathrm{i}$                  & $1, -1$             & $1, -1$                           \\ \hline
\end{tabular}
\caption{\label{tab: c4_ebrs2}Multiplicity-$2$ Wyckoff position-resolved EBRs of wallpaper group $p4$. The site-symmetry representations are labelled by their $\mathcal{C}_2$ eigenvalue $\lambda^{(2)}$. Depending on which Wyckoff position (WP) they are placed at, they give rise to different Bloch band $\mathcal{C}_4$ and $\mathcal{C}_2$ eigenvalues at the high-symmetry momenta $\mathbf{\Gamma},\mathbf{M}$ and $\mathbf{X}, \mathbf{X}'$, respectively.}
\end{table}

\subsection{General observations}
In the following we explicitly construct $\mathcal{C}_4$-symmetric compact Wannier states for all (spatially-obstructed) OAIs that have a compact representation when $\mathcal{C}_4$ symmetry is relaxed to $\mathcal{C}_2$ symmetry. (Recall that in Sec.~\ref{sec: nogo_p2symmetry}, it was shown that $\mathcal{C}_2$-symmetric compact Wannier bases exist only when the underlying lattice hosts at least as many $s \oplus p$ pairs of physical orbitals as there are obstructed orbitals in the OAI.)

From the outset, we can exclude from our consideration some OAIs that are guaranteed to have an OAI complement (these are compact by the results of Sec.~\ref{sec: oai_oai_compactness}): for instance, it follows from Tabs.~\ref{tab: c4_ebrs} and~\ref{tab: c4_ebrs2} that OAIs constructed from lattices hosting physical orbitals only at the $\mathrm{2c}$ position always have $N(\mathrm{AI}) \geq \bar{N}(\mathrm{OAI})$ (see Sec.~\ref{subsec: realspace_fragile}), and therefore also have an OAI complement. 
This is because the only combination of irreducible representations (irreps) from Tab.~\ref{tab: c4_ebrs} that reproduces the full set of momentum-resolved $\mathcal{C}_4$ eigenvalues of any irrep in Tab.~\ref{tab: c4_ebrs2} is one involving \emph{all} irreps of Tab.~\ref{tab: c4_ebrs}, which is a mobile cluster.

Up to the exchange $\mathrm{1a} \leftrightarrow \mathrm{1b}$, there are then two types of OAIs that may have a fragile complement: 
\begin{enumerate}[(1)]
\item{OAIs where $\mathrm{1a}$ is the only Wyckoff position carrying obstructed orbitals,}
\item{OAIs where $\mathrm{2c}$ is the only Wyckoff position carrying obstructed orbitals.}
\end{enumerate}
(Here we assume the OAI to be spatially obstructed, so that all obstructed Wyckoff positions are empty of physical orbitals -- see also the beginning of Sec.~\ref{sec: OAI_mc_generalities}.) We will show that both of these cases allow for compact representations if they are also $\mathcal{C}_2$-compact. The remaining types of $\mathcal{C}_4$-symmetric OAIs do not have a fragile complement (by the necessary condition of Sec.~\ref{subsec: realspace_fragile}), and so are guaranteed to have a compact representation (by the sufficient condition of Sec.~\ref{sec: oai_oai_compactness}). They clearly remain compact when $\mathcal{C}_4$ is relaxed to $\mathcal{C}_2$ symmetry, consistent with our general claim about $\mathcal{C}_4$ compactness.

\begin{figure}[t]
\centering
\includegraphics[width=1\textwidth,page=9]{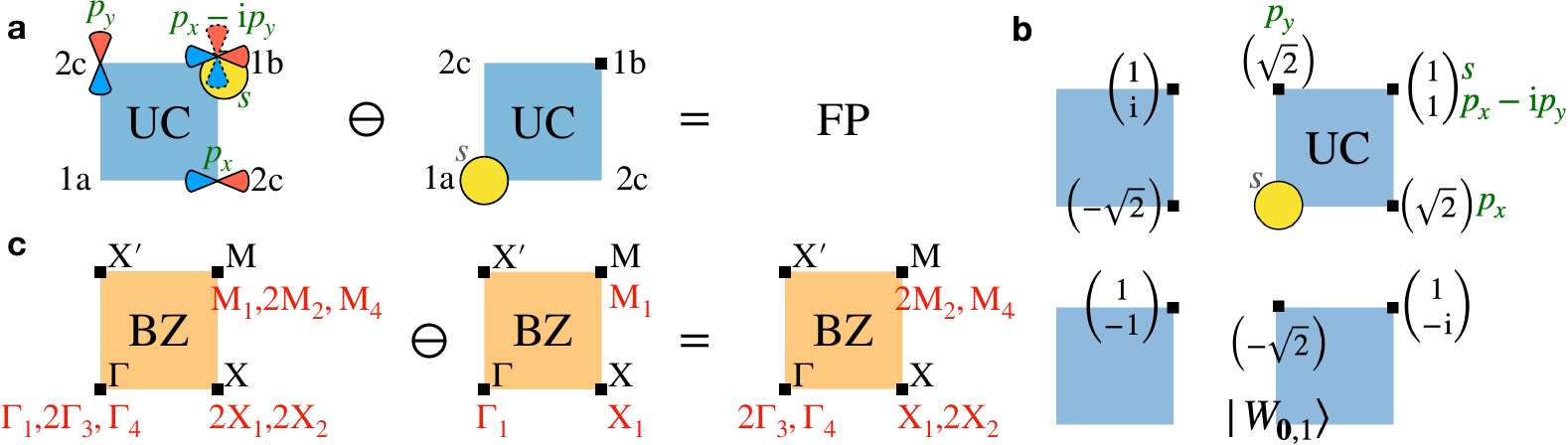}
\caption{(a)~Subtracting an OAI at Wyckoff position 1a to obtain a fragile phase with $\mathcal{C}_4$ rotational symmetry. (b)~Compact Wannier state for the OAI [also obtained from Eq.~\eqref{eq: compactC4state2} for $\lambda=1$]. The vectors shown at each site contain the (non-normalized) overlaps of $\ket{W_{\bs{0},\mathrm{1a}}}$ with all on-site orbitals (the corresponding orbital labels are indicated for the home unit cell). (c)~Brillouin zone decomposition of the relevant bands into irreducible representations. (Representation labels follow the Bilbao Crystallographic Server~\cite{Aroyo2011183}.)}
\label{fig: compactC4counterexampleWITH2C}
\end{figure}

\subsection{$1$-band OAIs} \label{sec: c4_1band}
To begin with, we discuss the case of OAIs composed of $1$ atomic band. 
First note that there are no $1$-band OAIs of type (2), because Wyckoff position $\mathrm{2c}$ has multiplicity $2$ and so can only be populated by $\mathcal{C}_4$-related pairs of bands. For $1$-band OAIs of type (1), where only $\mathrm{1a}$ carries an obstructed orbital with $\mathcal{C}_4$ eigenvalue $\lambda$, there are three inequivalent subtraction patterns that give rise to fragile complements: 
\begin{align} \label{eq: c4_1band_subtraction}
\left[(\lambda)_\mathrm{1b} \oplus (-\lambda)_\mathrm{1b} \oplus (\mathrm{i} \lambda)_\mathrm{1b}\right] \uparrow G \ominus (\lambda)_\mathrm{1a} \uparrow G= \mathrm{FP}, \\ \label{eq: c4_1band_subtraction2}
\left[(\lambda)_\mathrm{1b} \oplus (\mathrm{i}\lambda)_\mathrm{1b} \oplus (-\lambda^2)_\mathrm{2c}\right] \uparrow G \ominus (\lambda)_\mathrm{1a} \uparrow G= \mathrm{FP},
\\ \label{eq: c4_1band_subtraction3}
\left[(-\lambda)_\mathrm{1b} \oplus (\mathrm{i}\lambda)_\mathrm{1b} \oplus (\lambda^2)_\mathrm{2c}\right] \uparrow G \ominus (\lambda)_\mathrm{1a} \uparrow G= \mathrm{FP},
\end{align}
$\lambda = 1,-1,\mathrm{i},-\mathrm{i}$, where we label orbitals at Wyckoff positions $\mathrm{1b}$ and $\mathrm{2c}$ by their $\mathcal{C}_4$ and $\mathcal{C}_2$ eigenvalues, respectively. Because we want to treat multiple OAIs simultaneously and keep $\lambda$ a free variable, this notation is more convenient than the notation used on the Bilbao crystallographic server~\cite{Aroyo2011183}. Here, we do not include subtractions that are obtained by the exchange $\mathrm{i}\leftrightarrow -\mathrm{i}$ in Eqs.~\eqref{eq: c4_1band_subtraction}-~\eqref{eq: c4_1band_subtraction3} (while leaving $\lambda$ unchanged). Furthermore, we do not list configurations containing mobile clusters of orbitals (which are always compact), these contain (at least) one orbital of each $\mathcal{C}_4$ eigenvalue $1, -1, \mathrm{i}, -\mathrm{i}$. We also exclude configurations that do not host enough $s \oplus p$ pairs to guarantee $\mathcal{C}_2$ compactness, these do not contain an orbital pair with $\mathcal{C}_2$ eigenvalues $1$ and $-1$. We also do not consider unit cells that contain more physical orbitals than are needed to construct the OAI.

\begin{figure}[t]
\centering
\includegraphics[width=1\textwidth,page=10]{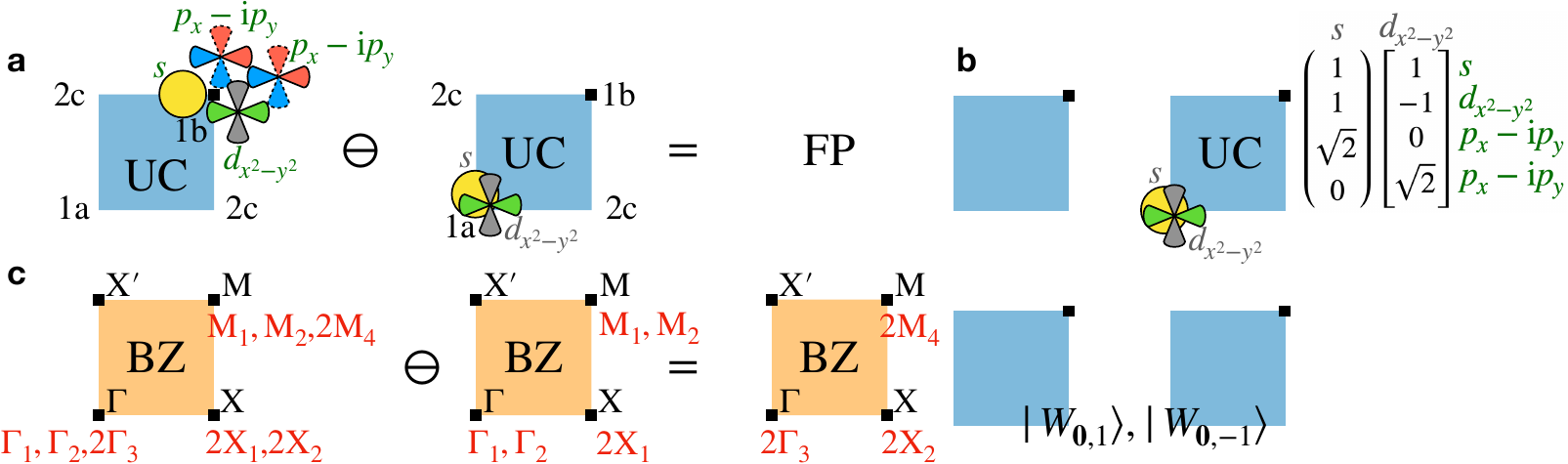}
\caption{(a)~Subtracting a 2-band OAI at Wyckoff position 1a to obtain a fragile root with $\mathcal{C}_4$ rotational symmetry. (b)~Compact Wannier states for the OAI [also obtained from Eq.~\eqref{eq: compactC4state3} for $\lambda=1$]. To minimize clutter, the vectors in this figure only show the asymmetric part of the (non-normalized) overlaps of the Wannier states with all on-site orbitals, the full states are obtained by applying $\mathcal{C}_4$. (c)~Brillouin zone decomposition of the relevant bands into irreducible representations. (Representation labels follow the Bilbao Crystallographic Server~\cite{Aroyo2011183}.)}
\label{fig: compactC4twoband}
\end{figure}

Any subtraction of the form listed above can be performed using strictly local Wannier states. For $\lambda = 1$, these are shown explicitly in Fig.~\ref{fig: compactC4counterexample}b for Eq.~\eqref{eq: c4_1band_subtraction}, and in Fig.~\ref{fig: compactC4counterexampleWITH2C}b for Eq.~\eqref{eq: c4_1band_subtraction2}. The compact Wannier states for $\lambda = -1, \mathrm{i}, -\mathrm{i}$, as well as for Eq.~\eqref{eq: c4_1band_subtraction3} follow from accordingly exchanging the underlying lattice orbitals. That is, for the OAI in Eq.~\eqref{eq: c4_1band_subtraction}, the compact Wannier state at $\bs{R} = \bs{0}$ reads
\begin{equation} \label{eq: compactC4state1} \begin{aligned}
\ket{W_{\bs{0} \lambda}} &= \frac{1}{4} \left[\ket{w_{\bs{0} \lambda}} + \lambda^* C_4|_{\bs{0},\mathrm{1a}} \ket{w_{\bs{0} \lambda}} + (\lambda^* C_4|_{\bs{0},\mathrm{1a}})^2 \ket{w_{\bs{0} \lambda}} + (\lambda^* C_4|_{\bs{0},\mathrm{1a}})^3 \ket{w_{\bs{0} \lambda}} \right], \\
\ket{w_{\bs{0} \lambda}} &= \ket{\bs{0}, (\lambda)_\mathrm{1b}} + \ket{\bs{0}, (-\lambda)_\mathrm{1b}} + \sqrt{2} \ket{\bs{0}, (\mathrm{i}\lambda)_\mathrm{1b}}.
\end{aligned} \end{equation}
Here $\ket{\bs{R}, (\mu)_i}$ denotes the basis state for the orbital with $\mathcal{C}_4$ eigenvalue $\mu$ at Wyckoff position $i$ of the unit cell at $\bs{R}$, and $C_4|_{\bs{0},\mathrm{1a}}$ implements a $\mathcal{C}_4$ rotation about Wyckoff position 1a at the origin. Furthermore, the $\bs{R} = \bs{0}$ compact Wannier state for the OAI in Eq.~\eqref{eq: c4_1band_subtraction2} is obtained from
\begin{equation} \label{eq: compactC4state2}
\ket{w_{\bs{0} \lambda}} = \ket{\bs{0}, (\lambda)_\mathrm{1b}} + \ket{\bs{0}, (\mathrm{i}\lambda)_\mathrm{1b}} + \sqrt{2} \ket{\bs{0}, (-\lambda^2)_\mathrm{2c}},
\end{equation}
where $\ket{\bs{0}, (-\lambda^2)_\mathrm{2c}}$ importantly only entails one of the two $\mathcal{C}_4$-related orbitals at 2c. The compact state for Eq.~\eqref{eq: c4_1band_subtraction3} follows directly by substituting $(\lambda)_\mathrm{1b} \rightarrow (-\lambda)_\mathrm{1b}$, $(-\lambda^2)_\mathrm{2c} \rightarrow (\lambda^2)_\mathrm{2c}$. 

It is straightforward to verify that the resulting states and their translates are orthonormal,
\begin{equation}
\braket{W_{\bs{0} \lambda} |W_{\bs{R} \lambda}} = \delta_{\bs{R},\bs{0}},
\end{equation} 
as required for a compact Wannier basis. We thus conclude that all spatially-obstructed $1$-band OAIs with $\mathcal{C}_4$ symmetry admit a compact Wannier basis as long as they do so when $\mathcal{C}_4$ symmetry is relaxed to $\mathcal{C}_2$ symmetry.

\subsection{$2$-band OAIs} \label{sec: c4_2band}

A straightforward way to obtain $2$-band OAIs is to stack (using the $\oplus$ operation) OAI subtractions of the form of Eqs.~\eqref{eq: c4_1band_subtraction}-\eqref{eq: c4_1band_subtraction3}. For instance, by stacking Eqs.~\eqref{eq: c4_1band_subtraction} and \eqref{eq: c4_1band_subtraction2}, we obtain
\begin{equation} \label{eq: stackingexample}
\left[2(\lambda)_\mathrm{1b} \oplus (-\lambda)_\mathrm{1b} \oplus 2(\mathrm{i} \lambda)_\mathrm{1b} \oplus (-\lambda^2)_\mathrm{2c}\right] \uparrow G \ominus 2(\lambda)_\mathrm{1a} \uparrow G= \mathrm{C},
\end{equation}
where $\mathrm{C}$ is the band complement of an OAI formed by two obstructed orbitals at 1a with equal $\mathcal{C}_4$ eigenvalue $\lambda$. Importantly, for general stacks, $\mathrm{C}$ need not be fragile, however, it is fragile in the present case.
If the individual OAIs forming the stack are all compact, then their combination will also be compact: since we not only stack OAIs but also their pertinent lattice orbitals, Wannier states belonging to different OAIs have disjoint support and are orthogonal to each other.

More interestingly, by inspecting Tabs.~\ref{tab: c4_ebrs} and~\ref{tab: c4_ebrs2}, we find that some $2$-band OAIs can also be constructed on unit cells that are not formed from stacks of the orbitals appearing in Eqs.~\eqref{eq: c4_1band_subtraction}-\eqref{eq: c4_1band_subtraction3}, where stacking is defined with respect to the $\oplus$ operation as in Eq.~\eqref{eq: stackingexample}. We need to study their compactness properties separately, because the support of Wannier states belonging to different OAI bands cannot be disjoint, so that the orthogonality between Wannier states of different bands is not guaranteed.

\begin{figure}[t]
\centering
\includegraphics[width=1\textwidth,page=11]{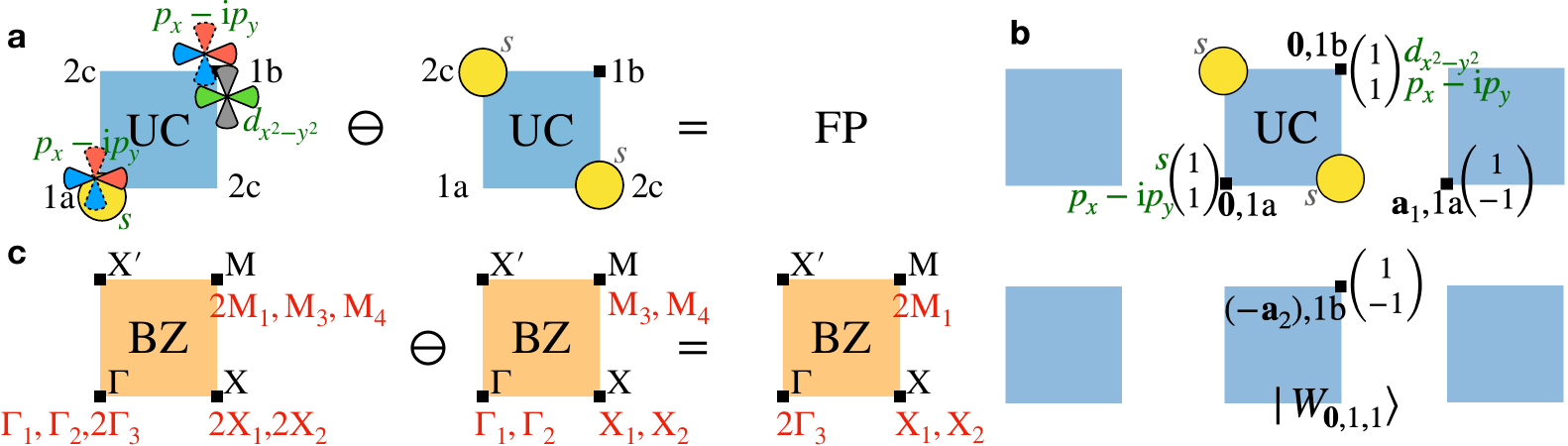}
\caption{(a)~Subtracting 2-band OAI at Wyckoff position 2c to obtain a fragile root with $\mathcal{C}_4$ rotational symmetry. (b)~Compact Wannier state for the OAI [also obtained from Eq.~\eqref{eq: wAt2cCompactState} for $\lambda=1$]. To minimize clutter, the vectors in this figure only show the (non-normalized) overlaps of one of the two compact Wannier states with all on-site orbitals, the other state is obtained by applying the $\mathcal{C}_4$ operation as defined in Eq.~\eqref{eq: wAt2cCompactState}. (c)~Brillouin zone decomposition of the relevant bands into irreducible representations. (Representation labels follow the Bilbao Crystallographic Server~\cite{Aroyo2011183}.)}
\label{fig: compactC4counterexampleAT2C}
\end{figure}

For $2$-band OAIs of type (1), where only $\mathrm{1a}$ carries obstructed orbitals, there is one such subtraction that gives rise to a fragile complement: 
\begin{align} \label{eq: c4_2band_subtraction}
\left[(\lambda)_\mathrm{1b} \oplus (-\lambda)_\mathrm{1b} \oplus (\mathrm{i} \lambda)_\mathrm{1b} \oplus (\mathrm{i} \lambda)_\mathrm{1b}\right] \uparrow G \ominus \left[(\lambda)_\mathrm{1a} \oplus (-\lambda)_\mathrm{1a} \right]\uparrow G = \mathrm{FP}.
\end{align}
Moreover, for OAIs of type (2), where only $\mathrm{2c}$ carries an obstructed orbital with $\mathcal{C}_2$ eigenvalue $\lambda^2$, there is another subtraction leading to a fragile complement: 
\begin{align} \label{eq: c4_1band_subtractionCASETWO}
\left[(\lambda)_\mathrm{1a} \oplus (\mathrm{i}\lambda)_\mathrm{1a} \oplus (-\lambda)_\mathrm{1b} \oplus (\mathrm{i}\lambda)_\mathrm{1b}\right] \uparrow G \ominus (\lambda^2)_\mathrm{2c} \uparrow G= \mathrm{FP}.
\end{align}
Here, we do not include subtractions that are obtained by the exchange $\mathrm{i}\leftrightarrow -\mathrm{i}$, or, in Eq.~\eqref{eq: c4_1band_subtractionCASETWO} only, by the exchange $\mathrm{1a}\leftrightarrow\mathrm{1b}$. Furthermore, we do not list configurations containing mobile clusters of orbitals (which are always compact), or those that do not host enough $s \oplus p$ pairs to guarantee $\mathcal{C}_2$ compactness (which are always non-compact). We also do not consider configurations containing more physical orbitals than are needed to construct the OAI.

Any subtraction of the form of Eqs.~\eqref{eq: c4_2band_subtraction} and~\eqref{eq: c4_1band_subtractionCASETWO} can be performed using strictly local and orthogonal Wannier states. For $\lambda = 1$, these are shown explicitly in Fig.~\ref{fig: compactC4twoband}b for Eq.~\eqref{eq: c4_2band_subtraction} and in Fig.~\ref{fig: compactC4counterexampleAT2C}b for Eq.~\eqref{eq: c4_1band_subtractionCASETWO}. The compact Wannier states for $\lambda = -1, \mathrm{i}, -\mathrm{i}$ follow from accordingly exchanging the underlying lattice orbitals. 
That is, for the OAIs in Eq.~\eqref{eq: c4_2band_subtraction}, the compact Wannier states at $\bs{R} = \bs{0}$ read
\begin{equation} \label{eq: compactC4state3} \begin{aligned}
\ket{W_{\bs{0} \lambda}} &= \frac{1}{4} \left[\ket{w_{\bs{0} \lambda}} + \lambda^* C_4|_{\bs{0},\mathrm{1a}} \ket{w_{\bs{0} \lambda}} + (\lambda^* C_4|_{\bs{0},\mathrm{1a}})^2 \ket{w_{\bs{0} \lambda}} + (\lambda^* C_4|_{\bs{0},\mathrm{1a}})^3 \ket{w_{\bs{0} \lambda}} \right], \\
\ket{W_{\bs{0} (-\lambda)}} &= \frac{1}{4} \left[\ket{w_{\bs{0} (-\lambda)}} -\lambda^* C_4|_{\bs{0},\mathrm{1a}} \ket{w_{\bs{0} (-\lambda)}} + (\lambda^* C_4|_{\bs{0},\mathrm{1a}})^2 \ket{w_{\bs{0} (-\lambda)}} - (\lambda^* C_4|_{\bs{0},\mathrm{1a}})^3 \ket{w_{\bs{0} (-\lambda)}} \right], \\
\ket{w_{\bs{0} \lambda}} &= \ket{\bs{0}, (\lambda)_\mathrm{1b}} + \ket{\bs{0}, (-\lambda)_\mathrm{1b}} + \sqrt{2} \ket{\bs{0}, (\mathrm{i}\lambda)_\mathrm{1b}}, \\
\ket{w_{\bs{0} (-\lambda)}} &= \ket{\bs{0}, (\lambda)_\mathrm{1b}} - \ket{\bs{0}, (-\lambda)_\mathrm{1b}} + \sqrt{2} \ket{\bs{0}, \tilde{(\mathrm{i}\lambda)}_\mathrm{1b}}.
\end{aligned} \end{equation}
Here $\ket{\bs{R}, (\mu)_i}$ denotes the basis state for the orbital with $\mathcal{C}_4$ eigenvalue $\mu$ at Wyckoff position $i$ of the unit cell at $\bs{R}$ [we denote the two inequivalent orbitals with $\mathcal{C}_4$ eigenvalue $\mathrm{i} \lambda$ by $(\mathrm{i} \lambda)_\mathrm{1b}$ and $\tilde{(\mathrm{i}\lambda)}_\mathrm{1b}$], and $C_4|_{\bs{0},\mathrm{1a}}$ implements a $\mathcal{C}_4$ rotation about Wyckoff position 1a at the origin. Finally, the compact states for Eq.~\eqref{eq: c4_1band_subtractionCASETWO} are given by 
\begin{equation} \label{eq: wAt2cCompactState} \begin{aligned}
\ket{W_{\bs{0} (\lambda^2),1}} =& \frac{1}{2\sqrt{2}}\big[\ket{{\bs{0}}, (\lambda)_{\mathrm{1a}}} + \ket{{\bs{0}}, (\mathrm{i} \lambda)_{\mathrm{1a}}} + \ket{{\bs{0}}, (-\lambda)_{\mathrm{1b}}} + \ket{{\bs{0}}, (\mathrm{i} \lambda)_{\mathrm{1b}}} \\ &+ \ket{\bs{a}_1, (\lambda)_{\mathrm{1a}}} - \ket{\bs{a}_1, (\mathrm{i} \lambda)_{\mathrm{1a}}} + \ket{(-\bs{a}_2), (-\lambda)_{\mathrm{1b}}} - \ket{(-\bs{a}_2), (\mathrm{i} \lambda)_{\mathrm{1b}}} \big], \\
\ket{W_{\bs{0} (\lambda^2),2}} =& C_4|_{\bs{0},\mathrm{1a}} \ket{W_{\bs{0} (\lambda^2),1}}.
\end{aligned} \end{equation}

It is straightforward to verify that the resulting states and their translates are orthonormal,
\begin{equation}
\braket{W_{\bs{0} \alpha} |W_{\bs{R} \beta}} = \delta_{\alpha \beta} \delta_{\bs{R},\bs{0}},
\end{equation} 
as required for a compact Wannier basis. We thus conclude that all spatially-obstructed $2$-band OAIs with $\mathcal{C}_4$ symmetry admit a compact Wannier basis, as long as they do so when $\mathcal{C}_4$ symmetry is relaxed to $\mathcal{C}_2$ symmetry.

\subsection{OAIs with an arbitrary number of atomic bands}
There are no $\mathcal{C}_2$-compact $n$-band OAIs with spatial obstruction and a fragile complement, where $n > 2$, that are not obtained by simply stacking (under the $\oplus$ operation) the $1$- and $2$-band OAI subtractions discussed in Secs.~\ref{sec: c4_1band} and~\ref{sec: c4_2band}. 
To see this, we note that any $1$- or $2$-band subset of such an $n$-band OAI must be of the form discussed in Secs.~\ref{sec: c4_1band} and~\ref{sec: c4_2band}. Adding a single orbital to the $1$-band OAIs in Eqs.~\eqref{eq: c4_1band_subtraction}-\eqref{eq: c4_1band_subtraction3}, together with the lattice orbitals required to support it, either produces a stack [Eq.~\eqref{eq: stackingexample}] or a $2$-band OAI of the form in Eqs.~\eqref{eq: c4_2band_subtraction} and~\eqref{eq: c4_1band_subtractionCASETWO}. 

Now, let us inspect the momentum-space irrep content of these $2$-band OAIs:

(1) Setting $\lambda = 1$ without loss of generality in Eq.~\eqref{eq: c4_2band_subtraction}, we find that $\mathrm{FP}$ has the momentum-space irreps shown in Fig.~\ref{fig: compactC4twoband}c. In order to subtract another OAI band at 1a with $\mathcal{C}_4$ eigenvalue $\pm 1$, we must add (at least) three supporting orbitals, giving rise to a stack of subtractions, because the existing momentum-space irreps are incompatible with the new band (Tab.~\ref{tab: c4_ebrs}). Alternatively, in order to subtract an OAI band at 1a with $\mathcal{C}_4$ eigenvalue $\pm \mathrm{i}$, we would have to add a supporting orbital with $\mathcal{C}_4$ eigenvalue $-\mathrm{i}$ at 1b, giving rise to a mobile cluster and preventing a fragile complement.

(2) Furthermore, setting $\lambda = 1$ without loss of generality in Eq.~\eqref{eq: c4_1band_subtractionCASETWO}, we find that $\mathrm{FP}$ has the momentum-space irreps shown in Fig.~\ref{fig: compactC4counterexampleAT2C}c. To add another obstructed orbital $(1)_{\mathrm{2c}}$, we must also stack on (under the $\oplus$) operation another copy of the supporting orbitals of Eq.~\eqref{eq: c4_1band_subtractionCASETWO}, giving rise to a stack of subtractions, because the existing momentum-space irreps are incompatible with the new band (Tab.~\ref{tab: c4_ebrs2}). On the other hand, to add an obstructed orbital $(-1)_{\mathrm{2c}}$, we need a mobile cluster, because the combination $(1)_{\mathrm{2c}} \oplus (-1)_{\mathrm{2c}}$ is itself a mobile cluster.

We therefore conclude that all $\mathcal{C}_4$-symmetric, $\mathcal{C}_2$-compact $(n > 2)$-band OAI subtractions that give fragile complements can be decomposed into stacks of disconnected $1$- or $2$-band OAI subtractions. All such stacks have a compact Wannier basis: the individual OAIs entering the stack were shown to be compact in Secs.~\ref{sec: c4_1band} and~\ref{sec: c4_2band}, and the Wannier states belonging to different OAIs have disjoint support by construction (recall that when stacking subtractions, we stack both OAI bands \emph{and} lattice orbitals). This implies that \emph{all} $\mathcal{C}_4$-protected OAIs with spatial obstruction admit a compact Wannier basis if and only if they do so when $\mathcal{C}_4$ symmetry is relaxed to $\mathcal{C}_2$ symmetry.

We note that this decomposition property of $\mathcal{C}_4$-symmetric OAI subtractions is markedly different from the case with $\mathcal{C}_3$-symmetry: As we will see in Sec.~\ref{sec: c3_3band_6_site_overarching}, the Wannier states of $\mathcal{C}_3$-symmetric $3$-band OAIs with fragile complement cannot all be chosen to have mutually disjoint support, that is, they must share some orbitals so that orthogonality becomes a nontrivial constraint.

\section{$\mathcal{C}_3$ symmetry} \label{sec: overarching_c3}
In this section, we study the compactness properties of OAIs with $\mathcal{C}_3$ rotational symmetry and a spatial obstruction. We define $\omega = e^{\mathrm{i}\frac{2\pi}{3}}$ and $\omega^* = e^{-\mathrm{i}\frac{2\pi}{3}}$. Since spinless $\mathcal{C}_3$ symmetry satisfies $(\mathcal{C}_3)^3 = \mathbb{1}$, its eigenvalues are taken from the set $\{1, \omega, \omega^*\}$. We use conventions where the lattice vectors of the $\mathcal{C}_3$-symmetric wallpaper group $p3$ are given by $\bs{a}_1 = \hat{x}$, $\bs{a}_2 = \hat{x}/2 + \sqrt{3}\hat{y}/2$. Correspondingly, the reciprocal lattice vectors are given by $\bs{b}_1 = 2\pi (\sqrt{3}\hat{x} - \hat{y})/\sqrt{3}$, $\bs{b}_2 = 4\pi \hat{y}/\sqrt{3}$. The maximal Wyckoff positions of the unit cell are $\bs{t}_\mathrm{1a} = \bs{0}$, $\bs{t}_\mathrm{1b} = (\hat{x}+\hat{y}/\sqrt{3})/2$, and $\bs{t}_\mathrm{1c} = (\hat{x}-\hat{y}/\sqrt{3})/2$. The high-symmetry momenta of the Brillouin zone are defined as $\bs{\Gamma} = \bs{0}$, $\bs{K} = 2\bs{b}_1/3 + \bs{b}_2/3$, and $\bs{K}' = \bs{b}_1/3 - \bs{b}_2/3$. For future reference, we also list the elementary band representations of wallpaper group $p3$ in Tab.~\ref{tab: c3_ebrs}.
\begin{table}[h] 
\begin{tabular}{|c|c|c|c|c|} 
\hline
$\lambda$                                	&WP& $\mathbf{\Gamma}$ & $\mathbf{K}$ & $\mathbf{K}'$ \\ \hline
\multirow{3}{*}{1}              		& 1a & $1$                  & $1$             & $1$              \\
                                			& 1b & $1$                  & $\omega^*$             & $\omega$              \\
                                			& 1c & $1$                  & $\omega$             & $\omega^*$              \\ \hline
\multirow{3}{*}{$\omega$}       	& 1a & $\omega$                  & $\omega$             & $\omega$              \\
                                			& 1b & $\omega$                  & $1$             & $\omega^*$              \\
                                			& 1c & $\omega$                  & $\omega^*$             & $1$              \\ \hline
\multirow{3}{*}{$\omega^*$} & 1a & $\omega^*$                  & $\omega^*$             & $\omega^*$              \\
                                			& 1b & $\omega^*$                  & $\omega$             & $1$              \\
                                			& 1c & $\omega^*$                  & $1$             & $\omega$             \\ \hline
\end{tabular}
\caption{\label{tab: c3_ebrs} Wyckoff position-resolved elementary band representations of wallpaper group $p3$. The site-symmetry representations are labelled by their $\mathcal{C}_3$ eigenvalue $\lambda$. Depending on which Wyckoff position (WP) they are placed at, they give rise to different Bloch band $\mathcal{C}_3$ eigenvalues at the high-symmetry momenta $\mathbf{\Gamma},\mathbf{K},\mathbf{K}'$ of the Brillouin zone.}
\end{table}

We may always assume the maximal Wyckoff position $\mathrm{1a}$ to be obstructed and empty of physical orbitals, this assumption does not lose generality: at least one site must be obstructed, and all $\mathcal{C}_3$-symmetric Wyckoff positions are equivalent. Moreover, if only $\mathrm{1b}$ or $\mathrm{1c}$ carries physical orbitals, it follows from Tab.~\ref{tab: c3_ebrs} that subtracting an $n$-band OAI necessitates the presence of $n$ mobile clusters of orbitals, which are each composed of three orbitals with $\mathcal{C}_3$-eigenvalues $1,\omega,\omega^*$. The presence of $n$ mobile clusters guarantees compactness and precludes a fragile complement. Therefore, we only need to investigate compactness on lattices where both $\mathrm{1b}$ and $\mathrm{1c}$ carry physical orbitals.

\subsection{$1$-band OAIs} \label{sec: c3_1band}

\begin{figure}[t]
\centering
\includegraphics[width=0.8\textwidth,page=3]{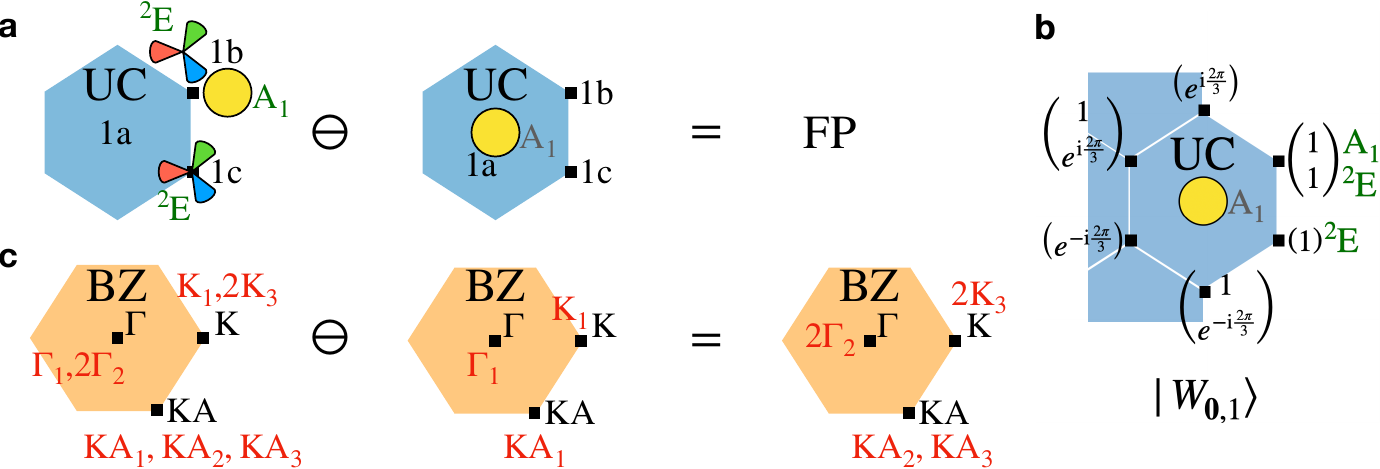}
\caption{(a)~Subtracting a 1-band OAI to obtain a fragile root with $\mathcal{C}_3$ rotational symmetry. (b)~Compact Wannier state for the OAI [also obtained from Eq.~\eqref{eq: compactC3state1} for $\lambda=1$]. Shown is the full symmetric Wannier state, which has overlap with orbitals in its home unit cell and three nearest-neighbor unit cells. (c)~Brillouin zone decomposition of the relevant bands into irreducible representations. (Representation labels follow the Bilbao Crystallographic Server~\cite{Aroyo2011183}.)}
\label{fig: onebandC3}
\end{figure}

All $1$-band atomic insulators with $\mathcal{C}_3$ symmetry can be represented by compact Wannier states. First, if they satisfy $N(\mathrm{AI}) \geq \bar{N}(\mathrm{OAI})$, their compactness follows automatically by the results of Sec.~\ref{sec: oai_oai_compactness}. Then, if they are complement to a fragile set of bands, we show that subtractability (the ability to project out the OAI from the unobstructed lattice bands with a gap at all momenta) is equivalent to compactness for all $1$ band OAIs.

Consulting Tab.~\ref{tab: c3_ebrs}, we find that all subtractions of an atomic band belonging to Wannier states with $\mathcal{C}_3$ eigenvalue $\lambda$, located at Wyckoff position $\mathrm{1a}$, are of a similar form:
\begin{align} \label{eq: c3_1band_subtraction}
\left[(\lambda)_\mathrm{1b} \oplus (\omega \lambda)_\mathrm{1b} \oplus (\omega \lambda)_\mathrm{1c}\right] \uparrow G \ominus (\lambda)_\mathrm{1a} \uparrow G= \mathrm{FP}, \\ \label{eq: c3_1band_subtraction2}
\left[(\omega \lambda)_\mathrm{1b} \oplus (\omega^*\lambda)_\mathrm{1b} \oplus (\lambda)_\mathrm{1c}\right] \uparrow G \ominus (\lambda)_\mathrm{1a} \uparrow G= \mathrm{FP}.
\end{align}
Here, we do not list the equivalent physical orbital configurations that can be obtained by exchanging eigenvalues $\omega \leftrightarrow \omega^*$ or Wyckoff positions $\mathrm{1b} \leftrightarrow \mathrm{1c}$. Furthermore, we do not list configurations containing mobile clusters of orbitals (which are always compact), these contain (at least) one orbital of each $\mathcal{C}_3$ eigenvalue $1, \omega, \omega^*$. We also do not consider unit cells that contain more physical orbitals than are needed for constructing the OAI. For Eq.~\eqref{eq: c3_1band_subtraction} and $\lambda = 1$, Fig.~\ref{fig: onebandC3}a depicts the real-space subtraction, while Fig.~\ref{fig: onebandC3}c shows the momentum-space symmetry representations of the corresponding Bloch bands.

\begin{figure}[t]
\centering
\includegraphics[width=0.8\textwidth,page=6]{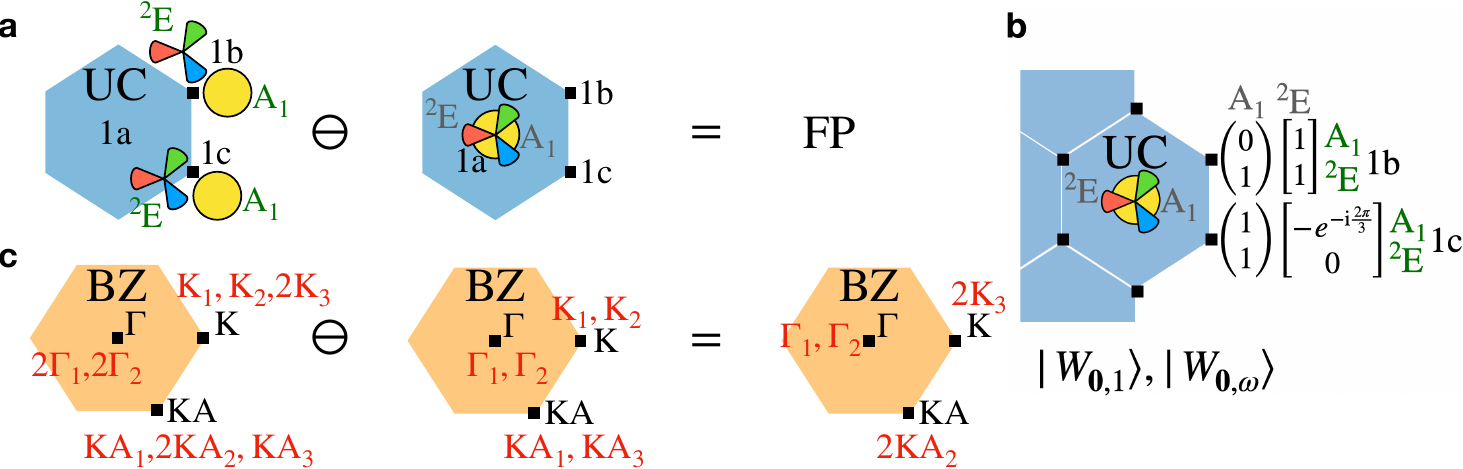}
\caption{(a)~Subtracting a 2-band OAI to obtain a fragile root with $\mathcal{C}_3$ rotational symmetry. (b)~Compact Wannier states for the OAI [also obtained from Eq.~\eqref{eq: compactC3state2} for $\lambda=1$]. To minimize clutter, we only show the asymmetric part of the Wannier state, the full state is obtained by applying $\mathcal{C}_3$. (c)~Brillouin zone decomposition of the relevant bands into irreducible representations. (Representation labels follow the Bilbao Crystallographic Server~\cite{Aroyo2011183}.)}
\label{fig: twobandC3}
\end{figure}

Any subtraction of the form of Eq.~\eqref{eq: c3_1band_subtraction},~\eqref{eq: c3_1band_subtraction2} can be performed using strictly local Wannier states. These are shown explicitly in Fig.~\ref{fig: onebandC3}b for Eq.~\eqref{eq: c3_1band_subtraction} in the case $\lambda = 1$. The compact Wannier states for $\lambda = \omega, \omega^*$, as well as for Eq.~\eqref{eq: c3_1band_subtraction2}, follow from accordingly exchanging the underlying lattice orbitals. That is, for the OAI in Eq.~\eqref{eq: c3_1band_subtraction}, the compact Wannier state at $\bs{R} = \bs{0}$ reads
\begin{equation} \label{eq: compactC3state1} \begin{aligned}
\ket{W_{\bs{0} \lambda}} &= \frac{1}{3} \left[\ket{w_{\bs{0} \lambda}} + \lambda^* C_3|_{\bs{0},\mathrm{a}} \ket{w_{\bs{0} \lambda}} + (\lambda^* C_3|_{\bs{0},\mathrm{a}})^2 \ket{w_{\bs{0} \lambda}} \right], \\
\ket{w_{\bs{0} \lambda}} &= \ket{\bs{0}, (\lambda)_\mathrm{1b}} + \ket{\bs{0}, (\omega \lambda)_\mathrm{1b}} + \ket{\bs{0}, (\omega \lambda)_\mathrm{1c}}.
\end{aligned} \end{equation}
Here $\ket{\bs{R}, (\mu)_i}$ denotes the basis state for the orbital with $\mathcal{C}_3$ eigenvalue $\mu$ at Wyckoff position $i$ of the unit cell at $\bs{R}$, and $C_3|_{\bs{0},\mathrm{a}}$ implements a $\mathcal{C}_3$ rotation about Wyckoff position 1a at the origin. Correspondingly, the $\bs{R} = \bs{0}$ compact Wannier state for the OAI in Eq.~\eqref{eq: c3_1band_subtraction2} is obtained from
\begin{equation}
\ket{w_{\bs{0} \lambda}} = \ket{\bs{0}, (\omega \lambda)_\mathrm{1b}} + \ket{\bs{0}, (\omega^* \lambda)_\mathrm{1b}} + \ket{\bs{0}, (\lambda)_\mathrm{1c}}.
\end{equation}
It is straightforward to verify that the resulting states and their translates are orthonormal,
\begin{equation}
\braket{W_{\bs{0} \lambda} |W_{\bs{R} \lambda}} = \delta_{\bs{R},\bs{0}},
\end{equation} 
as required for a compact Wannier basis.
We conclude that all 1-band atomic insulators with $\mathcal{C}_3$ symmetry allow for a compact representation, irrespective of whether they have an atomic or fragile complement.

\subsection{$2$-band OAIs} \label{sec: c3_2band}
A straightforward way to obtain $2$-band OAIs is to stack (using the $\oplus$ operation) OAI subtractions of the form of Eqs.~\eqref{eq: c3_1band_subtraction} and~\eqref{eq: c3_1band_subtraction2}. The resulting OAIs are clearly compact (in that they are spanned by a strictly local, orthonormal basis of Wannier states), because the respective Wannier states have disjoint support. (See also the discussion in Sec.~\ref{sec: c4_2band}.)

More interestingly, some spatially-obstructed $2$-band OAI subtractions cannot be obtained from stacking. We need to study their compactness properties separately, because the support of Wannier states belonging to different OAI bands must have at least some orbitals in common, so that the orthogonality between Wannier states of different bands is not guaranteed.
Consulting Tab.~\ref{tab: c3_ebrs}, such $2$-band atomic insulators with fragile complements can be obtained as subtractions of the form 
\begin{align} \label{eq: C3_twoband_subtraction}
\left[(\lambda)_\mathrm{1b} \oplus (\omega \lambda)_\mathrm{1b} \oplus (\lambda)_\mathrm{1c} \oplus (\omega \lambda)_\mathrm{1c}\right] \uparrow G \ominus \left[(\lambda)_\mathrm{1a} \oplus (\omega \lambda)_\mathrm{1a} \right] \uparrow G=& \mathrm{FP}, \\ \label{eq: C3_twoband_subtraction2}
\left[(\omega^* \lambda)_\mathrm{1b} \oplus (\lambda)_\mathrm{1b} \oplus (\omega \lambda)_\mathrm{1c} \oplus (\omega^* \lambda)_\mathrm{1c}\right] \uparrow G \ominus \left[(\lambda)_\mathrm{1a} \oplus (\omega \lambda)_\mathrm{1a} \right] \uparrow G=& \mathrm{FP},
\end{align}
where we do not list subtractions obtained by exchanging the eigenvalues $\omega \leftrightarrow \omega^*$ or Wyckoff positions $\mathrm{1b} \leftrightarrow \mathrm{1c}$. Moreover, we do not list the configurations containing mobile clusters of orbitals, or those that contain more physical orbitals than are needed to construct the OAI. (Here we assume the OAI to be spatially obstructed, so that the obstructed Wyckoff position is empty of physical orbitals -- see also the beginning of Sec.~\ref{sec: OAI_mc_generalities}.) For Eq.~\eqref{eq: C3_twoband_subtraction}, Fig.~\ref{fig: twobandC3}a and~c depict the $\lambda = 1$ real and momentum space subtractions, respectively. The two Wannier states with eigenvalues $\lambda$ and $\omega \lambda$ can be chosen to form a compact basis, as shown in Fig.~\ref{fig: twobandC3}b. The compact states for $\lambda = \omega, \omega^*$ in Eq.~\eqref{eq: C3_twoband_subtraction}, as well as those for all OAIs in Eq.~\eqref{eq: C3_twoband_subtraction2}, follow from accordingly exchanging the underlying lattice orbitals. That is, the $\bs{R} = \bs{0}$ compact Wannier state for the OAI in Eq.~\eqref{eq: C3_twoband_subtraction} reads 
\begin{equation} \label{eq: compactC3state2} \begin{aligned}
\ket{W_{\bs{0} \lambda}} &= \frac{1}{3} \left[\ket{w_{\bs{0} \lambda}} + \lambda^* C_3|_{\bs{0},\mathrm{a}} \ket{w_{\bs{0} \lambda}} + (\lambda^* C_3|_{\bs{0},\mathrm{a}})^2 \ket{w_{\bs{0} \lambda}} \right], \\
\ket{W_{\bs{0} (\omega \lambda)}} &= \frac{1}{3} \left[\ket{w_{\bs{0} \lambda}} + (\omega^* \lambda^*) C_3|_{\bs{0},\mathrm{a}} \ket{w_{\bs{0} \lambda}} + \left[(\omega^* \lambda^*) C_3|_{\bs{0},\mathrm{a}}\right]^2 \ket{w_{\bs{0} \lambda}} \right], \\
\ket{w_{\bs{0} \lambda}} &= \ket{\bs{0}, (\omega \lambda)_\mathrm{b}} + \ket{\bs{0}, (\lambda)_\mathrm{c}} + \ket{\bs{0}, (\omega \lambda)_\mathrm{c}}, 
\\ \ket{w_{\bs{0} (\omega \lambda)}} &= \ket{\bs{0}, (\lambda)_\mathrm{b}} + \ket{\bs{0}, (\omega \lambda)_\mathrm{b}} -\omega^* \ket{\bs{0}, (\lambda)_\mathrm{c}}.
\end{aligned} \end{equation}
Correspondingly, the compact basis for the OAI in Eq.~\eqref{eq: C3_twoband_subtraction2} is obtained from 
\begin{equation} \begin{aligned}
&\ket{w_{\bs{0} \lambda}} = \ket{\bs{0}, (\lambda)_\mathrm{1b}} + \ket{\bs{0}, (\omega \lambda)_\mathrm{1c}} + \ket{\bs{0}, (\omega^* \lambda)_\mathrm{1c}}, 
\quad \ket{w_{\bs{0} (\omega \lambda)}} = \ket{\bs{0}, (\omega^* \lambda)_\mathrm{1b}} + \ket{\bs{0}, (\lambda)_\mathrm{1b}} -\omega^* \ket{\bs{0}, (\omega \lambda)_\mathrm{1c}}.
\end{aligned} \end{equation}
It is straightforward to verify that the resulting states and their translates are orthonormal,
\begin{equation}
\braket{W_{\bs{0} \alpha} |W_{\bs{R} \beta}} = \delta_{\alpha \beta} \delta_{\bs{R},\bs{0}}, \quad \alpha,\beta \in \{\lambda, \omega \lambda\}.
\end{equation} 
We conclude that all spatially-obstructed 2-band atomic insulators with $\mathcal{C}_3$ symmetry allow for a compact representation, irrespective of whether they have an atomic or fragile complement.

\begin{figure}[t]
\centering
\includegraphics[width=0.7\textwidth,page=4]{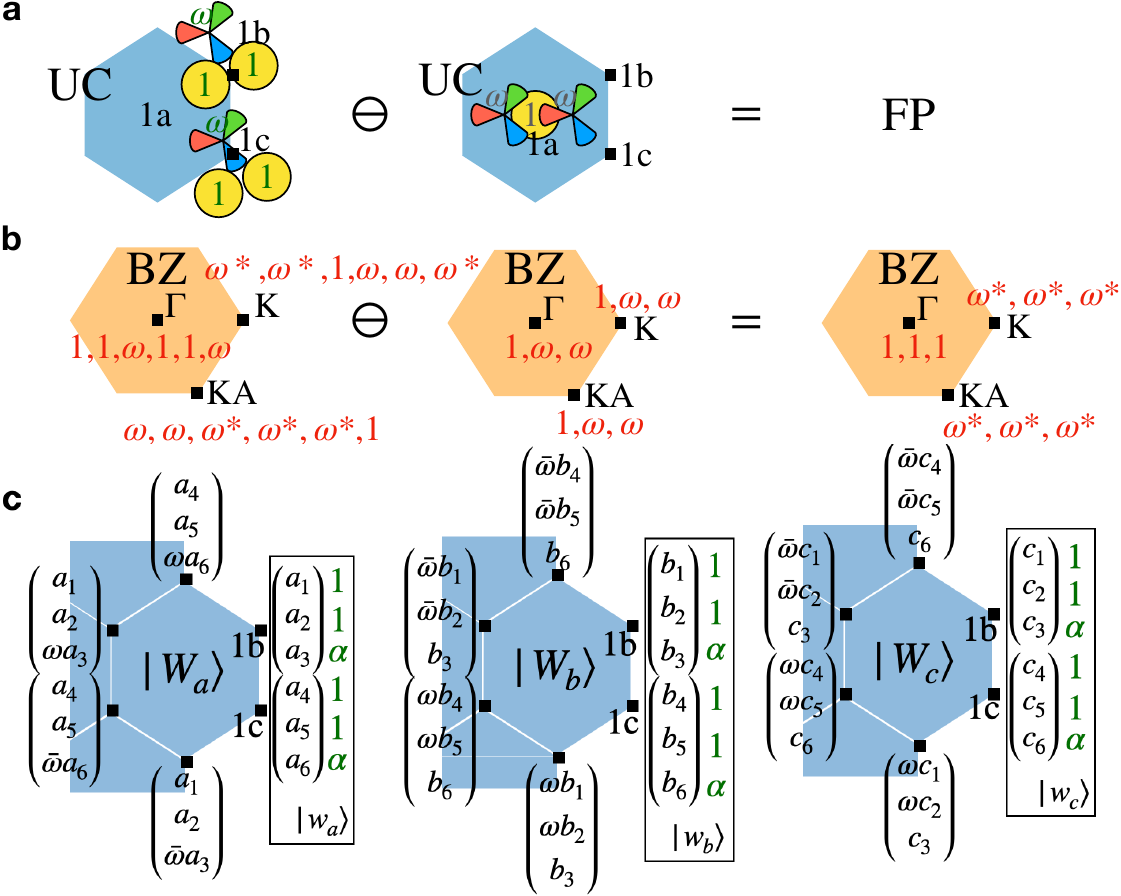}
\caption{(a)~Subtracting a 3-band OAI to obtain a fragile root with $\mathcal{C}_3$ rotational symmetry. (b)~Brillouin zone decomposition of the relevant bands into irreducible representations. (Representation labels follow the Bilbao Crystallographic Server~\cite{Aroyo2011183}.) (c)~Trial Wannier states for the OAI. Shown are the full symmetric Wannier states, which have overlap with orbitals in their home unit cell and three nearest-neighbor unit cells. As shown in Sec.~\ref{sec: c3_3band_6_site_overarching}, there is no compact parameter choice.}
\label{fig: noncompactC3}
\end{figure}

\subsection{$3$-band OAI: $6$-site non-compactness} \label{sec: c3_3band_6_site_overarching}
We next turn to spatially-obstructed $3$-band OAIs with $\mathcal{C}_3$ symmetry. Interestingly, and in contrast to the $1$- and $2$-band cases, we find compactness obstructions at least for small trial Wannier states. These arise by a mechanism that is entirely different from the $\mathcal{C}_2$ obstructions governing OAIs with $\mathcal{C}_2$ and $\mathcal{C}_4$ symmetry.

\subsubsection{Compactness constraints}
Consider a 3-band OAI whose Wannier states are located on the 1a position and have $\mathcal{C}_3$ eigenvalues $\lambda_a = 1$, $\lambda_b = \omega$, $\lambda_c = \omega$, where $\omega = e^{\mathrm{i} \frac{2\pi}{3}}$. We can construct such an OAI from a lattice that has 3 orbitals on the 1b position (with eigenvalues $\{1,1,\omega\}$), and another 3 orbitals on the 1c position (also with eigenvalues $\{1,1,\omega\}$). For spatial obstructions, this is the smallest number of orbitals giving rise to a 3-band OAI. These orbitals do not form a mobile cluster and so the complement of the OAI is necessarily fragile by the results of Sec.~\ref{subsec: realspace_fragile}. Using the representation labels from the Bilbao crystallographic server~\cite{Aroyo2011183}, the corresponding subtraction is given by,
\begin{equation} \label{eq: 3bandC3noncompact_subtraction}
\left[2(A_1)_\mathrm{1b} \oplus ({}^{2}E)_\mathrm{1b} \oplus 
2(A_1)_\mathrm{1c} \oplus ({}^{2}E)_\mathrm{1c} \right] \uparrow G \ominus \left[(A_1)_\mathrm{1a} \oplus
2({}^{2}E)_\mathrm{1a} \right] \uparrow G
= \mathrm{FP}
\end{equation}
and pictorially represented in Fig.~\ref{fig: noncompactC3}a,b.

We now set out to find a compact Wannier basis for this OAI, where for the moment we restrict our attention to Wannier states that have support on only $6$ physical sites arranged in a hexagonal shape (recall from Secs.~\ref{sec: c3_1band} and~\ref{sec: c3_2band} that this support is sufficient for establishing the compactness of all $\mathcal{C}_3$-symmetric 1- and 2-band OAIs). We then define trial Wannier states as shown in Fig.~\ref{fig: noncompactC3}c, where $\ket{W_a}$ denotes the state with $\mathcal{C}_3$ eigenvalue $\lambda_a = 1$, while $\ket{W_b}$, $\ket{W_c}$ are the states with $\mathcal{C}_3$ eigenvalue $\lambda_b = \lambda_c = \omega$. By default, these are chosen to transform under $\mathcal{C}_3$ symmetry according to their respective eigenvalue, leaving a total of $3 \times 6 = 18$ remaining free complex variables that are encoded in the three $6$-dimensional vectors $\ket{w_a}$, $\ket{w_b}$, and $\ket{w_c}$, defined in Eq.~\eqref{eq: symmWannierAnsatz}. Following the notation of Fig.~\ref{fig: noncompactC3}c, we write
\begin{equation}
\ket{w_a} = \begin{pmatrix}a_1 \\ a_2\\ a_3 \\ a_4\\ a_5 \\ a_6 \end{pmatrix}, \quad \ket{w_b} = \begin{pmatrix}b_1 \\ b_2 \\ b_3 \\ b_4 \\ b_5 \\ b_6 \end{pmatrix}, \quad \ket{w_c} = \begin{pmatrix}c_1 \\ c_2 \\ c_3 \\ c_4 \\ c_5 \\ c_6 \end{pmatrix}.
\end{equation}
Orthogonality with all translates implies the constraints
\begin{equation} \label{eq: constraint0-1}
\bra{w_\mu} \tilde{C}_{3 \nu} \ket{w_\nu} = 0, 
\end{equation}
where we defined the $6 \times 6$ matrices
\begin{equation}
\tilde{C}_{3,a} = \begin{pmatrix} 1 & & & & & \\ & 1 & & & & \\ & & \omega & & & \\ & & & 1 & & \\ & & & & 1 & \\ & & & & & \omega^* \end{pmatrix}, 
\quad \tilde{C}_{3,b} = \begin{pmatrix} \omega^* & & & & & \\ & \omega^* & & & & \\ & & 1 & & & \\ & & & \omega & & \\ & & & & \omega & \\ & & & & & 1 \end{pmatrix} = \tilde{C}_{3,c}.
\end{equation}
These constraints have to be supplemented by on-site orthogonality,
\begin{equation} \label{eq: constraint0-2}
\braket{w_b | w_c} = 0.
\end{equation}
(The orthogonality with $\ket{w_a}$ is already guaranteed by $\lambda_a \neq \lambda_b = \lambda_c$.)
The diagonal constraints give 
\begin{equation} \label{eq: constraint4}
\begin{aligned}
\bra{w_a} \tilde{C}_{3,a} \ket{w_a} = 0 \quad &\rightarrow \quad |a_1|^2 + |a_2|^2 + |a_4|^2 + |a_5|^2 = |a_3|^2 = |a_6|^2 = 1, \\
\bra{w_b} \tilde{C}_{3,b} \ket{w_b} = 0 \quad &\rightarrow \quad |b_1|^2 + |b_2|^2 = |b_3|^2 + |b_6|^2 = |b_4|^2 + |b_5|^2 = 1, \\
\bra{w_c} \tilde{C}_{3,c} \ket{w_c} = 0 \quad &\rightarrow \quad |c_1|^2 + |c_2|^2 = |c_3|^2 + |c_6|^2 = |c_4|^2 + |c_5|^2 = 1,
\end{aligned}
\end{equation}
where we have fixed the normalization to $\braket{w_\mu | w_\mu} = 3$.
We can rephrase the off-diagonal constraints in simpler terms by taking linear combinations. For instance, 
\begin{equation} \label{eq: constraint1}
\begin{aligned}
\braket{w_b | w_c} + \braket{w_b | \tilde{C}_{3,c} | w_c} + \left(\braket{w_c | \tilde{C}_{3,b} | w_b}\right)^* \quad &\propto \quad c_3 b^*_3+c_6 b^*_6 = 0 \\
\braket{w_b | w_c} + \omega \braket{w_b | \tilde{C}_{3,c} | w_c} + \omega^* \left(\braket{w_c | \tilde{C}_{3,b} | w_b}\right)^* \quad &\propto \quad c_1 b^*_1+c_2 b^*_2 = 0, \\
\braket{w_b | w_c} + \omega^* \braket{w_b | \tilde{C}_{3,c} | w_c} + \omega \left(\braket{w_c | \tilde{C}_{3,b} | w_b}\right)^* \quad &\propto \quad c_4 b^*_4+c_5 b^*_5 = 0,
\end{aligned}
\end{equation}
where $\propto$ indicates proportionality.
We also have that 
\begin{equation} \label{eq: constraint3}
\begin{aligned}
\bra{w_a} \tilde{C}_{3,b} \ket{w_b} - \omega \left(\bra{w_b} \tilde{C}_{3,a} \ket{w_a} \right)^* \quad &\propto \quad a^*_6 b_6 + \omega^*(a^*_1 b_1 + a^*_2 b_2) = 0, \\
\bra{w_a} \tilde{C}_{3,b} \ket{w_b} - \omega^* \left(\bra{w_b} \tilde{C}_{3,a} \ket{w_a} \right)^* \quad &\propto \quad a^*_3 b_3 + \omega(a^*_4 b_4 + a^*_5 b_5) = 0, \\ 
\bra{w_a} \tilde{C}_{3,c} \ket{w_c} - \omega \left(\bra{w_c} \tilde{C}_{3,a} \ket{w_a} \right)^* \quad &\propto \quad a^*_6 c_6 + \omega^*(a^*_1 c_1 + a^*_2 c_2) = 0, \\
\bra{w_a} \tilde{C}_{3,c} \ket{w_c} - \omega^* \left(\bra{w_c} \tilde{C}_{3,a} \ket{w_a} \right)^* \quad &\propto \quad a^*_3 c_3 + \omega(a^*_4 c_4 + a^*_5 c_5) = 0.
\end{aligned}
\end{equation}
Eqs~\eqref{eq: constraint4},~\eqref{eq: constraint1}, and~\eqref{eq: constraint3} are $10$ equations in total, they exhaust all constraints contained in Eqs.~\eqref{eq: constraint0-1} and~\eqref{eq: constraint0-2}. 
\subsubsection{Proving inconsistency directly} \label{subsec: direct_inconsistency_hexagon}
We next show that the combined system of all constraints does not allow for a solution. For this, we first solve Eqs~\eqref{eq: constraint4} and~\eqref{eq: constraint1} for $\ket{w_b}$ and $\ket{w_c}$. The most general solution is
\begin{equation}
\ket{w_b} = \begin{pmatrix}b_1 \\ b_2 \\ b_3 \\ b_4 \\ b_5 \\ b_6 \end{pmatrix} = 
\begin{pmatrix} 
\rho_g e^{\mathrm{i}\phi_g} \\ 
\sqrt{1-\rho^2_g} e^{\mathrm{i}\phi_h} \\ 
\rho_i e^{\mathrm{i}\phi_i} \\ 
\rho_j e^{\mathrm{i}\phi_j} \\ 
\sqrt{1-\rho^2_j} e^{\mathrm{i}\phi_k} \\ 
\sqrt{1-\rho^2_i} e^{\mathrm{i}\phi_l}
\end{pmatrix}, \quad 
\ket{w_c} = \begin{pmatrix}c_1 \\ c_2 \\ c_3 \\ c_4 \\ c_5 \\ c_6 \end{pmatrix} = 
\begin{pmatrix}
\sqrt{1-\rho^2_g} e^{\mathrm{i}(\phi_g + \phi_m)} \\
-\rho_g e^{\mathrm{i}(\phi_h + \phi_m)} \\
\sqrt{1-\rho^2_i} e^{\mathrm{i}(\phi_i + \phi_o)} \\
\sqrt{1-\rho^2_j} e^{\mathrm{i}(\phi_j + \phi_p)} \\ 
-\rho_j e^{\mathrm{i}(\phi_k + \phi_p)} \\ 
-\rho_i e^{\mathrm{i}(\phi_l + \phi_o)}
\end{pmatrix},
\end{equation}
where $\rho_\mu \in \mathbb{R}$, $0 \leq \rho_\mu \leq 1$, and $\phi_\mu \in \mathbb{R}$.
Using Eq.~\eqref{eq: constraint3}, we then compute 
\begin{equation}
\begin{aligned}
&1 = |b_6|^2 + |c_6|^2 = |a^*_1 b_1 + a^*_2 b_2|^2 + |a^*_1 c_1 + a^*_2 c_2|^2 = |a_1|^2 + |a_2|^2 + 2\mathrm{Re} [a^*_1 a_2 (b_1 b^*_2 + c_1 c^*_2)] = |a_1|^2 + |a_2|^2, \\
&1 = |b_3|^2 + |c_3|^2 = |a^*_4 b_4 + a^*_5 b_5|^2 + |a^*_4 c_4 + a^*_5 c_5|^2 = |a_4|^2 + |a_5|^2 + 2\mathrm{Re} [a^*_4 a_5 (b_4 b^*_5 + c_4 c^*_5)] = |a_4|^2 + |a_5|^2, \\
\end{aligned}
\end{equation}
so that we arrive at
\begin{equation}
2 = |a_1|^2 + |a_2|^2 + |a_4|^2 + |a_5|^2,
\end{equation}
in contradiction to the first line of Eq.~\eqref{eq: constraint4}. This implies that no solutions exist.

\subsubsection{Proving inconsistency via the Cauchy–Schwarz inequality}
Observe that Eqs.~\eqref{eq: constraint4} and~\eqref{eq: constraint3}  imply
\begin{equation}
\left|\begin{pmatrix}a_1\\ a_2\end{pmatrix}^\dagger \begin{pmatrix} b_1 \\ b_2 \end{pmatrix} \right|^2+ \left|\begin{pmatrix} a_4\\ a_5 \end{pmatrix}^\dagger \begin{pmatrix} b_4 \\ b_5 \end{pmatrix}\right|^2 = |a^*_6 b_6 |^2 + |a^*_3 b_3|^2 = 1.
\end{equation}
At the same time, we know from the Cauchy–Schwarz inequality that 
\begin{equation}
\left|\begin{pmatrix}a_1\\ a_2\end{pmatrix}^\dagger \begin{pmatrix} b_1 \\ b_2 \end{pmatrix} \right|^2 = \left|\begin{pmatrix}a_1\\ a_2\end{pmatrix}\right|^2 \left|\begin{pmatrix} b_1 \\ b_2 \end{pmatrix} \right|^2 - \epsilon, \quad
\left|\begin{pmatrix} a_4\\ a_5 \end{pmatrix}^\dagger \begin{pmatrix} b_4 \\ b_5 \end{pmatrix}\right|^2 = 
\left|\begin{pmatrix} a_4\\ a_5 \end{pmatrix}\right|^2 \left| \begin{pmatrix} b_4 \\ b_5 \end{pmatrix}\right|^2 - \delta,
\end{equation}
where $\epsilon \geq 0$, $\delta \geq 0$.
Then, \begin{equation}
1 = \left|\begin{pmatrix}a_1\\ a_2\end{pmatrix}^\dagger \begin{pmatrix} b_1 \\ b_2 \end{pmatrix} \right|^2+ \left|\begin{pmatrix} a_4\\ a_5 \end{pmatrix}^\dagger \begin{pmatrix} b_4 \\ b_5 \end{pmatrix}\right|^2 = 
\left|\begin{pmatrix}a_1\\ a_2\end{pmatrix}\right|^2 \left|\begin{pmatrix} b_1 \\ b_2 \end{pmatrix} \right|^2+ \left|\begin{pmatrix} a_4\\ a_5 \end{pmatrix}\right|^2 \left| \begin{pmatrix} b_4 \\ b_5 \end{pmatrix}\right|^2 - \epsilon - \delta = 1 - \epsilon - \delta,
\end{equation}
from which we deduce $\epsilon = \delta = 0.$
Since the Cauchy–Schwarz inequality is only an equality in the case of linear dependence, it follows that
\begin{equation}
\begin{pmatrix}a_1\\ a_2\end{pmatrix} \propto \begin{pmatrix} b_1 \\ b_2 \end{pmatrix}, \quad \begin{pmatrix} a_4\\ a_5 \end{pmatrix} \propto \begin{pmatrix} b_4 \\ b_5 \end{pmatrix}
\end{equation}
are proportional to each other. In the very same way, we obtain
\begin{equation}
\begin{pmatrix}a_1\\ a_2\end{pmatrix} \propto \begin{pmatrix} c_1 \\ c_2 \end{pmatrix}, \quad \begin{pmatrix} a_4\\ a_5 \end{pmatrix} \propto \begin{pmatrix} c_4 \\ c_5 \end{pmatrix}.
\end{equation}
We find 
\begin{equation}
\begin{pmatrix} b_1 \\ b_2 \end{pmatrix} \propto \begin{pmatrix} c_1 \\ c_2 \end{pmatrix}, \quad \begin{pmatrix} b_4 \\ b_5 \end{pmatrix} \propto \begin{pmatrix} c_4 \\ c_5 \end{pmatrix},
\end{equation}
in direct contradiction to Eq.~\eqref{eq: constraint1}.

\begin{figure}[t]
\centering
\includegraphics[width=0.93\textwidth,page=14]{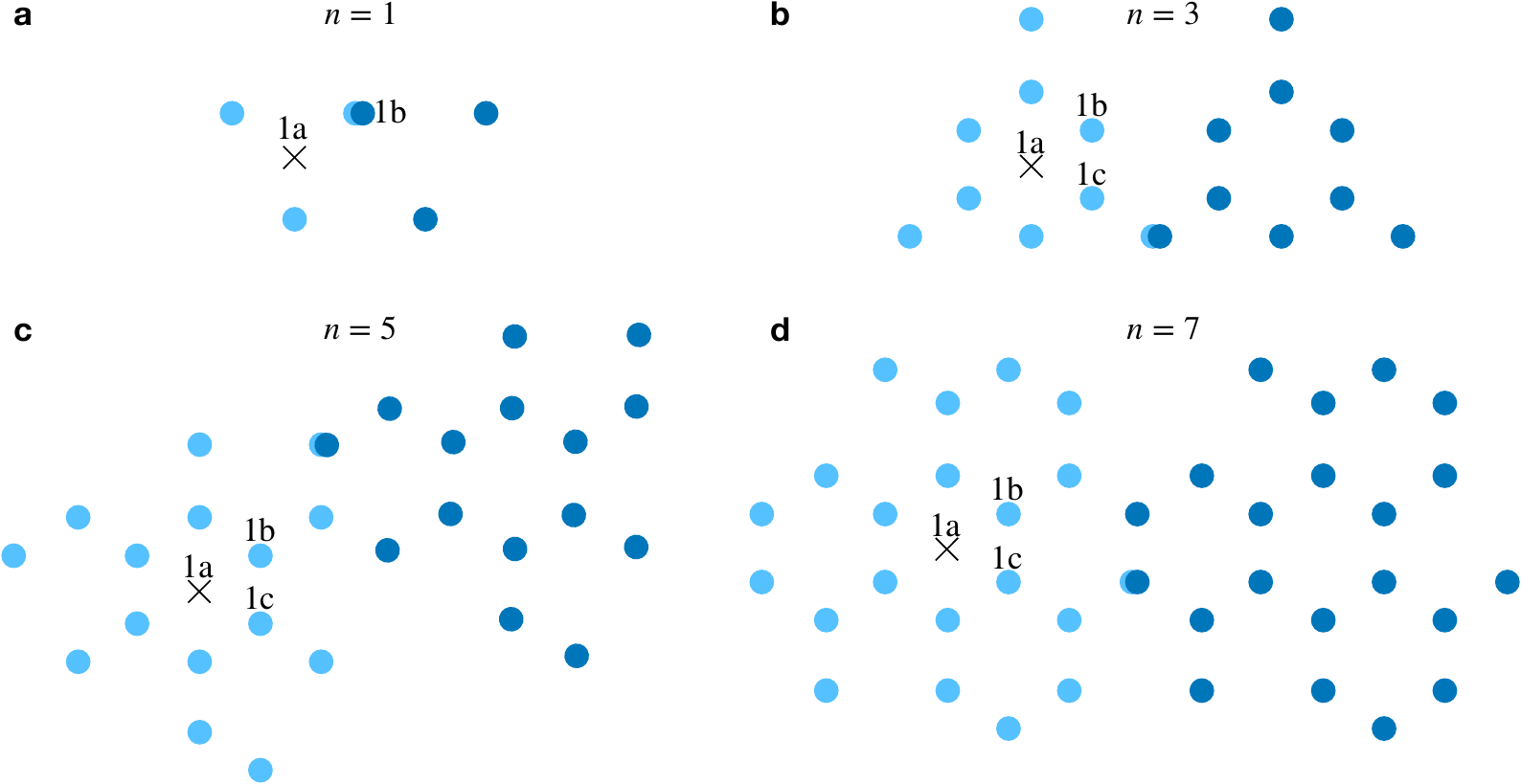}
\caption{Trial Wannier state overlaps for the $\mathcal{C}_3$-symmetric $3$-band OAI in Eq.~\eqref{eq: 3bandC3noncompact_subtraction}. By $\mathcal{C}_3$ symmetry, all trial states must have overlap with $3n$, $n \in \mathbb{Z}$, sites. The even cases $n=2$ and $n=4$ are extensively treated in Secs.~\ref{sec: c3_3band_6_site_overarching} and~\ref{sec: c3_12site_problem}, respectively. (a)~$n=1$. The trial Wannier state (light blue) centered about the 1a position (black x) has overlap with one of its translates (dark blue) on a single site. Since the translated trial state can also be obtained from a $\mathcal{C}_3$ rotation about the overlap site, this situation is equivalent to that depicted in Fig.~\ref{fig: compactInvObstruction} for $\mathcal{C}_2$ symmetry: for such a $\mathcal{C}_2$-symmetric overlap, we have shown in Eq.~\eqref{eq: c2_sp_requirement} that a mobile cluster of orbitals (i.e., a $s \oplus p$ pair) must be present on the overlap site to ensure orthogonality. By the same reasoning, to ensure orthogonality in the present case, the overlap site must host a $\mathcal{C}_3$-symmetric mobile cluster. Since this is not the case -- instead, each physical site carries the $\mathcal{C}_3$ representations $\{1,1,\omega\}$ of the site-symmetry group, see Eq.~\eqref{eq: 3bandC3noncompact_subtraction} -- there is no orthogonal solution. (b)~$n=3$ and (c)~$n=5$. Again, we can find a translation that corresponds to a $\mathcal{C}_3$ rotation about a single overlap site. Hence, there is no orthogonal solution in either case. (d)~$n=7$. The only translated states that overlap with the trial state on a single site do not correspond to a $\mathcal{C}_3$ rotation about that site. Therefore, we cannot immediately conclude that a compact Wannier basis with support on $3n=21$ sites does not exist.}
\label{fig: oddwanniertrials}
\end{figure}

\subsubsection{Proving inconsistency via symmetries} \label{sec: 6site_symmetries}
We can also show inconsistency directly, and without first solving a part of the equations as done in Sec.~\ref{subsec: direct_inconsistency_hexagon}, by exploiting the symmetries of the problem.
\begin{enumerate}[(I)]
\item{There are two ``internal" $U(2)$ symmetries: on each of the two inequivalent sites of the unit cell, $\mathrm{1b}$ and $\mathrm{1c}$, we may freely rotate between the two equivalent orbitals of $\mathcal{C}_3$ eigenvalue $1$.}
\item{There is one ``external" $U(2)$ symmetry that rotates between $\ket{w_b}$ and $\ket{w_c}$ (these have the same $\mathcal{C}_3$ eigenvalue $\lambda_b = \lambda_c = \omega$). The unitarily rotated Wannier states are guaranteed to satisfy all constraints, Eqs.~\eqref{eq: constraint4}-\eqref{eq: constraint3}.}
\end{enumerate}
We first use Symmetry (I) to set $a_2= a_5 = 0$ without loss of generality (that is, we orient our basis along the respective subvectors of $\ket{w_a}$). Eq.~\eqref{eq: constraint3} then gives
\begin{equation}
\begin{aligned}
|a_6|^2 |b_6|^2 &= |a_1|^2 |b_1|^2, \quad |a_3|^2 |b_3|^2 = |a_4|^2 |b_4|^2, \\
|a_6|^2 |c_6|^2 &= |a_1|^2 |c_1|^2, \quad |a_3|^2 |c_3|^2 = |a_4|^2 |c_4|^2.
\end{aligned}
\end{equation}
Together with Eq.~\eqref{eq: constraint4}, these imply
\begin{equation}
\begin{aligned}
|a_1|^2 |b_2|^2 + |a_4|^2 |b_5|^2 = 0, \\
|a_1|^2 |c_2|^2 + |a_4|^2 |c_5|^2 = 0.
\end{aligned}
\end{equation}
Since $a_1$ and $a_4$ cannot both be zero [Eq.~\eqref{eq: constraint4}], we get that at least $b_2 = c_2 = 0$ or $b_5 = c_5 = 0$. In either case, we may now use Symmetry (II) to ensure $b_1 c_1 = 0$ or $b_4 c_4 = 0$, respectively (in general, if two vectors are linearly dependent, we can unitarily rotate to a basis where one of the new vectors is the zero vector). But any solution of $b_1 c_1 = 0$ or $b_4 c_4 = 0$ gives a contradiction with Eq.~\eqref{eq: constraint4}.

\begin{figure}[t]
\centering
\includegraphics[width=1\textwidth,page=5]{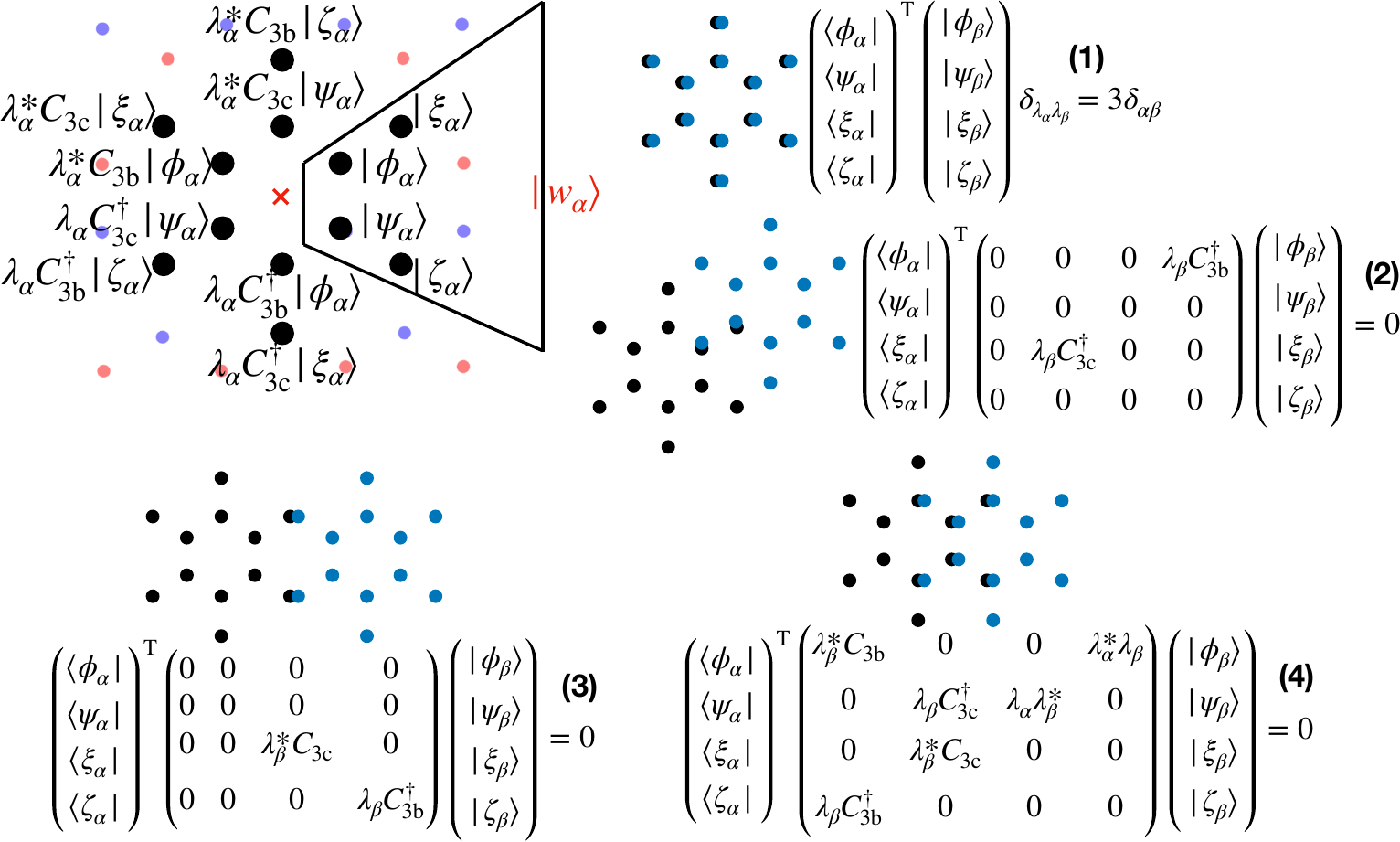}
\caption{$12$-site compactness constraints for the OAI shown in Fig.~\ref{fig: noncompactC3}a,b. Here, each site carries 3 orbitals with $\mathcal{C}_3$ eigenvalues $\{1,1,\omega\}$, so that the $\mathcal{C}_3$ matrices used in the constraint equations are given by $C_{3 \mathrm{b}} = C_{3 \mathrm{c}} = \mathrm{diag}(1,1,\omega)$. Then, we decompose the asymmetric part of the Wannier states as $\ket{w_\alpha} = (\ket{\phi_\alpha},\ket{\psi_\alpha},\ket{\xi_\alpha},\ket{\zeta_\alpha})$, where each component is a $3$-dimensional complex vector. The resultant symmetric Wannier states have $\mathcal{C}_3$ eigenvalues $\lambda_a = 1$ and $\lambda_b = \lambda_c = \omega$, respectively. The choice of state normalization to $\braket{\phi_\alpha | \phi_\alpha}+\braket{\psi_\alpha | \psi_\alpha}+\braket{\xi_\alpha | \xi_\alpha}+\braket{\zeta_\alpha | \zeta_\alpha}=3$ is made for convenience. This list of constraints is exhaustive, further translates only yield linearly dependent constraints.}
\label{fig: C3_star_wannier}
\end{figure}

\subsection{$3$-band OAI: $12$-site non-compactness} \label{sec: c3_12site_problem}
We have proven that an orthonormal $6$-site (hexagonal) Wannier basis for the $3$-band OAI depicted in Fig.~\ref{fig: noncompactC3}a,b does not exist. In generalization, we now investigate the $12$-site problem. By $\mathcal{C}_3$ symmetry, all trial states for a compact Wannier basis have overlap with exactly $3n$, $n \in \mathbb{Z}$, physical sites. We note that compact $9$-site Wannier states cannot exist: some translates would have a shared overlap on a single site, and can be obtained from a $\mathcal{C}_3$ rotation about that site. Then, as shown in Fig.~\ref{fig: oddwanniertrials}, orthogonality with translates necessitates a mobile set of atoms, which is not present here -- instead, each physical site carries the $\mathcal{C}_3$ representations $\{1,1,\omega\}$ of the site-symmetry group. Similarly, compact $3$-site and $15$-site Wannier states cannot exist. However, the same argument cannot be used to exclude all candidate Wannier states overlapping with an odd number of sites: starting at $21$ sites, the minimal shared overlap between translates is not anymore equivalent to a $\mathcal{C}_3$ rotation (see Fig.~\ref{fig: oddwanniertrials}).

\subsubsection{Compactness constraints}

Referring to Fig.~\ref{fig: C3_star_wannier}, we make the identification
\begin{equation}
\begin{aligned}
&\ket{w_\alpha} = (\ket{\phi_\alpha},\ket{\psi_\alpha},\ket{\xi_\alpha},\ket{\zeta_\alpha}), \\
&\ket{\phi_\alpha} = (\alpha_1, \alpha_2, \alpha_3)^\mathrm{T}, \quad \ket{\psi_\alpha} = (\alpha_4, \alpha_5, \alpha_6)^\mathrm{T}, \quad \ket{\xi_\alpha} = (\alpha_7, \alpha_8, \alpha_9)^\mathrm{T}, \quad \ket{\zeta_\alpha} = (\alpha_{10}, \alpha_{11}, \alpha_{12})^\mathrm{T}
\end{aligned}
\end{equation}
for $\alpha = a,b,c$. Here, $\ket{w_\alpha}$ denotes the asymmetric part out of which the $\mathcal{C}_3$-symmetric Wannier states are constructed. (The $\mathcal{C}_3$ eigenvalues are given by $\lambda_a = 1$ and $\lambda_b = \lambda_c = \omega$.)
Using the results of Sec.~\ref{sec: real_space_constraints_general}, the compactness constraints depicted in Fig.~\ref{fig: C3_star_wannier} can then be simplified to yield the following system of equations:

\begin{equation} \label{eq: eq11}
\begin{aligned}
3 = &|a_{1}|^2+|a_{2}|^2+|a_{3}|^2+|a_{4}|^2+|a_{5}|^2+|a_{6}|^2+3 |a_{9}|^2, \\
1 = &|a_{3} + \omega^* a_{12}|^2, \\
1 = &|a_{6} + \omega a_{9}|^2, \\
0 = &a_{4} a_{7}^*+a_{5} a_{8}^*+a_{10} a_{1}^*+a_{11} a_{2}^*+ \omega^* \left(a_{12} a_{3}^*+a_{6} a_{9}^*\right), \\
|a_{9}|^2 = &|a_{7}|^2+|a_{8}|^2+|a_{10}|^2+|a_{11}|^2 = |a_{12}|^2,
\end{aligned}
\end{equation}
In Eq.~\eqref{eq: eq11}, the last line follows from constraint (3) of Fig.~\ref{fig: C3_star_wannier} for $\alpha=\beta=a$, and the first line follows from this and constraint (1). The second to fourth lines follow from combining these with constraints (2) and (4).
\begin{equation} \label{eq: eq22}
\begin{aligned}
3 = &|b_{1}|^2+|b_{2}|^2+|b_{3}|^2+|b_{4}|^2+|b_{5}|^2+|b_{6}|^2+3 \left(|b_{7}|^2+|b_{8}|^2\right), \\
1 = &|b_{3}+b_{12}|^2+|b_{6}+b_{9}|^2, \\
1 = &|b_{4} + \omega^* b_{7}|^2+|b_{5} + \omega^* b_{8}|^2, \\
0 = &b_{4} b_{7}^*+b_{5} b_{8}^*+b_{10} b_{1}^*+b_{11} b_{2}^*+ \omega^* \left(b_{12} b_{3}^*+ b_{6} b_{9}^*\right), \\
|b_{7}|^2+|b_{8}|^2 = &|b_{10}|^2+|b_{11}|^2 = |b_{9}|^2+|b_{12}|^2,
\end{aligned}
\end{equation}
In Eq.~\eqref{eq: eq22}, the last line follows from constraint (3) of Fig.~\ref{fig: C3_star_wannier} for $\alpha=\beta=b$, and the first line follows from this and constraint (1). The second to fourth lines follow from combining these with constraints (2) and (4).
\begin{equation} \label{eq: eq33}
\begin{aligned}
3 = &|c_{1}|^2+|c_{2}|^2+|c_{3}|^2+|c_{4}|^2+|c_{5}|^2+|c_{6}|^2+3 \left(|c_{7}|^2+|c_{8}|^2\right), \\
1 = &|c_{3}+c_{12}|^2+|c_{6}+c_{9}|^2, \\
1 = &|c_{4} + \omega^* c_{7}|^2+|c_{5} + \omega^* c_{8}|^2, \\
0 = &c_{4} c_{7}^*+c_{5} c_{8}^*+c_{10} c_{1}^*+c_{11} c_{2}^*+ \omega^* \left(c_{12} c_{3}^*+ c_{6} c_{9}^*\right), \\
|c_{7}|^2+|c_{8}|^2 = &|c_{10}|^2+|c_{11}|^2 = |c_{9}|^2+|c_{12}|^2,
\end{aligned}
\end{equation}
In Eq.~\eqref{eq: eq33}, the last line follows from constraint (3) of Fig.~\ref{fig: C3_star_wannier} for $\alpha=\beta=c$, and the first line follows from this and constraint (1). The second to fourth lines follow from combining these with constraints (2) and (4).
\begin{equation} \label{eq: eq12}
\begin{aligned}
0 = &\left[b_1 (a_1-2 a_{10})^* + b_2 (a_2 - 2 a_{11})^* - b_{12} a_3^* \right] + \omega \left[(b_6 - 2b_9) a_6^* -b_4 a_7^* -b_5 a_8^* \right], \\
0 = &\left[b_4 (a_4 -2 a_7)^* + b_5 (a_5 - 2 a_8)^* -b_9 a_6^* \right] + \omega^* \left[(b_3 - 2 b_{12}) a_3^* - b_1 a_{10}^* - b_2 a_{11}^* \right], \\
0 = &b_{4} a_{7}^*+b_{5} a_{8}^*+b_{10} a_{1}^*+b_{11} a_{2}^*+ \omega^* \left(b_{12} a_{3}^*+b_{6} a_{9}^*\right), \\
0 = &b_{7} a_{4}^*+b_{8} a_{5}^*+b_{1} a_{10}^*+b_{2} a_{11}^*+ \omega \left(b_{3} a_{12}^*+b_{9} a_{6}^*\right) \\
0 = &b_{7} a_{7}^*+b_{8} a_{8}^*+\omega b_{12} a_{12}^*, \\
0 = &b_{10} a_{10}^*+b_{11} a_{11}^*+\omega^*b_{9} a_{9}^*,
\end{aligned}
\end{equation}
In Eq.~\eqref{eq: eq12}, the last two lines follow from combining constraint (3) of Fig.~\ref{fig: C3_star_wannier} for $(\alpha,\beta)=(a,b)$ with the complex-conjugated version of constraint (3) for $(\alpha,\beta)=(b,a)$. The first four lines follow from combining these with the constraints (2) and (4). Constraint (1) is satisfied trivially because the full Wannier states corresponding to $a$ and $b$ have different $\mathcal{C}_3$ eigenvalues.
\begin{equation} \label{eq: eq13}
\begin{aligned}
0 = &\left[c_1 (a_1-2 a_{10})^* + c_2 (a_2 - 2 a_{11})^* - c_{12} a_3^* \right] + \omega \left[(c_6 - 2 c_9) a_6^* - c_4 a_7^* - c_5 a_8^* \right], \\
0 = &\left[c_4 (a_4 -2 a_7)^* + c_5 (a_5 - 2 a_8)^* -c_9 a_6^* \right] + \omega^* \left[(c_3 - 2 c_{12}) a_3^* - c_1 a_{10}^* - c_2 a_{11}^* \right], \\
0 = &c_{4} a_{7}^*+c_{5} a_{8}^*+c_{10} a_{1}^*+c_{11} a_{2}^*+\omega^* \left(c_{12} a_{3}^*+c_{6} a_{9}^*\right), \\
0 = &c_{7} a_{4}^*+c_{8} a_{5}^*+c_{1} a_{10}^*+c_{2} a_{11}^*+\omega \left(c_{3} a_{12}^*+c_{9} a_{6}^*\right) \\
0 = &c_{7} a_{7}^*+c_{8} a_{8}^*+\omega c_{12} a_{12}^*, \\
0 = &c_{10} a_{10}^*+c_{11} a_{11}^*+\omega^*c_{9} a_{9}^*,
\end{aligned}
\end{equation}
In Eq.~\eqref{eq: eq13}, the last two lines follow from combining constraint (3) of Fig.~\ref{fig: C3_star_wannier} for $(\alpha,\beta)=(a,c)$ with the complex-conjugated version of constraint (3) for $(\alpha,\beta)=(c,a)$. The first four lines follow from combining these with the constraints (2) and (4). Constraint (1) is satisfied trivially because the full Wannier states corresponding to $a$ and $c$ have different $\mathcal{C}_3$ eigenvalues.
\begin{equation} \label{eq: eq23}
\begin{aligned}
0 = &c_{1} b_{1}^*+c_{2} b_{2}^*+c_{3} b_{3}^*+c_{4} b_{4}^*+c_{5} b_{5}^*+c_{6} b_{6}^*+3\left(c_{7} b_{7}^*+c_{8} b_{8}^*\right), \\
0 = &(c_{3} + c_{12})(b_{3} + b_{12})^*+(c_{6} + c_{9})(b_{6} + b_{9})^*, \\
0 = &(c_{4} + \omega^* c_{7})(b_{4} + \omega^* b_{7})^*+(c_{5} + \omega^* c_{8})(b_{5} + \omega^* b_{8})^*, \\
0 = &c_{7} b_{4}^*+c_{8} b_{5}^*+c_{1} b_{10}^*+c_{2} b_{11}^*+ \omega \left(c_{3} b_{12}^*+c_{9} b_{6}^*\right) \\
0 = &c_{4} b_{7}^*+c_{5} b_{8}^*+c_{10} b_{1}^*+c_{11} b_{2}^*+ \omega^* \left(c_{12} b_{3}^*+ c_{6} b_{9}^*\right), \\
c_{7} b_{7}^*+c_{8} b_{8}^* = &c_{10} b_{10}^*+c_{11} b_{11}^* = c_{9} b_{9}^*+c_{12} b_{12}^*.
\end{aligned}
\end{equation}
In Eq.~\eqref{eq: eq23}, the last line follows from combining constraint (3) of Fig.~\ref{fig: C3_star_wannier} for $(\alpha,\beta)=(b,c)$ with the complex-conjugated version of constraint (3) for $(\alpha,\beta)=(c,b)$. The first line follows from constraint (1). The second to fifth lines follow from combining these with constraints (2) and (4). 

We now want to determine if this system of equations has a solution.

\subsubsection{Symmetries of the problem}
The compactness constraints can be formally written as
\begin{equation}
\sum_{\alpha \beta} \bra{w_\alpha} N^{\lambda}_{\alpha \beta} \ket{w_\beta} = 0,
\end{equation}
where $N^{\lambda}_{\alpha \beta}$ captures the set of constraints [each line in Eqs.~\eqref{eq: eq11}-\eqref{eq: eq23} corresponds to at least one $\lambda$].

Let us recall the symmetries of the problem, which are the same as for the $6$-site constraints (Sec.~\ref{sec: 6site_symmetries}):

\begin{enumerate}[(I)]
\item{There are two ``internal" $U(2)$ symmetries, that is, operations commuting with all $N^{\lambda}_{\alpha \beta}$: on each of the two inequivalent sites of the unit cell, $\mathrm{1b}$ and $\mathrm{1c}$, we may freely rotate between the two equivalent orbitals of $\mathcal{C}_3$ eigenvalue $1$. It then follows that we can always choose a basis where two overlaps entering $\ket{w_\alpha}$ are set to zero from the outset, one for each set of equivalent orbitals.}
\item{There is one ``external" $U(2)$ symmetry that rotates between the Wannier states labelled by $b$ and $c$ (these have the same $\mathcal{C}_3$ eigenvalue $\lambda_b = \lambda_c = \omega$). The unitarily rotated Wannier states are guaranteed to satisfy the same constraints. This symmetry allows us to set an arbitrary element of $\ket{w_b}$ to zero, given that the corresponding element of $\ket{w_c}$ is nonzero, and only when this does not interfere with the basis choice made for (1).}
\end{enumerate}

\subsubsection{Proving inconsistency}
We first show that one of $a_{7 \dots 12}$, $b_{7 \dots 12}$, or $c_{7 \dots 12}$ must vanish. (The same method proves inconsistency of the $6$-site system, see Sec.~\ref{sec: 6site_symmetries}.) We use Symmetry (I) to set $a_{8} = a_{11} = 0$ from the outset. Then, Eqs.~\eqref{eq: eq11},~\eqref{eq: eq22}, and~\eqref{eq: eq12} give
\begin{equation} \label{eq: outzeroproof}
\begin{aligned}
|a_{9}|^2 = &|a_{7}|^2+|a_{10}|^2 = |a_{12}|^2 \\
|b_{7}|^2+|b_{8}|^2 = &|b_{10}|^2+|b_{11}|^2 = |b_{9}|^2+|b_{12}|^2 \\
|b_{7}|^2 |a_{7}|^2 = & |b_{12}|^2 |a_{12}|^2, \\
|b_{10}|^2 |a_{10}|^2 = &|b_{9}|^2 |a_{9}|^2.
\end{aligned}
\end{equation}
These can be manipulated to yield
\begin{equation}
|a_7|^2 |b_8|^2 + |a_{10}|^2 |b_{11}|^2 = 0.
\end{equation}
By the unspoiled Symmetry (II) between $b$ and $c$, we also have
\begin{equation}
|a_7|^2 |c_8|^2 + |a_{10}|^2 |c_{11}|^2 = 0.
\end{equation}
There are two options:
\begin{enumerate}[a.]
\item{$a_7 = a_{10} = 0$. In this case, all $a_{7 \dots 12} = 0$ have to vanish by Eq.~\eqref{eq: outzeroproof}.}
\item{$b_8 = c_8 = 0$ or $b_{11} = c_{11} = 0$. In this case, we can use Symmetry (II) to also set $b_7 = 0$ or $b_{10} = 0$ without loss of generality. But in both cases, $b_{7 \dots 12} = 0$ follows from Eq.~\eqref{eq: outzeroproof}.}
\end{enumerate}
We conclude that one of $a_{7 \dots 12}$ or $b_{7 \dots 12}$ must vanish (if instead $c_{7 \dots 12} = 0$, we can always relabel $c \leftrightarrow b$).

\paragraph{$\mathbf{a_{7 \dots 12} = 0}$}
Let us now use Symmetry (I) to set $a_2 = a_5 = 0$. Then, Eqs.~\eqref{eq: eq11},~\eqref{eq: eq22}, and~\eqref{eq: eq12} give
\begin{equation} \label{eq: aoutzero_followup}
\begin{aligned}
1 = &|a_{1}|^2+|a_{4}|^2=|a_{3}|^2= |a_{6}|^2, \\
|b_{7}|^2+|b_{8}|^2 = &|b_{10}|^2+|b_{11}|^2 = |b_{9}|^2+|b_{12}|^2 \\
|b_{10}|^2 |a_{1}|^2 = &|b_{12}|^2 |a_{3}|^2, \\
|b_{7}|^2 |a_{4}|^2= & |b_{9}|^2 |a_{6}|^2,
\end{aligned}
\end{equation}
from which we deduce
\begin{equation}
|a_4|^2 |b_8|^2 + |a_1|^2 |b_{11}|^2 = 0.
\end{equation}
By the unspoiled symmetry between $b \leftrightarrow c$, we also get
\begin{equation}
|a_4|^2 |c_8|^2 + |a_1|^2 |c_{11}|^2 = 0.
\end{equation}
Now, $a_1$ and $a_4$ cannot both be zero by Eq.~\eqref{eq: aoutzero_followup}, and so there are two inequivalent solutions: either (1) $b_{8} = c_{8} = b_{11} = c_{11} = 0$, or (2) $a_4 = b_{11} = c_{11} = 0$ (the remaining solution $a_1 = b_{8} = c_{8} = 0$ is clearly related). In both cases, we can use Symmetry (II) and Eq.~\eqref{eq: aoutzero_followup} to set $b_{7 \dots 12} = 0$ without loss of generality.

From Eqs.~\eqref{eq: eq11},~\eqref{eq: eq22}, and~\eqref{eq: eq12} we then get
\begin{equation} \label{eq: aoutzeroboutzero_followup}
\begin{aligned}
1 = &|a_{1}|^2+|a_{4}|^2=|a_{3}|^2= |a_{6}|^2, \\
1 = &|b_{1}|^2+|b_{2}|^2= |b_{3}|^2+|b_{6}|^2= |b_{4}|^2+|b_{5}|^2, \\
|b_{1}|^2 |a_{1}|^2 =& |b_{6}|^2 |a_{6}|^2, \\
|b_{4}|^2 |a_{4}|^2 =& |b_{3}|^2 |a_{3}|^2,
\end{aligned}
\end{equation}
which enforces
\begin{equation} \label{eq: aoutzeroboutzero_followup2}
|a_1|^2 |b_2|^2 + |a_4|^2 |b_5|^2 = 0.
\end{equation}

Now, we can start treating the two cases mentioned earlier separately:

(1) $c_{8} = c_{11} = 0$: Eq.~\eqref{eq: aoutzeroboutzero_followup2} implies $b_2 = b_5 = 0$ [or else $a_4 =0$, which is treated in the following paragraph-- recall that the case $a_1 =0$ is related, and that both $a_4$ and $a_1$ cannot be zero by Eq.~\eqref{eq: aoutzeroboutzero_followup}]. Then, we observe that Eqs.~\eqref{eq: eq12}-\eqref{eq: eq23}, together with Eq.~\eqref{eq: aoutzeroboutzero_followup}, now imply
\begin{equation}
\begin{aligned}
|a_{4}|^2 =& |b_3|^2 = 1-|b_6|^2 \\
|c_9|^2 =& |c_7|^2 |a_{4}|^2 \\
|c_7|^2 =& |c_9|^2 |b_6|^2.
\end{aligned}
\end{equation}
We deduce that $|c_7|^2 = |c_7|^2 (1-|b_6|^2) |b_6|^2$, and so $c_7 = 0$ must hold [the function $(1-x^2)x^2$ is strictly smaller than $1$ for all real $x\leq0$]. Together with $c_{8} = 0$, this implies $c_{7 \dots 12} = 0$ by Eq.~\eqref{eq: eq33}. In conclusion, we have $a_{7 \dots 12} = 0$, $b_{7 \dots 12} = 0$, and $c_{7 \dots 12} = 0$, the problem therefore reduces to the $6$-site system that we know to be inconsistent. 

(2) $a_{4} = c_{11} = 0$:
From Eqs.~\eqref{eq: aoutzeroboutzero_followup} and~\eqref{eq: aoutzeroboutzero_followup2}, it follows that $b_2 = b_3 = 0$. But then Eq.~\eqref{eq: eq23} enforces $c_{10} = 0$ in its second-to-last line, which together with $c_{11} = 0$ implies $c_{7 \dots 12} = 0$ by Eq.~\eqref{eq: eq33}. The problem again reduces to the $6$-site system that we know to be inconsistent. 

\paragraph{$\mathbf{b_{7 \dots 12} = 0}$}
We keep $a_{8} = a_{11} = 0$. Then, Eqs.~\eqref{eq: eq11}-\eqref{eq: eq13} give
\begin{equation} \label{eq: bout_followup}
\begin{aligned}
|a_{9}|^2 = &|a_{7}|^2+|a_{10}|^2 = |a_{12}|^2 \\
1 = &|b_{1}|^2+|b_{2}|^2= |b_{3}|^2+|b_{6}|^2= |b_{4}|^2+|b_{5}|^2, \\
|b_{4}|^2 |a_{7}|^2=&|b_{6}|^2 |a_{9}|^2, \\
|b_{1}|^2 |a_{10}|^2=&|b_{3}|^2 |a_{12}|^2 \\
|c_{7}|^2+|c_{8}|^2 = &|c_{10}|^2+|c_{11}|^2 = |c_{9}|^2+|c_{12}|^2 \\
|c_{7}|^2 |a_{7}|^2 = & |c_{12}|^2 |a_{12}|^2, \\
|c_{10}|^2 |a_{10}|^2 = &|c_{9}|^2 |a_{9}|^2.
\end{aligned}
\end{equation}

These equations can be manipulated to yield
\begin{equation}
\begin{aligned}
|a_{7}|^2 |b_5|^2 + |a_{10}|^2 |b_2|^2 &= 0, \\
|a_{7}|^2 |c_8|^2 + |a_{10}|^2 |c_{11}|^2 &= 0.
\end{aligned}
\end{equation}
We already know that setting $a_7 = a_{10} = 0$ leads to a contradiction. So there are two possibilities: either (1) $b_5 = c_8 = b_2 = c_{11} = 0$, or (2) $a_7 = b_2 = c_{11} = 0$ (the remaining solution $a_{10} = b_{5} = c_{8} = 0$ is clearly related). Note that here we cannot use Symmetry (II) anymore to make progress, because we have already used it to set $b_{7 \dots 12} = 0$.

(1) $b_5 = c_8 = b_2 = c_{11} = 0$:
Eqs.~\eqref{eq: bout_followup} and~\eqref{eq: eq23} give
\begin{equation} \label{eq: bout_followup2}
\begin{aligned}
|a_{10}|^2 = &|b_3|^2 |a_9|^2,\\
|c_9|^2 |a_9|^2 = &|c_7|^2 |a_{10}|^2, \\
|c_7|^2 = &|c_9|^2 |b_6|^2, \\
\end{aligned}
\end{equation}
from which we deduce $|c_9|^2 |a_9|^2 = |c_9|^2 |a_9|^2 (1-|b_6|^2) |b_6|^2$, and so $|c_9|^2 |a_9|^2 = 0$. Since $a_9 = 0$ sets $a_{7 \dots 12} = 0$ via Eq.~\eqref{eq: bout_followup}, we must choose $c_9 = 0$. Then, by Eqs.~\eqref{eq: bout_followup} and~\eqref{eq: bout_followup2}, we also have that $c_7 = c_{12} = 0$, and so $c_{7 \dots 12} = 0$. Using Eq.~\eqref{eq: eq23}, it follows that $c_1 = c_4 = 0$. Finally, Eq.~\eqref{eq: eq13} yields $c_3 = c_6 = 0$ unless $a_{7 \dots 12} = 0$, in glaring inconsistency with Eq.~\eqref{eq: eq33}.

(2) $a_7 = b_2 = c_{11} = 0$:
From Eq.~\eqref{eq: eq12} and~\eqref{eq: eq13} it follows that $c_{12} = b_6 = 0$ (again, unless $a_{7 \dots 12} = 0$, which we know to yield an inconsistency), and so $c_{7 \dots 12} = 0$ by the second-to-last line of Eq.~\eqref{eq: eq23}. Eq.~\eqref{eq: eq23} also enforces $c_1 = 0$. From Eq.~\eqref{eq: eq13} we again get $c_3 = c_6 = 0$ (unless $a_{7 \dots 12} = 0$), in contradiction with Eq.~\eqref{eq: eq33}.

\subsubsection{Conclusion}
To summarize, we have shown that the combination of all $12$-site compactness constraints inevitably leads to a contradiction. If it exists at all, any compact Wannier basis of the 3-band OAI shown in Fig.~\ref{fig: noncompactC3}a,b must therefore have overlap with at least $18$ physical sites.

\section{Representation obstructions} \label{sec: repob}
In this section, we discuss representation-obstructed OAIs (defined at the beginning of Sec.~\ref{sec: OAI_mc_generalities}). The Wannier centers of such OAIs coincide with the positions of the physical orbitals of the crystal. However, the Wannier site-symmetry representation does not coincide with that of any physical orbital on the same site -- otherwise, the OAI would be unobstructed. First, we note that the proof for $\mathcal{C}_2$ symmetry in Sec.~\ref{sec: nogo_p2symmetry} does not distinguish between spatial and representation obstructions, and so applies equally well to both cases. However, the analyses of Secs.~\ref{sec: overarching_p4symmetry} and~\ref{sec: overarching_c3}, treating $\mathcal{C}_4$- and $\mathcal{C}_3$-symmetric OAIs, respectively, both exclusively apply to spatial obstruction. Here, we point out some salient examples where the compactness properties of $\mathcal{C}_4$- and $\mathcal{C}_3$-symmetric OAIs with spatial and representation obstructions differ.

We note that it is not possible to determine whether an atomic limit has a spatial or representation obstruction from symmetry data alone -- we must also know the orbitals of the crystalline lattice $\Lambda$. Then, we can find the symmetry data of the atomic insulator $A$ from the relation
\begin{equation}
B[A] = B[\Lambda] - B[\mathrm{FP}],
\end{equation}
where $\mathrm{FP}$ is the fragile band complement of the atomic insulator $A$ and $B[X]$ is the symmetry data vector of the set of bands $X$~\cite{ZhidaFragileTwist2}, which encodes its BZ irrep multiplicities. Now, for the case of spinless time-reversal symmetry which we focus on in this work, knowledge of $B[A]$ uniquely determines the Wannier centers of $A$ (\href{https://journals.aps.org/prb/abstract/10.1103/PhysRevB.99.245151}{PRB 99, 245151}). If these do not align with the atomic positions of the crystalline lattice, then $A$ has a spatial obstruction. If they do align, then $A$ must have a representation obstruction because otherwise $\mathrm{FP}$ would not be fragile (we could simply subtract $A$ from $\Lambda$ to obtain another a lattice with fewer orbitals in its unit cell). 

\subsection{$\mathcal{C}_4$ symmetry}
Consider the family of $\mathcal{C}_4$-symmetric fragile root states
\begin{equation} \label{eq: c4simpleREPobstruction}
\left[(\gamma)_\mathrm{1b} \oplus (-\gamma)_\mathrm{1b} \oplus (-\gamma)_\mathrm{1a}\right] \uparrow G \ominus (\gamma)_\mathrm{1a} \uparrow G= \mathrm{FP},
\end{equation}
where $\gamma \in \{1,-1,\mathrm{i},-\mathrm{i}\}$ is a free parameter. Here, $\mathrm{FP}$ is the band complement of a representation-obstructed OAI at the 1a Wyckoff position. The OAI $(\gamma)_\mathrm{1a} \uparrow G$ is non-compact: since the physical lattice does not host any $s \oplus p$ pairs at the same Wyckoff position (recall that both $\mathcal{C}_4$ eigenvalues $\gamma$ and $-\gamma$ correspond to the same $\mathcal{C}_2$ eigenvalue $\gamma^2$), the overlap between translated copies of the OAI Wannier state, which necessarily has non-zero range due to the representation obstruction, cannot be made to vanish (see Fig.~\ref{fig: compactInvObstruction} and Sec.~\ref{subsec: c2_sp_realspace_crit}). Nevertheless, when relaxing $\mathcal{C}_4$ symmetry to $\mathcal{C}_2$ symmetry, the orbitals $(-\gamma)_\mathrm{1a}$ and $(\gamma)_\mathrm{1a}$ both turn into the same orbital $(\gamma^2)_\mathrm{1a}$ (now labelled by its $\mathcal{C}_2$ eigenvalue), so that we can choose the OAI Wannier state to be located on just a single orbital (zero range): the OAI in Eq.~\eqref{eq: c4simpleREPobstruction} hence becomes unobstructed. Consequently, we can find $\mathcal{C}_2$-symmetric compact Wannier states for the OAI $(\gamma)_\mathrm{1a} \uparrow G$ [these are given in Eq.~\eqref{eq: c2compact_covalents}], but not $\mathcal{C}_4$-symmetric ones.
This finding is in stark contrast to the case of spatially-obstructed OAIs with $\mathcal{C}_4$ symmetry: recall that for these, we showed in Sec.~\ref{sec: overarching_p4symmetry} that they are only non-compact as long as they remain so upon the relaxation of $\mathcal{C}_4$ symmetry to $\mathcal{C}_2$ symmetry.

It is interesting to note that the fragile phase $\mathrm{FP}$ appearing in Eq.~\eqref{eq: c4simpleREPobstruction} is directly related to the fragile phase appearing in Eq.~\eqref{eq: c4_2band_subtraction}: we have
\begin{equation}
\begin{aligned}
\mathrm{FP} &= \left[(\gamma)_\mathrm{1b} \oplus (-\gamma)_\mathrm{1b} \oplus (-\gamma)_\mathrm{1a}\right] \uparrow G \ominus (\gamma)_\mathrm{1a} \uparrow G \\
&= \left[(\gamma)_\mathrm{1b} \oplus (-\gamma)_\mathrm{1b} \oplus (-\gamma)_\mathrm{1a} \oplus (\mathrm{i}\gamma)_\mathrm{1b} \oplus (-\mathrm{i}\gamma)_\mathrm{1b} \right] \uparrow G \ominus \left[(\mathrm{i}\gamma)_\mathrm{1b} \oplus (-\mathrm{i}\gamma)_\mathrm{1b} \oplus (\gamma)_\mathrm{1a}\right] \uparrow G\\
&= \left[(\gamma)_\mathrm{1a} \oplus (-\gamma)_\mathrm{1a} \oplus (-\gamma)_\mathrm{1a} \oplus (\mathrm{i}\gamma)_\mathrm{1a} \oplus (-\mathrm{i}\gamma)_\mathrm{1a} \right] \uparrow G \ominus \left[(\mathrm{i}\gamma)_\mathrm{1b} \oplus (-\mathrm{i}\gamma)_\mathrm{1b} \oplus (\gamma)_\mathrm{1a}\right] \uparrow G
\\
&= \left[(-\gamma)_\mathrm{1a} \oplus (-\gamma)_\mathrm{1a} \oplus (\mathrm{i}\gamma)_\mathrm{1a} \oplus (-\mathrm{i}\gamma)_\mathrm{1a} \right] \uparrow G \ominus \left[(\mathrm{i}\gamma)_\mathrm{1b} \oplus (-\mathrm{i}\gamma)_\mathrm{1b}\right] \uparrow G,
\end{aligned}
\end{equation}
which is equal to Eq.~\eqref{eq: c4_2band_subtraction} upon the exchange $\mathrm{1a} \leftrightarrow \mathrm{1b}$ and the substitution $\gamma \rightarrow -\mathrm{i}\lambda$. Here, we have made use of the Wyckoff position-independence of mobile cluster orbitals (Sec.~\ref{subsec: mobile_clusters}), which for $\mathcal{C}_4$ symmetry implies that 
\begin{equation}
\left[(\gamma)_\mathrm{1a} \oplus (-\gamma)_\mathrm{1a} \oplus (\mathrm{i}\gamma)_\mathrm{1a} \oplus (-\mathrm{i}\gamma)_\mathrm{1a} \right] \uparrow G = \left[(\gamma)_\mathrm{1b} \oplus (-\gamma)_\mathrm{1b} \oplus (\mathrm{i}\gamma)_\mathrm{1b} \oplus (-\mathrm{i}\gamma)_\mathrm{1b} \right] \uparrow G,
\end{equation}
for any choice of $\gamma$.
We therefore find that the same fragile state can have both a non-compact, representation-obstructed $1$-band OAI complement, or a compact, spatially-obstructed $2$-band OAI complement.

One might ask if every representation-obstructed OAI with a fragile complement must be non-compact. That this is not the case can be seen by considering
\begin{equation} \label{eq: c4compactREPobstruction}
\left[(\gamma)_\mathrm{1b} \oplus (-\gamma)_\mathrm{1b} \oplus (-\gamma)_\mathrm{1a} \oplus (\mathrm{i}\gamma)_\mathrm{1a}\right] \uparrow G \ominus (\gamma)_\mathrm{1a} \uparrow G= \mathrm{FP} \oplus (\mathrm{i}\gamma)_\mathrm{1a} \uparrow G,
\end{equation}
where we added the orbital $(\mathrm{i}\gamma)_\mathrm{1a} \uparrow G$ on both sides of Eq.~\eqref{eq: c4simpleREPobstruction}. Here, the OAI $(\gamma)_\mathrm{1a} \uparrow G$ is still representation-obstructed, while its complement is still fragile (although it is not anymore a fragile root state). However, we can now find a compact Wannier state that (at $\bs{R} = \bs{0}$) reads 
\begin{equation} \begin{aligned}
\ket{W_{\bs{0} \gamma}} &= \frac{1}{4} \left[\ket{w_{\bs{0} \gamma}} + \gamma^* C_4|_{\bs{0},\mathrm{1a}} \ket{w_{\bs{0} \gamma}} + (\gamma^* C_4|_{\bs{0},\mathrm{1a}})^2 \ket{w_{\bs{0} \gamma}} + (\gamma^* C_4|_{\bs{0},\mathrm{1a}})^3 \ket{w_{\bs{0} \gamma}} \right], \\
\ket{w_{\bs{0} \gamma}} &= \ket{\bs{0}, (\gamma)_\mathrm{1b}} + \ket{\bs{0}, (-\gamma)_\mathrm{1b}} + \ket{\bs{a}_1, (-\gamma)_\mathrm{1a}} + \ket{\bs{a}_1, (\mathrm{i}\gamma)_\mathrm{1a}}.
\end{aligned} \end{equation}
Here $\ket{\bs{R}, (\mu)_i}$ denotes the basis state for the orbital with $\mathcal{C}_4$ eigenvalue $\mu$ at Wyckoff position $i$ of the unit cell at $\bs{R}$, and $C_4|_{\bs{0},\mathrm{1a}}$ implements a $\mathcal{C}_4$ rotation about Wyckoff position 1a at the origin.

In conclusion, the requirement for representation-obstructed OAIs to be compact generally imposes stronger constraints than for spatially-obstructed OAIs: the same set of fragile bands may be the complement of a compact spatially-obstructed OAI, but not that of a compact representation-obstructed OAI. However, we have also seen that some fragile phases admit compact OAI complements of both types, while others admit no compact complement of any type (an example for the latter class are fragile phases with $\mathcal{C}_2$ symmetry).

\begin{figure}[t]
\centering
\includegraphics[width=0.7\textwidth,page=13]{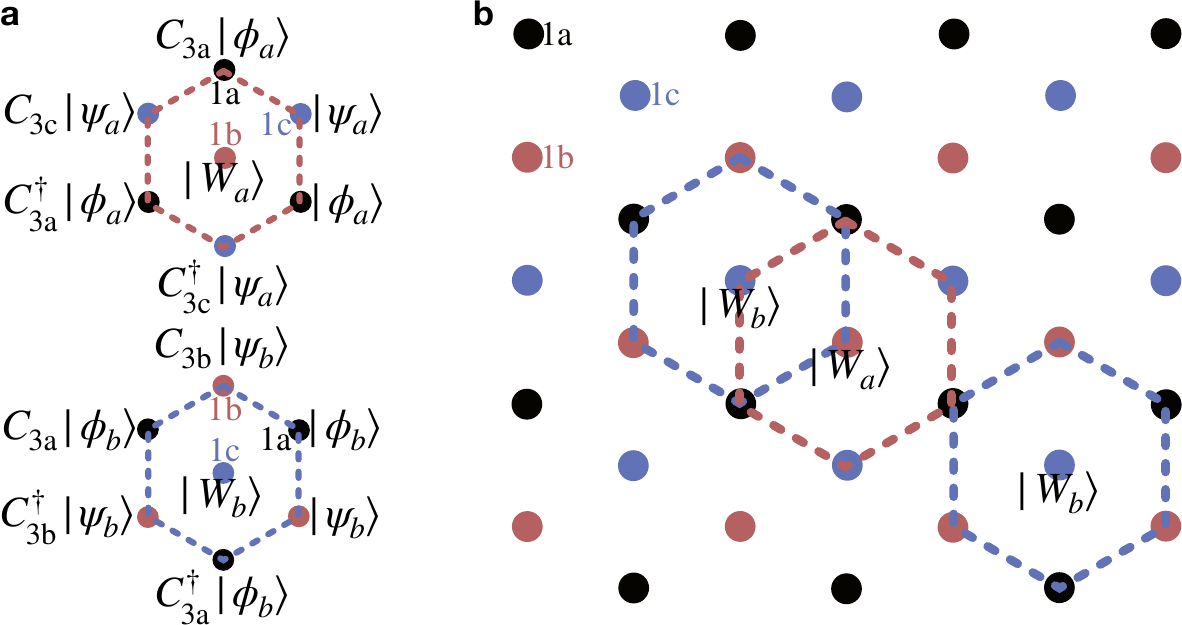}
\caption{Trial Wannier states for a $2$-band representation-obstructed OAI with $\mathcal{C}_3$ symmetry, overlapping with $6$ lattice sites. (a)~The two trial states, $\ket{W_a}$ and $\ket{W_b}$, are centered at the 1b and 1c Wyckoff positions of the unit cell, respectively, and both carry $\mathcal{C}_3$ eigenvalue $1$. When acting on the atomic orbitals, $\mathcal{C}_3$ is represented by the matrices $C_{3 \mathrm{a}} = \mathrm{diag}(1,1,\omega)$, $C_{3 \mathrm{b}} = C_{3 \mathrm{c}} = \mathrm{diag}(\omega)$ [see Eq.~\eqref{eq: repobC32bandsubtraction} for $\gamma = 1$]. (b)~Two examples of translated copies of $\ket{W_b}$ that have a nonzero overlap with $\ket{W_a}$. These two overlaps, together with the constraint that each individual trial state be orthogonal to its translates, are sufficient to rule out a compact Wannier basis.}
\label{fig: C3RO}
\end{figure}

\subsection{$\mathcal{C}_3$ symmetry}
We next discuss representation obstructions with $\mathcal{C}_3$ symmetry. Here, there are no representation-obstructed $1$-band OAIs: changing the $\mathcal{C}_3$ eigenvalue of an orbital at any maximal Wyckoff position also changes the $\mathcal{C}_3$ eigenvalue of the induced band at all high-symmetry momenta of the Brillouin zone. Therefore, lattice bands induced from any given Wyckoff position cannot be used to construct $\mathcal{C}_3$-symmetric OAIs at the same position. [Contrast this with the case of $\mathcal{C}_4$ symmetry: an $s$ and a $d_{x^2-y^2}$ orbital at 1a (1b) both yield a $\mathcal{C}_2$ eigenvalue $+1$ at the high-symmetry momentum $\bs{X}$, this fact allows for the construction of the representation-obstructed $1$-band OAI of Eq.~\eqref{eq: c4simpleREPobstruction} (when $\gamma = 1$).]

Consider then the family of representation-obstructed $2$-band OAIs
\begin{equation} \label{eq: repobC32bandsubtraction}
\left[(\omega \gamma)_\mathrm{1b} \oplus (\omega \gamma)_\mathrm{1c} \oplus 2(\gamma)_\mathrm{1a} \oplus (\omega \gamma)_\mathrm{1a}\right] \uparrow G \ominus \left[(\gamma)_\mathrm{1b} \oplus (\gamma)_\mathrm{1c} \right] \uparrow G= \mathrm{FP},
\end{equation}
Let us set $\gamma = 1$ for clarity and without loss of generality. We define the trial Wannier states of the OAI as $\ket{W_\alpha}$, where $\alpha = a$ corresponds to the orbital $(1)_\mathrm{1b}$ while $\alpha = b$ corresponds to the orbital $(1)_\mathrm{1c}$. Assuming trial states that overlap with $6$ lattice sites and implementing $\mathcal{C}_3$ symmetry, we obtain the states shown in Fig~\ref{fig: C3RO}~(a), where $\ket{\Psi_\alpha}$ are one-dimensional vectors pertaining to the overlap of $\ket{W_\alpha}$ with the orbital $(\omega)_\mathrm{1c}$ for $\alpha = a$, or $(\omega)_\mathrm{1b}$ for $\alpha = b$. Furthermore,
\begin{equation}
\ket{\phi_\alpha} = (\alpha_1,\alpha_2,\alpha_3)
\end{equation}
are $3$-dimensional vectors containing the overlaps of $\ket{W_\alpha}$ with the two $(1)_\mathrm{1a}$ orbitals (equal to $\alpha_1$ and $\alpha_2$) and the orbital $(\omega)_\mathrm{1a}$ (equal to $\alpha_3$).
We know from Secs.~\ref{sec: c3_1band}-\ref{sec: c3_3band_6_site_overarching}, first line of Eq.~\eqref{eq: constraint4}, that for $\ket{W_\alpha}$ to be individually compact, it is required that
\begin{equation} \label{eq: repobindividualC3compactnesscondition}
|\alpha_1|^2+|\alpha_2|^2 = |\alpha_3|^2 = \braket{\Psi_\alpha |\Psi_\alpha}.
\end{equation}
Then, without loss of generality, we can set $\ket{\phi_a} = (1,0,1)$. Here, we fixed the state normalization, exploited the $U(2)$ symmetry that rotates between the two $(\gamma)_\mathrm{1a}$ orbitals [see also Symmetry (I) in Sec.~\ref{sec: 6site_symmetries}] to set $a_2 = 0$, and absorbed all phase factors by a redefinition of the Hilbert space basis states. The condition that the two overlaps shown in Fig~\ref{fig: C3RO}~(b) vanish, as required by Wannier state orthogonality, then translates to 
\begin{equation}
b_1 + \omega b_3 = 0, \quad 2b_1 + (1+\omega^*) b_3 = 0 \quad \rightarrow \quad b_1 = b_3 = 0,
\end{equation}
where we used the $\mathcal{C}_3$ representation matrix $C_{3 \mathrm{a}}=\mathrm{diag}\{1,1,\omega\}$. But this result is incompatible with Eq.~\eqref{eq: repobindividualC3compactnesscondition} (Wannier states cannot have zero norm), so that the OAI $\left[(\gamma)_\mathrm{1b} \oplus (\gamma)_\mathrm{1c} \right] \uparrow G$ in Eq.~\eqref{eq: repobC32bandsubtraction} does not admit compact Wannier states that overlap with $6$ lattice sites. This finding is in stark contrast to the case of spatially obstructed $2$-band OAIs with $\mathcal{C}_3$ symmetry, all of which were shown in Sec.~\ref{sec: c3_2band} to admit a $6$-site Wannier basis. The system of equations constraining $12$-site Wannier states (see Sec.~\ref{sec: c3_12site_problem}) for the representation-obstructed $2$-band OAI in Eq.~\eqref{eq: repobC32bandsubtraction} becomes much more involved, and we have not yet found a solution to it, nor proven its inconsistency.

Interestingly, the fragile state $\mathrm{FP}$ in Eq.~\eqref{eq: repobC32bandsubtraction} can be related to the fragile state in Eq.~\eqref{eq: 3bandC3noncompact_subtraction}: we have
\begin{equation}
\begin{aligned}
\mathrm{FP} =& \left[(\omega \gamma)_\mathrm{1b} \oplus (\omega \gamma)_\mathrm{1c} \oplus 2(\gamma)_\mathrm{1a} \oplus (\omega \gamma)_\mathrm{1a}\right] \uparrow G \ominus \left[(\gamma)_\mathrm{1b} \oplus (\gamma)_\mathrm{1c} \right] \uparrow G \\
=& \left[(\omega \gamma)_\mathrm{1b} \oplus (\omega \gamma)_\mathrm{1c} \oplus 2(\gamma)_\mathrm{1a} \oplus 2(\omega \gamma)_\mathrm{1a} \oplus 2(\omega^* \gamma)_\mathrm{1a}\right] \uparrow G \ominus \left[(\gamma)_\mathrm{1b} \oplus (\gamma)_\mathrm{1c} \oplus (\omega \gamma)_\mathrm{1a} \oplus 2(\omega^* \gamma)_\mathrm{1a} \right] \uparrow G \\
=& \left[(\omega \gamma)_\mathrm{1b} \oplus (\omega \gamma)_\mathrm{1c} \oplus (\gamma)_\mathrm{1b} \oplus (\omega \gamma)_\mathrm{1b} \oplus (\omega^* \gamma)_\mathrm{1b} \oplus (\gamma)_\mathrm{1c} \oplus (\omega \gamma)_\mathrm{1c} \oplus (\omega^* \gamma)_\mathrm{1c}\right] \uparrow G \\&\ominus \left[(\gamma)_\mathrm{1b} \oplus (\gamma)_\mathrm{1c} \oplus (\omega \gamma)_\mathrm{1a} \oplus 2(\omega^* \gamma)_\mathrm{1a} \right] \uparrow G \\
=& \left[2(\omega \gamma)_\mathrm{1b} \oplus (\omega^* \gamma)_\mathrm{1b} \oplus 2(\omega \gamma)_\mathrm{1c} \oplus (\omega^* \gamma)_\mathrm{1c}\right] \uparrow G \ominus \left[(\omega \gamma)_\mathrm{1a} \oplus 2(\omega^* \gamma)_\mathrm{1a} \right] \uparrow G,
\end{aligned}
\end{equation}
which is equal to Eq.~\eqref{eq: 3bandC3noncompact_subtraction} for the choice $\gamma = \omega^*$. Here, we have made use of the Wyckoff position-independence of mobile cluster orbitals (Sec.~\ref{subsec: mobile_clusters}). 

We conclude that the same fragile state can be expressed as the complement of a representation-obstructed $2$-band OAI, or a spatially-obstructed $3$-band OAI. In both cases, there is an obstruction to find a compact Wannier basis for the OAI, at least for short-ranged Wannier states.

\bibliography{references}